\documentclass[twocolumn,amsmath,amssymb,floatfix,pre,floatfix]{revtex4-1}

\usepackage{graphicx,color}
\definecolor{brown}{rgb}{0.63,0.27,0.18}
\definecolor{orange}{rgb}{1.00,0.65,0.00}

\usepackage{afterpage} 

\usepackage{moresize}
\usepackage{dcolumn}
\usepackage{bm,multirow}

\makeatletter
\newcommand*{\balancecolsandclearpage}{%
  \close@column@grid
  \twocolumngrid
}
\makeatother

\newcommand{\be}{\begin{equation}}
\newcommand{\ee}{\end{equation}}
\usepackage{color}

\begin{document}

\newcommand {\rsq}[1]{\left< R^2 (#1)\right.}
\newcommand {\rsqL}{\left< R^2 (L) \right>}
\newcommand {\rsqbp}{\left< R^2 (N_{bp}) \right>}
\newcommand {\Nbp}{N_{bp}}
\newcommand {\etal}{{\em et al.}}
\newcommand{\Ham}{{\cal H}}
\newcommand{\AngeloComment}[1]{\textcolor{red}{(AR) #1}}
\newcommand{\AndreaComment}[1]{\textcolor{blue}{A: #1}}
\newcommand{\JanComment}[1]{\textcolor{green}{(JS) #1}}

\newcommand{\NewText}[1]{\textcolor{orange}{#1}}
\newcommand{\scs}{\ssmall}

\newcommand{\Tau}{\mathrm{T}}



\title{Computer simulations of melts of ring polymers with non-conserved topology: A dynamic Monte Carlo lattice model} 

\author{Mattia Alberto Ubertini}
\email{mubertin@sissa.it}
\affiliation{Scuola Internazionale Superiore di Studi Avanzati (SISSA), Via Bonomea 265, 34136 Trieste, Italy}

\author{Angelo Rosa}
\email{anrosa@sissa.it}
\affiliation{Scuola Internazionale Superiore di Studi Avanzati (SISSA), Via Bonomea 265, 34136 Trieste, Italy}

\date{\today}

\begin{abstract}
We present computer simulations of a dynamic Monte Carlo algorithm for polymer chains on the FCC lattice which takes explicitly into account the possibility to overcome topological constraints by controlling the rate at which nearby polymer strands may cross through each other.
By applying the method to systems of interacting ring polymers at melt conditions, we characterize their structure and dynamics by measuring, in particular, the amounts of knots and links which are formed during the relaxation process.
In comparison to standard melts of unknotted and unconcatenated rings, our simulations demonstrate that the mechanism of strand crossing is responsible for fluidizing the melt provided the time scale of the process is faster than the internal relaxation of the chain,
in agreement with recent experiments employing solutions of DNA rings in the presence of the type II topoisomerase enzyme.
In the opposite case of slow rates the melt is shown to become slower, and this prediction may be easily validated experimentally.
\end{abstract}

\maketitle

\section{Introduction}\label{sec:Intro}
In dense polymer liquids and melts, the local Brownian motion of each polymer chain is subject to long-lived topological constraints (a.k.a. entanglements) imposed by the presence of the other chains.
Well documented manifestations of entangled polymer chain behavior include
chains reptative motion in monodisperse melts of linear polymers~\cite{DoiEdwardsBook,deGennes1971,KremerGrest-JCP1990}
and
chains spatial segregation in monodisperse melts of unknotted and unconcatenated ring polymers~\cite{CatesDeutschJPhysFrance1986,halverson2011molecular-statics,GrosbergSM2014,RosaEveraersPRL2014}.

Polymer chains under typical dense conditions become mutually entangled because they are effectively uncrossable to each other~\cite{DoiEdwardsBook,RubinsteinColbyBook}.
In recent years, direct ``manipulation'' of entanglements in single chain molecules has opened new routes to fine-tune the mechanical properties of polymeric materials.
This is for instance the case of the so called {\it smart materials} like polycatenanes and polyrotaxanes~\cite{dePabloRowanScience2017,Rowan-Review-NatMat2021}, which are made of interlocked components whose internal degrees of freedom and mobility shape the unique conformational space of the molecule.

Interlocking and other topology manipulations are not exclusive to synthetic molecules, in fact they take also a prominent role in the organization of the long DNA molecules which constitute the genomes of many organisms.
For instance, in eukaryotic nuclei in normal cell conditions (interphase) the cm-long filament of DNA of each chromosome is densely packed into a corresponding $\mu$m-sized ``territory''~\cite{CremerBros2001,RosaEveraersPlos2008}.
In this situation, tight confinement may result in an ``excess'' of entanglements which may be detrimental~\cite{BrahmachariMarko2019} at the later stage of cell division:
a specific class of enzymes, the topoisomerases~\cite{Champoux2001} and in particular the type II topoisomerase (hereafter, topoII), removes the entanglement~\cite{SikoravJannink1994} between two nearby DNA strands by cutting one strand, moving the other through the cut and ligating the broken strand back.

Recently, the Spakowitz's group at Stanford~\cite{SpakowitzPRL2018} showed that the ``strand crossing'' action performed by topoII is capable of ``fluidizing'' concentrated solutions of unknotted and unconcatenated DNA rings.
Moreover, by blocking the activity of the enzyme, the once free rings become permanently linked with each other:
under these conditions, the DNA solution becomes equivalent to a so called ``Olympic'' gel.
Such materials, theorized first by P.-G. de Gennes~\cite{DeGennes1979,RaphaelDeGennes1997} and notable for their theoretical~\cite{LangMacromolecules2012,LangSommerPRL2014,LangSommer2015,FischerSommerJCP2015} as well as biological implications ({\it e.g.}, the kinetoplast DNA of certain protozoa~\cite{RengerWolstenholme-JCB1972,KlotzDoylePNAS2020} can be modeled~\cite{michieletto2015kinetoplast} as an ``Olympic'' gel), are maintained together by topological bonds and not by chemical cross-links as in the case of traditional gels~\cite{RubinsteinColbyBook}.
Studies like the one from the Spakowitz's group demonstrate that it is indeed possible to bend polymer topology to produce materials capable of switching from liquid-like to more solid-like behavior.

In this paper, we present the results of extensive numerical simulations describing the formation of linked networks of ring polymers in melt conditions.
The work generalizes the efficient Monte Carlo scheme for lattice polymers described in Refs.~\cite{Hugouvieux2009,Schram-LatticeModel2018} by expanding the set of stochastic moves in order to take explicitly into account the random occurrence of strand crossings between nearby polymer fibers.

By studying the behaviors of melts of rings at different chain monomer numbers $N$ and by comparing the systems in presence and absence of strand crossings,
we confirm that the strand crossing mechanism is capable of relaxing the effects of the topological constraints between different rings and enhance the mobility of the fluid.
On the other hand, this comes at the price of increasing the topological ``complexity'' of the chains in terms of links and knots.
We find that single chains swell with respect to the unknotted and unconcatenated ensemble, their average size increasing $\propto N^{1/2}$ as in ideal Gaussian rings:
yet, we demonstrate that the stationary chain size is not equivalent to Gaussian and analyze in detail its structural and dynamical properties.

The paper is organized as follows.
In Section~\ref{sec:ModelMethods}, we describe the polymer model, the numerical details of the algorithm and its the computational cost and summarize the relevant length scales of the polymer melts.
In Sec.~\ref{sec:Results}, we present the main results of the work.
Then, in Sec.~\ref{sec:Discussion}, we discuss an effect related to the efficiency of the strand crossing mechanism that may be tested in experiments 
employing DNA rings.
The material presented here is complemented by additional figures in the Supplemental Material (SM) file.

\section{The polymer model: simulation protocol, length scales, methods}\label{sec:ModelMethods}

\subsection{The kinetic Monte Carlo algorithm}\label{sec:PolymerModel}
We employ a kinetic Monte Carlo (MC) algorithm on the three-dimensional FCC lattice with lattice spacing $=a$ corresponding to our unit of length,
and we model solutions of ring polymers with {\it excluded volume} interactions.

\begin{figure}
\includegraphics[width=0.42\textwidth]{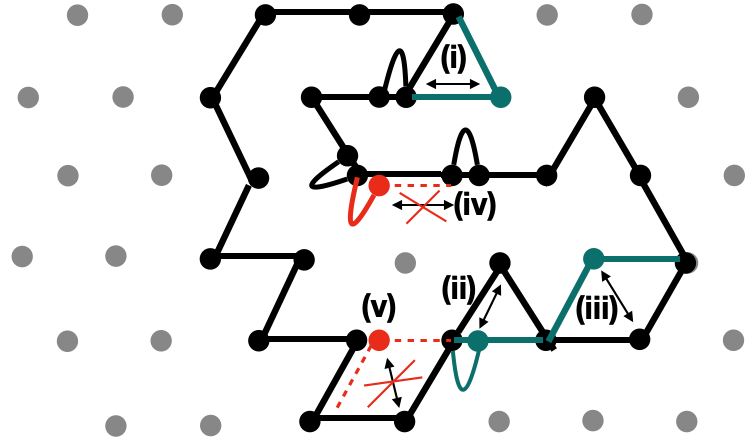} 
\caption{
$2d$ representation of the lattice polymer model with topology-preserving moves.
Each monomer (black dot) occupy a lattice site (grey dot) and two nearest neighbor monomers are joined by a black line representing the polymer bond between them.
For two nearest neighbor monomers occupying the same lattice site the bond between them (a black arc in the figure) makes a unit of stored length.
The green lines and dots are examples of MC {\it allowed} moves:
(i) a unit of stored length unfolding to a normal bond;
(ii) a bond folding into a unit of stored length;
(iii) a Rouse-like move.
The red lines and dots are examples of MC {\it forbidden} moves: (iv) three consecutive monomers along the chain occupying the same lattice site; (v) two non-nearest neighbor monomers violating the excluded volume constraint.
}
\label{fig:MC_move}
\end{figure}

The core of the algorithm is based on the elastic lattice polymer model inspired by the Rubinstein's~\cite{Rubinstein-Repton-PRL1987} repton model and developed in~\cite{Hugouvieux2009,Schram-LatticeModel2018}.
In this scheme (illustrated for simplicity in $2d$ in Figure~\ref{fig:MC_move}) two consecutive monomers along the chain either sit on nearest neighbour lattice sites or they can be on the same lattice site:
no more than two consecutive monomers may occupy the same lattice site, while non-consecutive monomers are never allowed to occupy the same lattice site due to excluded volume.
The bond length $b$ between nearest neighbor monomers takes then two possible values, $=a$ or $=0$:
in the latter case the bond is said to host a unit of {\it stored length}.
For a polymer with $N$ bonds, the total contour length $L \equiv N\langle b\rangle < Na$ where $\langle b\rangle$ is the average bond length.
This numerical trick makes the polymer elastic.

The dynamic evolution of the chains is implemented by combining two kinds of MC moves:
(i) topology-preserving (Sec.~\ref{sec:Algorithm-TopologyPreservingMoves})
and
(ii) topology-changing by stochastic strand crossing (Sec.~\ref{sec:Algorithm-TopologyChangingMoves}).

\subsubsection{Topology-preserving moves}\label{sec:Algorithm-TopologyPreservingMoves}
The first two moves are the same as in the original model~\cite{Hugouvieux2009,Schram-LatticeModel2018} and, by construction, they preserve the overall topological state of the system.
They consist in randomly picking a monomer of one of the chains in the system and attempting its displacement towards one of the nearest lattice sites (see Fig.~\ref{fig:MC_move} for a schematic illustration of these moves). 
The move is accepted if chain connectivity is preserved and with the additional constraints that (1) either the destination lattice site is empty or (2) the lattice site is occupied by only one of the nearest neighbor monomers along the chain.
In analogy with classical~\cite{DoiEdwardsBook,RubinsteinColbyBook} polymer dynamics,
case (1) is an example of {\it Rouse}-like move while case (2) is a {\it reptation}-like move (essentially the move produces mass drift along the contour length of the chain, as occurring in reptation dynamics).
It is easy to see that at low polymer densities most of lattice sites are empty and Rouse moves prevail over reptation,
while in the opposite case of high polymer densities reptation becomes the dominant mode through which polymer chains relax.
Therefore the algorithm reproduces known~\cite{DoiEdwardsBook,RubinsteinColbyBook} features of polymer dynamics and, thanks to the stored length ``trick'' which integrates local fluctuations of the chain density, remains efficient even when it is applied to the equilibration of very large systems~\cite{Schram-LatticeModel2018}.

\subsubsection{Topology-changing (strand crossing) moves}\label{sec:Algorithm-TopologyChangingMoves}

%
\begin{figure}
$$
\begin{array}{cc}
\multicolumn{2}{c}{\mbox{\underline{No} strand crossing}} \\
\includegraphics[width=0.21\textwidth]{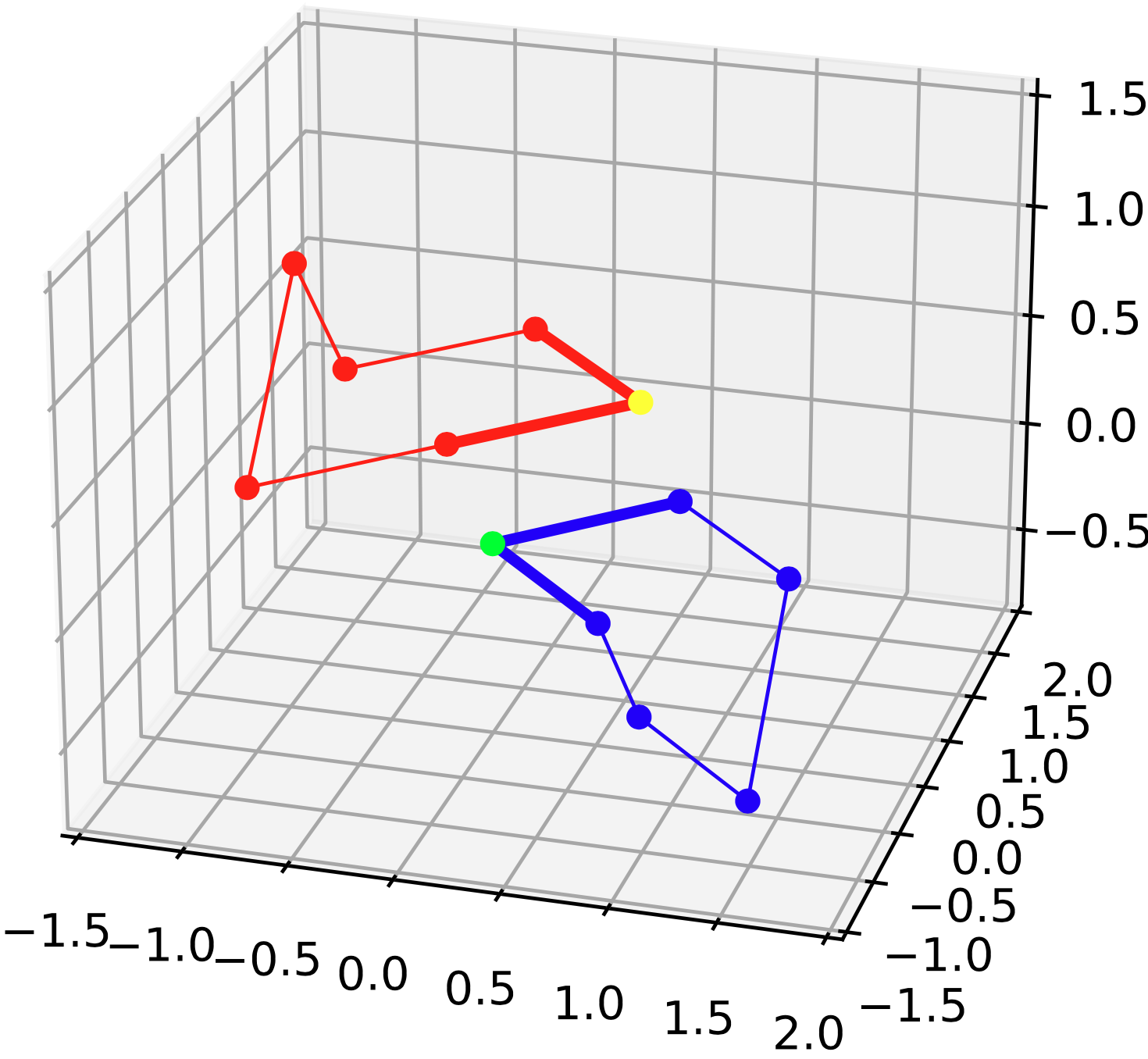} &
\includegraphics[width=0.21\textwidth]{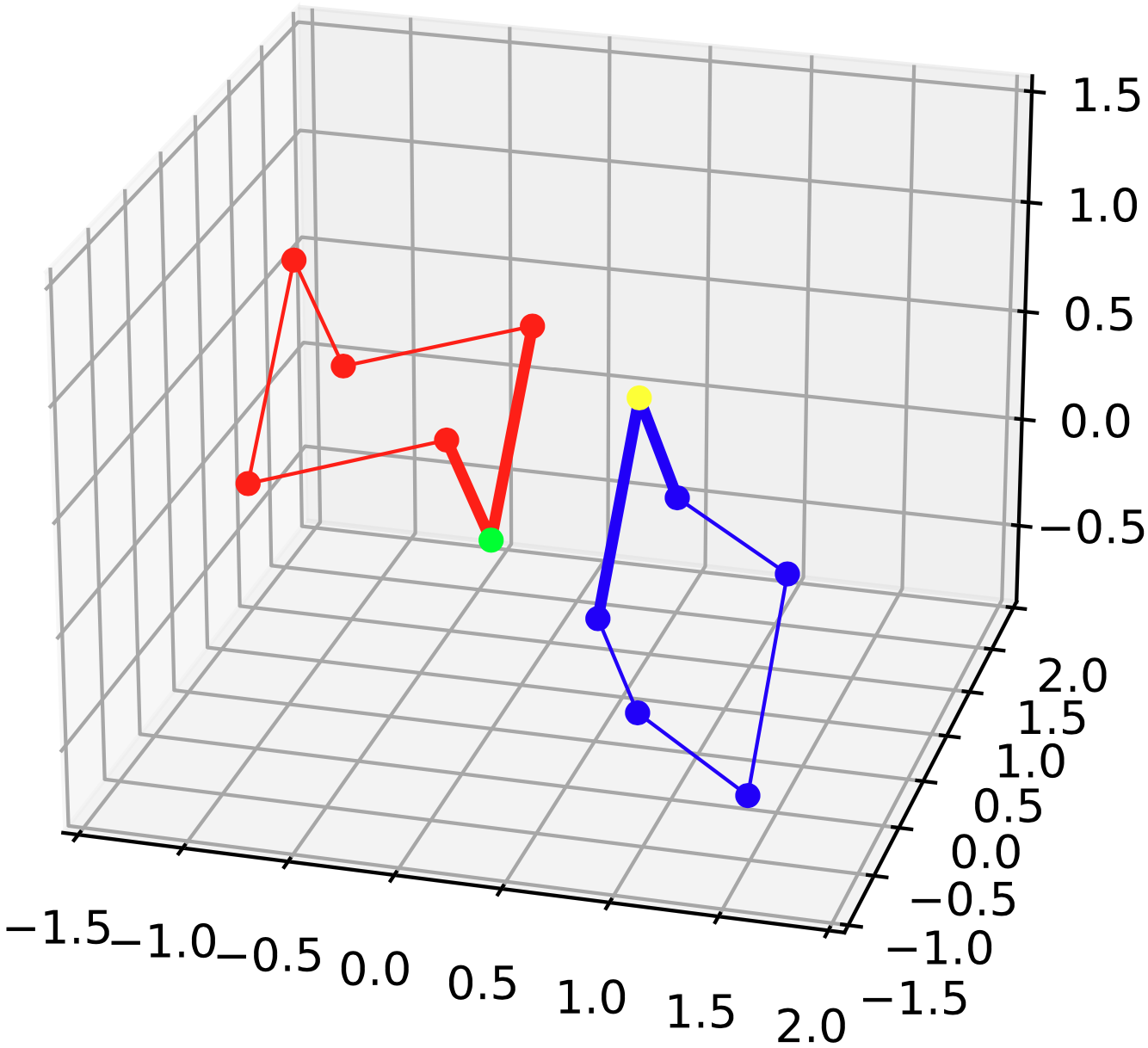} \\
\\
\multicolumn{2}{c}{\mbox{\underline{With} strand crossing}} \\
\includegraphics[width=0.22\textwidth]{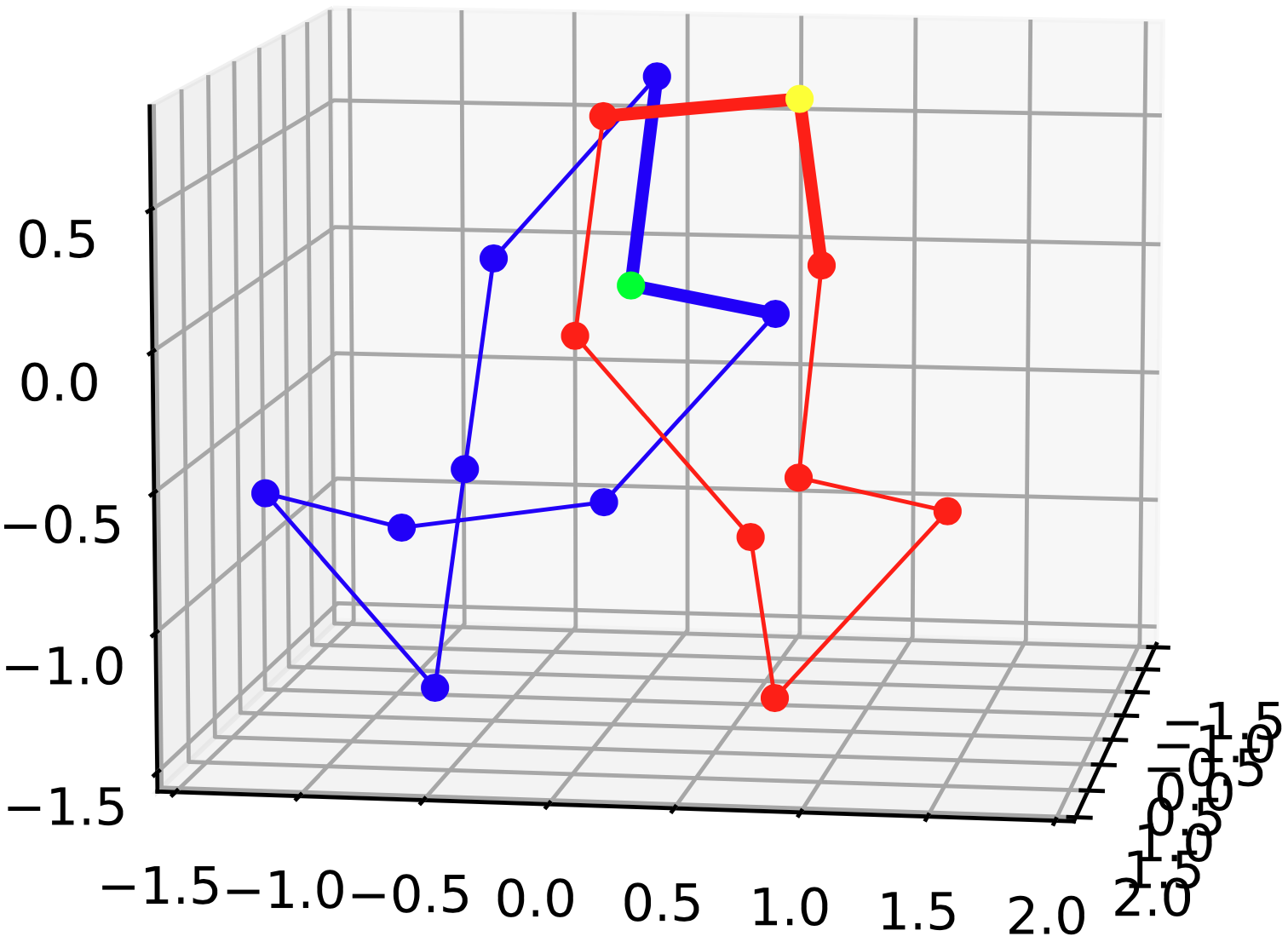} &
\includegraphics[width=0.22\textwidth]{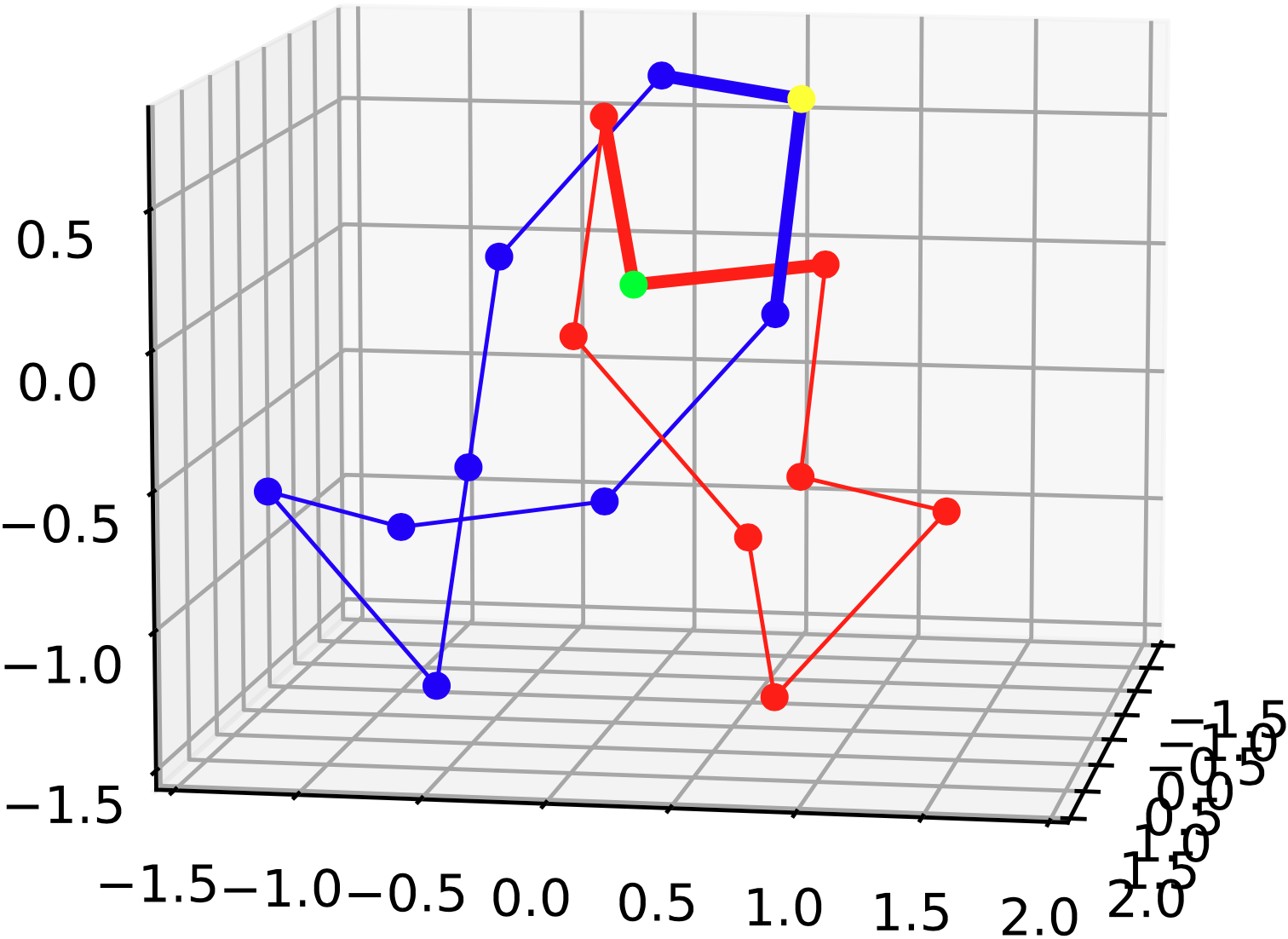} \\
\end{array}
$$
\caption{
Illustration of Monte Carlo moves for strand crossing.
On each pair of ring polymers we identify those strands (thick red and blue lines in the figure) of contour length $=2a$ whose central monomers (in green and yellow) are one lattice site far apart.
The two strands can be ``transformed'', such that the original central monomer of one strand turns into into the central monomer of the other strand (l.h.s. panels {\it vs.} r.h.s. panels) and viceversa, without violating the chain connectivity constraint.
In some cases ({\it e.g.}, as in the top row) this operation does not lead to strand crossing, in others ({\it e.g.}, as in the bottom row) it does.
We list all cases ($12$ in total) leading to strand crossing and implement them in our MC algorithm.
The configurations in the bottom row constitute an example of a linking (left $\rightarrow$ right) or an unlinking (right $\rightarrow$ left) event.
}
\label{fig:swapping_moves_polymer}
\end{figure}

Here we are interested in studying melt of ring polymers where topology changes are induced over time. 
In particular we consider the basic mechanism of {\it strand crossing} (hereafter, SC) involving a pair of nearby polymer filaments, similar to the action triggered by the enzyme topoII in DNA rings solutions~\cite{SpakowitzPRL2018}.

As explained in Sec.~\ref{sec:Algorithm-TopologyPreservingMoves}, the original~\cite{Hugouvieux2009,Schram-LatticeModel2018} lattice model does not include such feature:
we show here that it is however possible to remove this ``constraint'' and we describe the simplest possible MC move capable of inducing a single crossing between two nearby polymer strands.
These polymer strands may either belong to the same chain ({\it intra-chain} SC) or they can stay on two distinct chains ({\it inter-chain} SC).

The new move (which is also one of the main contributions of this paper) is illustrated with the help of the two examples in Fig.~\ref{fig:swapping_moves_polymer}.
Take two distinct polymer strands of effective contour length $=2a$ like, {\it e.g.}, the two thick segments of the red and blue curves. 
The two segments are chosen with the constraints that the corresponding central monomers (in yellow and green) 
(i)
are positioned at lattice site distance $=a$
and
(ii)
one can switch position with the other and being reconnected to the other chain with no violation of polymer connectivity and preserving the contour length. 
By exhaustive search, we have then compiled the list of all possible ($36$ in total) swapping moves compatible with these constraints.
By closer examination,
we verify that $24$ of them do not produce SC (essentially the two chain strands remain on parallel planes even after swapping, see the polymer configurations in the top row in Fig.~\ref{fig:swapping_moves_polymer}),
while the remaining $12$ moves effectively lead to a single SC (as in the polymer configurations in the bottom row in Fig.~\ref{fig:swapping_moves_polymer}).
The successful SC has been verified by looking at the {\it variation}, $|\Delta \mathcal G| = +1$, of the {\it Gauss linking number} $\mathcal G$ (see definition, Eq.~\eqref{eq:GaussLN}) 
relative to the piecewise closed curves formed by the triplets of monomers involved in the MC swapping move.

The implementation of this move in the kinetic MC algorithm is as the following.
We pick randomly two polymer strands of effective contour length $=2a$, then check whether they belong to the set compatible with a SC and, if so, we swap the corresponding central monomers.
When the two involved strands belong to the same ring the move is introducing {\it knots} in the chain (Sec.~\ref{sec:Knots}),
while on two separate rings it will induce the formation of {\it links} (Sec.~\ref{sec:Network-Str+Dyn}).

\subsection{Simulation details}\label{sec:SimulationDetails}
We have considered bulk solutions of $M$ closed (ring) polymer chains, each chain made of $N$ monomers or bonds.
With values $N\times M = \left[ 40\times5120, 80\times2560, 160\times1280, 320\times640, 640\times320 \right]$
each system contains a fixed number of monomers $=204800$.
Bulk conditions are implemented through the enforcement of periodic boundary conditions in a simulation box of total volume $V=L^3$,
where the linear sizes of the box, $L$, has been fixed based on the monomer number density 
$\rho a^3 \equiv \frac{NM}Va^3 = 1.23$ 
corresponding to melt conditions~\cite{Schram-LatticeModel2018,SchramRosaEveraers2019}.

We have studied and compared structure and dynamics for different set-up's:
\begin{itemize}
\item
Ring polymer melts with {\it non-conserved} chain topology.
Here, the topological state of the system changes in time according to the SC mechanism.
Therefore, the MC scheme includes the whole set of dynamic moves described in Secs.~\ref{sec:Algorithm-TopologyPreservingMoves} and~\ref{sec:Algorithm-TopologyChangingMoves}.
\item
Ring polymer melts with {\it conserved} chain topology.
Here, only moves from Sec.~\ref{sec:Algorithm-TopologyPreservingMoves} are included.
Since now topology can not relax the choice of the initial state is crucial.
The following two options have been considered:
(i)
Equilibrated melts of {\it unknotted and unconcatenated} or (for brevity and as in Ref.~\cite{SchramRosaEveraers2019}) {\it untangled} rings.
(ii)
Equilibrated melts of {\it permanently catenated} rings, corresponding to the equilibrated polymer conformations obtained at the end of the simulations with non-conserved chain topology.
The name anticipates some properties of the rings (catenation and linking) that will be discussed in depth in Sec.~\ref{sec:Network-Str+Dyn}.
\item
For additional comparison, we have also considered ideal ({\it i.e.}, no excluded volume and no topological interactions) rings. 
\end{itemize}

At each MC time step, monomers are picked at random and time is measured in MC units of $\tau_{\rm MC} \equiv N\cdot M$.
For polymer solutions with non-constrained topology (Sec.~\ref{sec:Algorithm-TopologyChangingMoves}) one needs to specify also the {\it rate} 
at which SC's occur.
In principle, this rate is a free parameter of our model that we must tune.
For the typical experimental conditions described in the work by Spakowitz {\it et al.}~\cite{SpakowitzPRL2018},
it was estimated that the action rate of topoII on DNA rings is close to its {\it intrinsic} rate of ${\mathcal O}(1{\rm s}^{-1})$ and $\approx 10^4$ times slower than the mean diffusion time of a single DNA persistence length.
Considering that our polymers are pretty flexible (see Sec.~\ref{sec:Properties-PolymerSols}), we take here one SC move (modeled according to Sec.~\ref{sec:Algorithm-TopologyChangingMoves}) each $10^4$ MC time steps with only topology-preserving moves.
Notice that this choice implies that the enzyme topoII is assumed to be immediately available for the reaction, {\it i.e.} the process of SC is intrinsically {\it reaction-limited}.
Nonetheless, we will also discuss (see Sec.~\ref{sec:Discussion}) smaller values of $\lambda_{\rm SC}$ corresponding to ``less efficient'' topoII.

\subsection{Comparison to other simulation methods}\label{sec:Comparison2PreviousWorks}
In this section, we discuss briefly two computational methods which appeared in the past dealing with the formation of linked gels in melts of entangled rings.

In Refs.~\cite{LangMacromolecules2012,LangSommerPRL2014,LangSommer2015,FischerSommerJCP2015}, Lang and coworkers adapted the bond fluctuation model~\cite{CarmesinKremer1988,Paul1991} to construct ``Olympic'' gels from untangled melts of rings.
To achieve this task though, they had to introduce a set of ``diagonal'' moves which temporarily remove entanglements consenting polymer bonds to overlap (the so called ``x-traps''~\cite{Tanaka2000}).
More or less in the same period, Michieletto and coworkers~\cite{michieletto2015kinetoplast} used classical Brownian dynamics simulations of a bead-spring polymer model to construct model ``Olympic'' gel conformations for the DNA kinetoplast. In this case, entanglements were removed by switching off the non-bonded monomer-monomer interactions of the system, letting the system to equilibrate and reintroducing the interactions back again.
Both protocols, then, do not look suitable to study the dynamics of the linking process because they temporarily switch off entanglements and excluded volume interactions. 

On the contrary, the linking protocol introduced here avoids the unphysical bond-bond overlaps and preserves the excluded volume interactions.
For these reasons, the protocol can be used to model not only the structure (Sec.~\ref{sec:SingleChainBehavior}) but also the dynamics (Sec.~\ref{sec:Network-Str+Dyn}) of gel formation employing DNA rings in the presence of topoII.

\subsection{Computing observables for polymer structure}\label{sec:Observables}

%
\begin{table}
\begin{tabular}{cccc}
\hline
\hline
\, $N$ \, & \, $M$ \, & \, $T_{\rm run} \, [\tau_{\rm MC}]$ \, & $\tau_d \, [\tau_{\rm MC}]$ \\
\hline
\\
\multicolumn{4}{c}{Ideal rings} \\
\hline
40 & 100 & $\simeq 6.0 \cdot 10^6$ & $\simeq 7.0 \cdot 10^3$ \\
80 & 100 & $\simeq 6.0 \cdot 10^6$ & $\simeq 7.0 \cdot 10^3$ \\
160 & 100 & $\simeq 6.0 \cdot 10^6$ & $\simeq 1.0 \cdot 10^4$ \\
320 & 100 & $\simeq 9.0 \cdot 10^6$ & $\simeq 4.0 \cdot 10^4$ \\
640 & 200 & $\simeq 1.5\cdot 10^7$ & $\simeq 2.0 \cdot 10^5$ \\
\\
\multicolumn{4}{c}{Melts of untangled rings} \\
\hline
40 & 5120 & $\simeq 2.0 \cdot 10^6$ & $\simeq 1.0 \cdot 10^4$ \\
80 & 2560 & $\simeq 2.0 \cdot 10^6$ & $\simeq 3.0 \cdot 10^4$ \\
160 & 1280 & $\simeq 2.0 \cdot 10^6$ & $\simeq 2.0 \cdot 10^5$ \\
320 & 640 & $\simeq 4.0 \cdot 10^6$ & $\simeq 7.0 \cdot 10^5$ \\
640 & 320 & $\simeq 1.5\cdot 10^7$ & $\simeq 3.0 \cdot 10^6$ \\
\\
\multicolumn{4}{c}{Melts of rings with strand crossings} \\
\hline
40 & 5120 & $\simeq 2.0 \cdot 10^6$ & $\simeq 1.0 \cdot 10^4$ \\
80 & 2560 & $\simeq 2.0 \cdot 10^6$ & $\simeq 3.0 \cdot 10^4$ \\
160 & 1280 & $\simeq 2.0 \cdot 10^6$ & $\simeq 1.0 \cdot 10^5$ \\
320 & 640 & $\simeq 5.0 \cdot 10^6$ & $\simeq 6.0 \cdot 10^5$ \\
640 & 320 & $\simeq 1.4 \cdot 10^7$ & $\simeq 2.0 \cdot 10^6$ \\
\\
\multicolumn{4}{c}{Melts of permanently catenated rings} \\
\hline
40 & 5120 & $\simeq 2.0 \cdot 10^6$ & $\simeq 1.0 \cdot 10^4$ \\
80 & 2560 & $\simeq 2.0 \cdot 10^6$ & $\simeq 5.0 \cdot 10^4$ \\
160 & 1280 & $\simeq 7.0 \cdot 10^6$ & $\simeq 5.0 \cdot 10^5$ \\
320 & 640 & $\simeq 9.8 \cdot 10^7$ & -- \\
640 & 320 & $\simeq 1.9 \cdot 10^8$ & -- \\
\hline
\hline
\end{tabular}
\caption{
Computational cost of MC runs.
In interacting systems (melts) $M$ is the total number of chains, whereas for ideal systems with no excluded volume interactions it represents the number of single independent runs.
(i)
$T_{\rm run}$: length of the single MC run. 
(ii)
$\tau_d$: ring self-diffusion time.
Values for permanently catenated rings with $N=320$ and $N=640$ are not defined because the corresponding time mean-square displacements of the centre of mass ($g_3(\tau)$, Eq.~\eqref{eq:g3}) attain the characteristic plateaus for stacked dynamics (see Fig.~\ref{fig:SingleChainDynamics}(c)). 
$\tau_{\rm MC}=N\cdot M$ is the Monte Carlo time unit (see Sec.~\ref{sec:SimulationDetails} for details).
}
\label{tab:ComputationalCost}
\end{table}

The ensemble average value, $\langle {\mathcal O} \rangle$, for the generic single-chain observable $\mathcal O$ is given by the mathematical expression:
\begin{equation}\label{eq:DefineAvObs}
\langle {\mathcal O} \rangle \equiv \frac1M \, \sum_{m=1}^M \, \frac1{\tau_d} \, \int_{T_{\rm run}-\tau_d}^{T_{\rm run}} {\mathcal O}_m (t) \, dt \, ,
\end{equation}
where
${\mathcal O}_m (t)$ is the value of the observable calculated for the $m$-th ring at time $t$ and
$T_{\rm run}$ is the total runtime of the MC trajectory (values for $T_{\rm run}$, illustrating the computational cost of our simulations, are reported in Table~\ref{tab:ComputationalCost}).
The time average in Eq.~\eqref{eq:DefineAvObs} is calculated by discarding the initial portion of each trajectory which is of the order of the self-diffusion time ($\tau_d(N)$, see values in Table~\ref{tab:ComputationalCost}) of the polymers.
$\tau_d(N)$ corresponds to the time scale for the polymer to diffuse of a distance the size of its own mean gyration radius, $g_3(\tau_d(N)) \equiv \langle R_g^2(N)\rangle$,
where $g_3(\tau)$ is the time mean-square displacement of the chain centre of mass (see definition, Eq.~\eqref{eq:g3}).

\subsection{Polymer model: length scales}\label{sec:Properties-PolymerSols}

%
\begin{table}
\begin{tabular}{ccccc}
\hline
\hline
\\
$\langle b \rangle / a $ \, & \, $\ell_K / a$ \, & \, $\rho_K \ell_K^3$ \, & \, $L_e / \ell_K$ \, & \, $N_e$ \, \\
\hline
0.74 & 1.48 & $1.9937$ & $100.633$ & $201.266$ \\
\hline
\hline
\end{tabular}
\caption{
Values of physical parameters for melts of ring polymers on the FCC lattice with unit distance $=a$ and monomer number density $\rho a^3 = 1.23$: 
(i) $\langle b \rangle$, mean bond length;
(ii) $\ell_K$, Kuhn length (Eq.~\eqref{eq:Define-Lk}); 
(iii) $\rho_K \ell_K^3$, number of Kuhn segments per Kuhn volume~\cite{UchidaJCP2008};
(iv) $L_e$, entanglement length (Eq.~\eqref{eq:KavassalisNoolandi-ii});
(v) $N_e \equiv L_e / \langle b\rangle$, number of bonds per entanglement length.
}
\label{tab:PolymerModel-LengthScales}
\end{table}

For completeness, here we give a few additional details about the relevant length scales (summarized in Table~\ref{tab:PolymerModel-LengthScales}) used to characterize the local bending and the entanglement properties of polymer melts.
We remind the reader that, as in Refs.~\cite{Schram-LatticeModel2018,SchramRosaEveraers2019}, we chose the monomer number density $\rho =1.23a^{-3}$ where $a$ is the FCC lattice unit distance (Sec.~\ref{sec:PolymerModel}). 

{\it Average bond length}, $\langle b\rangle$ -- 
Due to excluded volume effects and chain packing the average bond length $\langle b \rangle = 0.74 a < a$. 

{\it Kuhn length}, $\ell_K$ --
The Kuhn length is used to quantify the flexibility of polymer chains~\cite{DoiEdwardsBook,RubinsteinColbyBook}.
Given the mean-square end-to-end distance, $\langle R^2(\ell) \rangle$, between monomers at contour length separation $\ell$ on {\it linear} chains,
$\ell_K$ is defined as~\cite{DoiEdwardsBook,RubinsteinColbyBook}:
\begin{equation}\label{eq:Define-Lk}
\ell_K \equiv \lim_{\ell\rightarrow\infty} \frac{\langle R^2(\ell) \rangle}{\ell} \, ,
\end{equation}
provided that such limit exists~\cite{HsuBinder-DefineLk2010}.
In order to determine the polymer Kuhn length of our polymer chains, 
we have simulated systems of $M=640$ linear chains with $N=320$ monomers per chain and with chain dynamics as described in Sec.~\ref{sec:Algorithm-TopologyPreservingMoves}.
After equilibration, we have computed the ratio (Eq.~\eqref{eq:Define-Lk}) $\frac{\langle R^2(\ell)\rangle}{\ell}$ where $\ell=n\langle b\rangle$ is the contour length separation between any two monomers separated by $n$ bonds along the chain. 
We have found that this quantity reaches a plateau in the region $\ell=[200\langle b\rangle, 300\langle b\rangle]$ which has been then fitted to a constant value. 

{\it Entanglement length}, $L_e$ --
Dense untangled rings are known~\cite{CatesDeutschJPhysFrance1986,RosaEveraersPRL2014,PanyukovRubinsteinMacromolecules2016} to compact above a characteristic length scale, the entanglement length $L_e$ of the chains.
According to the classical packing argument by Lin~\cite{Lin1987} and by Kavassalis and Noolandi~\cite{KavassalisNoolandiPRL1987},
the number of entanglement strands inside the volume spanned by a single entanglement volume,
\begin{equation}\label{eq:KavassalisNoolandi-i}
\frac{\rho_K}{L_e/\ell_K} \langle R^2(L_e) \rangle^{3/2} \simeq 20 \, ,
\end{equation}
is a universal constant.
In Eq.~\eqref{eq:KavassalisNoolandi-i}, $\rho_K$ is the number density of Kuhn segments and $\langle R^2(L_e) \rangle = \ell_K L_e$ is the mean-square end-to-end distance of a linear polymer chain of contour length $=L_e$ (Eq.~\eqref{eq:Define-Lk}).
Eq.~\eqref{eq:KavassalisNoolandi-i} is then equivalent to:
\begin{equation}\label{eq:KavassalisNoolandi-ii}
\frac{L_e}{\ell_K} \simeq \left(\frac{20}{\rho_K \ell_K^{3}}\right)^2 \, ,
\end{equation}
{\it i.e.}, the ratio $L_e / \ell_K$ is a function of the number of Kuhn segments inside the (Kuhn) volume $=\ell_K^3$.
By using Eq.~\eqref{eq:KavassalisNoolandi-ii} it is a simple exercise to extract $L_e/\ell_K$ and the corresponding number of monomers per entanglement length, $N_e$. 
In particular, we notice that 
the largest rings with $N=640\approx3N_e$ are above the entanglement threshold and are expected~\cite{CatesDeutschJPhysFrance1986,RosaEveraersPRL2014,PanyukovRubinsteinMacromolecules2016} to crumple due to topological constraints.

\section{Results}\label{sec:Results}

\subsection{Single-chain structure}\label{sec:SingleChainBehavior}

\subsubsection{Ring size}\label{sec:RingSize}

%
\begin{table}
\begin{tabular}{ccc}
\hline
\hline
\, $N$ \, & $\langle R_g^2 \rangle / a^2$ & $P_{\rm knot}$ \\
\hline
\\
\multicolumn{3}{c}{Ideal rings} \\
\hline
40 & $3.178 \pm 0.002$ & -- \\
80 & $6.284 \pm 0.006$ & -- \\
160 & $12.52 \pm 0.02$ & -- \\
320 & $24.9 \pm 0.1$ & -- \\
640 & $50.0 \pm 0.3$ & -- \\
\\
\multicolumn{3}{c}{Melts of untangled rings} \\
\hline
40 & $3.3334 \pm 0.0004$ & -- \\
80 & $6.361 \pm 0.003$ & -- \\
160 & $11.88 \pm 0.02$ & -- \\
320 & $21.9 \pm 0.1$ & -- \\
640 & $38.1 \pm 0.5$ & -- \\
\\
\multicolumn{3}{c}{Melts of rings with strand crossings} \\
\hline
40 & $3.5093 \pm 0.0005$ & $0$ \\
80 & $7.088 \pm 0.004$ & $1 \cdot 10^{-3}$ \\
160 & $14.30 \pm 0.02$ & $6 \cdot 10^{-3}$ \\
320 & $28.6 \pm 0.1$ & $3 \cdot 10^{-2}$ \\
640 & $57.6 \pm 0.4$ & $9 \cdot 10^{-2}$ \\
\\
\multicolumn{3}{c}{Melts of permanently catenated rings} \\
\hline
40 & $3.521 \pm 0.004$ & -- \\
80 & $7.10 \pm 0.01$ & -- \\
160 & $14.36 \pm 0.07$ & -- \\
320 & $29.3 \pm 0.3$ & -- \\
640 & $59 \pm 1$ & -- \\
\hline
\hline
\end{tabular}
\caption{
Single-chain properties in melts of $N$-monomer rings.
(i)
$\langle R_g^2 \rangle$: ring mean-square gyration radius, expressed in lattice units.
(ii)
$P_{\rm knot}$: mean knotting probability per chain (only for melts of rings with strand crossings).
}
\label{tab:ChainStructure}
\end{table}
\begin{figure}
$$
\begin{array}{c}
\includegraphics[width=0.45\textwidth]{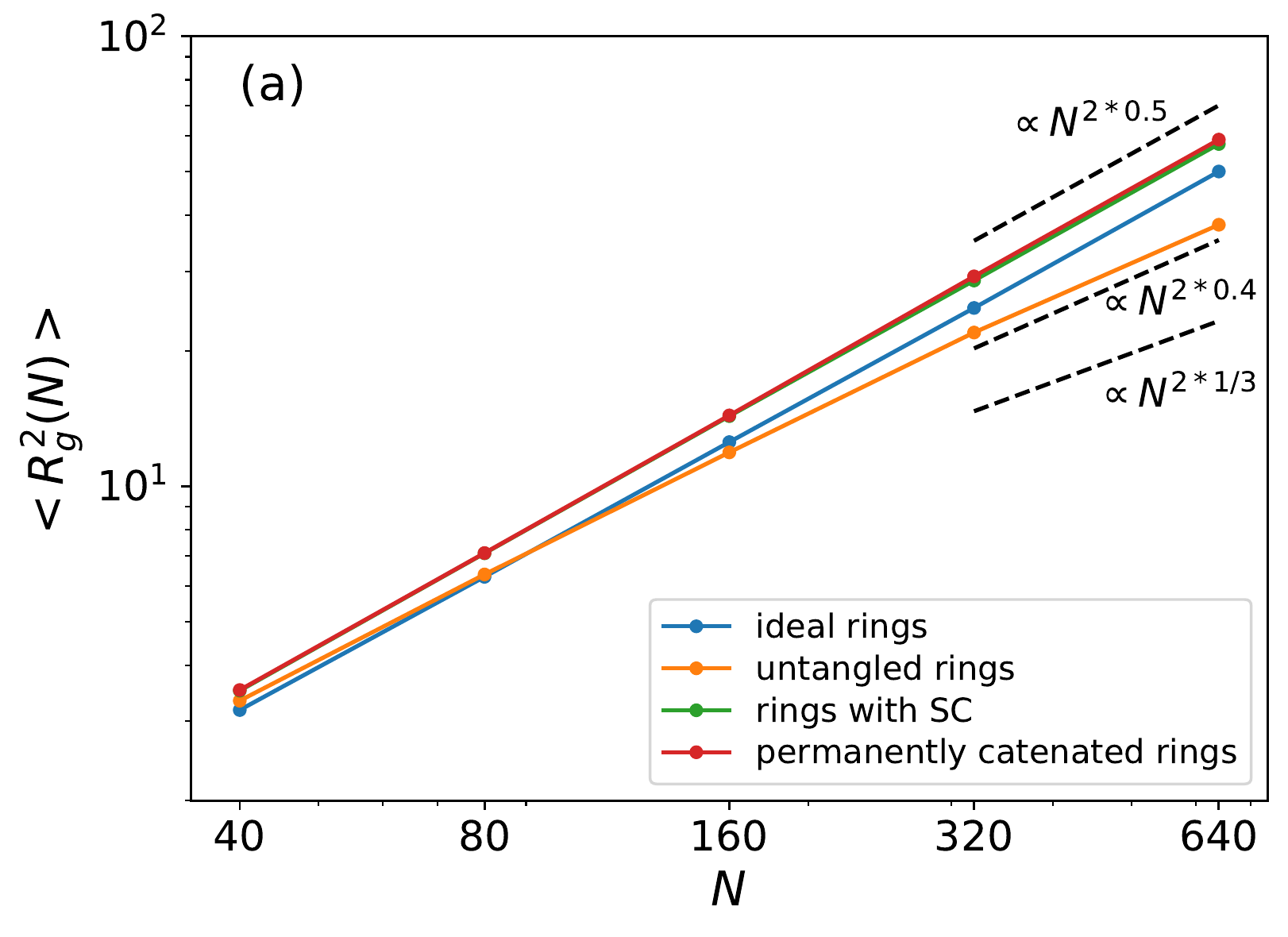} \\
\includegraphics[width=0.45\textwidth]{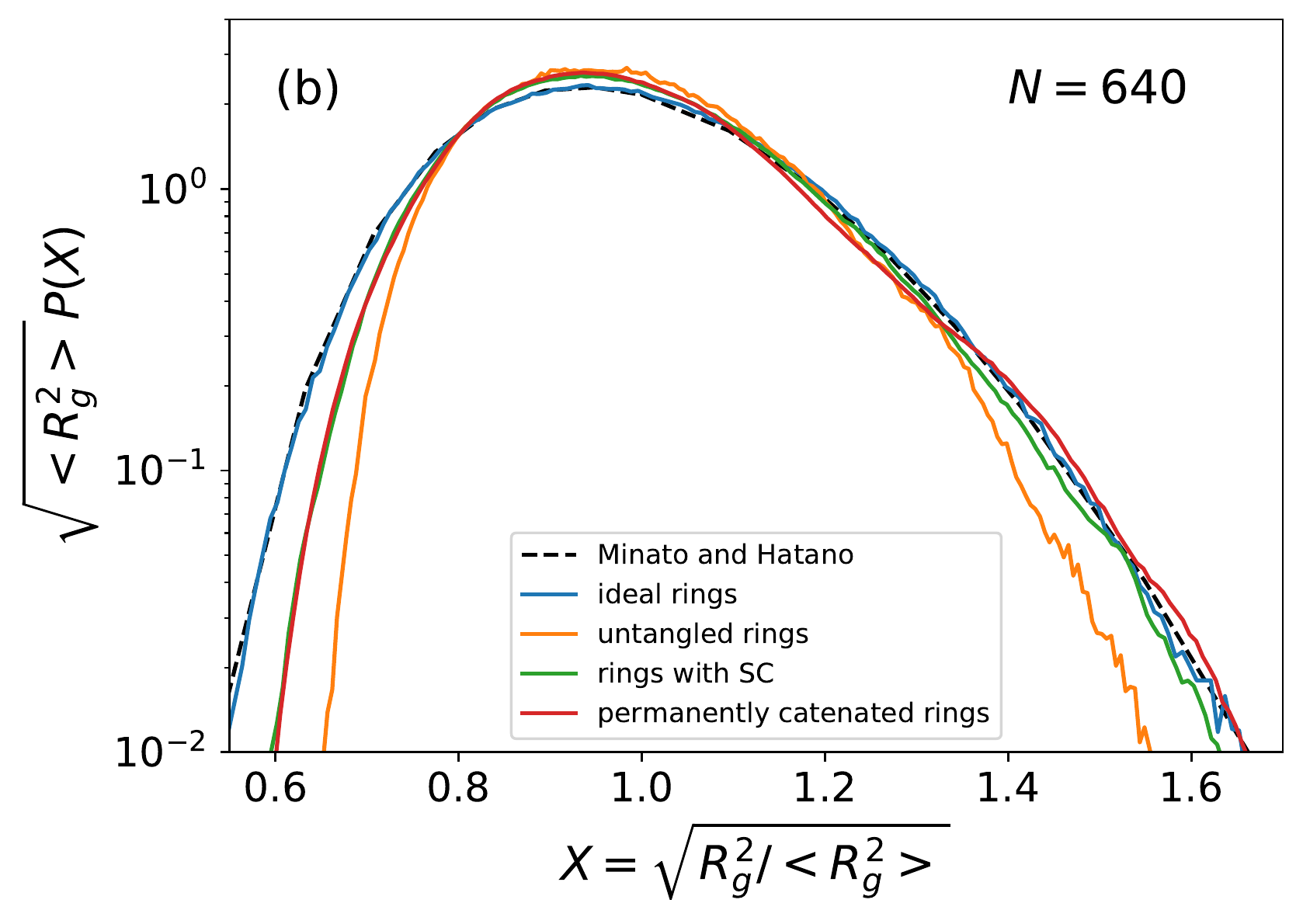}
\end{array}
$$
\caption{
(a)
Mean-square gyration radius of ring polymers, $\langle R_g^2(N) \rangle$, as a function of the number of bonds, $N$ (for detailed values at each $N$, see Table~\ref{tab:ChainStructure}).
Symbols of different colors are for different polymer ensembles and lines correspond to asymptotic behaviors (see the legend for details).
(b)
Distribution functions of gyration radius, $P(R_g / \sqrt{\langle R_g^2\rangle})$, in the different ring ensembles and for the largest chains ($N=640$).
The dashed line corresponds to the exact analytical result for ideal rings~\cite{minato1977distribution}.
}
\label{fig:gyration_radius}
\end{figure}

We have studied first how the average ring size, or the polymer mean-square gyration radius
\begin{equation}\label{eq:Rg2}
\langle R_g^2(N) \rangle \equiv \frac1N \sum_{i=1}^N \langle ( \vec r_i - \vec r_{\rm cm} )^2 \rangle \sim N^{2\nu} \, ,
\end{equation}
scales as a function of $N$.
In Eq.~\eqref{eq:Rg2}, $\vec r_i$ are the monomer coordinates, $\vec r_{\rm cm}$ is the chain centre of mass and $\nu$ is the Flory scaling exponent~\cite{RubinsteinColbyBook}.
The results for the different ensembles are summarized in Table~\ref{tab:ChainStructure} and plotted in Fig.~\ref{fig:gyration_radius}(a).

In agreement with Refs.~\cite{halverson2011molecular-statics,RosaEveraersPRL2014,SchramRosaEveraers2019}, topological constraints in untangled melts are ineffective below $N\lesssim N_e$ where $N_e \approx 200$ is the total number of monomers per entanglement length (see Sec.~\ref{sec:Properties-PolymerSols}).
Above $N_e$, the mutual topological constraints between nearby rings let the chains to deviate from the ideal behavior $\langle R^2_{g}(N) \rangle \sim N^{2\cdot1/2}$ and to become more compact:
in particular here we report the scaling $\langle R^2_{g}(N) \rangle \sim N^{2\cdot0.4}$, which describes~\cite{CatesDeutschJPhysFrance1986} the slow crossover to the asymptotic compact regime $\langle R^2_{g}(N) \rangle \sim N^{2\cdot1/3}$~\cite{GrosbergSM2014,RosaEveraersPRL2014}.

In the presence of active SC's the rings swell again, $\langle R_g^2 \rangle \sim N^{2\cdot1/2}$,
and their behavior does match the one obtained once SC's are frozen again and rings turn permanently catenated (overlying red and violet symbols in Fig.~\ref{fig:gyration_radius}(a)):
we argue that this is a consequence of the fact that the SC time scale $\lambda^{-1}_{\rm SC}$ is much larger than the typical diffusion time of a single monomer, therefore the polymer has the time to rearrange itself between two consecutive SC's and to attain a state which does not undergo further changes once SC's are turned off.

Interestingly, although the scaling behavior appears compatible with the one of ideal chains with $\nu=1/2$, rings structure remains non-ideal even in the presence of SC's.
To show that, we have computed the complete distribution function, $P(R_g)$, of the gyration radius and compared their shapes 
in the different ensembles (see Fig.~\ref{fig:gyration_radius}(b) for the particular case $N=640$ and Fig.~\ref{fig:Rg-PDFs} in SM for rings of different $N$'s).
As expected from their gyration radii, rings with active SC's and permanently catenated rings have the same $P(R_g)$ (green and red lines in Fig.~\ref{fig:gyration_radius}(b) and Fig.~\ref{fig:Rg-PDFs}(c, d) in SM).
For small $R_g$'s though, these curves deviate substantially from the one describing ideal rings (blue line in Fig.~\ref{fig:gyration_radius}(b) and Fig.~\ref{fig:Rg-PDFs}(a) in SM).
The latter, notably, fits to the analytical function (dashed lines in Fig.~\ref{fig:gyration_radius}(b) and Fig.~\ref{fig:Rg-PDFs}(a) in SM) by Minato and Hatano~\cite{minato1977distribution}.

\subsubsection{Knot statistics in ring polymers with SC's}\label{sec:Knots}
Topological constraints in untangled melts make the chains more compact with respect to the ideal case, a situation which is radically altered in the presence of active SC's (Fig.~\ref{fig:gyration_radius}(a)).
SC's act in the same way regardless the two strands are on the same or on different rings (see Sec.~\ref{sec:Algorithm-TopologyChangingMoves}):
for this reason they change both, single-chain topology by forming {\it knots} and inter-chain topology by forming {\it links} (studied in Sec.~\ref{sec:Network-Str+Dyn}).

There exists conspicuous literature on the effects of physical knots (and links) in polymer filaments and soft matter, see the review work~\cite{Micheletti-PhysRep2011}.
Knots in closed curves can be classified based on the number of unresolvable crossings they present when one tries to smoothly deform the curve so to force it to lie on a plane~\cite{OrlandiniWhittington-RMP2007,Micheletti-PhysRep2011}.
So, there exist one single knot with three crossings ($3_1$, the trefoil knot), one with four ($4_1$, the figure-eight knot), two with five crossings ($5_1$ and $5_2$).
At increasing knot complexity, the same number of crossings correspond to several knot types.

In general, knots classification is operated by means of suitable {\it topological invariants}.
One of the simplest and most popular of the knot invariants, which we also adopt in the present work, is given by the so called Alexander polynomial~\cite{OrlandiniWhittington-RMP2007,Micheletti-PhysRep2011} of the knot,
which provides a mathematically tractable representation of the smallest number of chain crossings occurring in the closed curve.
Here, we have used the open package KymoKnot~\cite{Tubiana2018} to detect and classify the knots which form in our polymer chains by SC's. 

\begin{figure}
\includegraphics[width=0.45\textwidth]{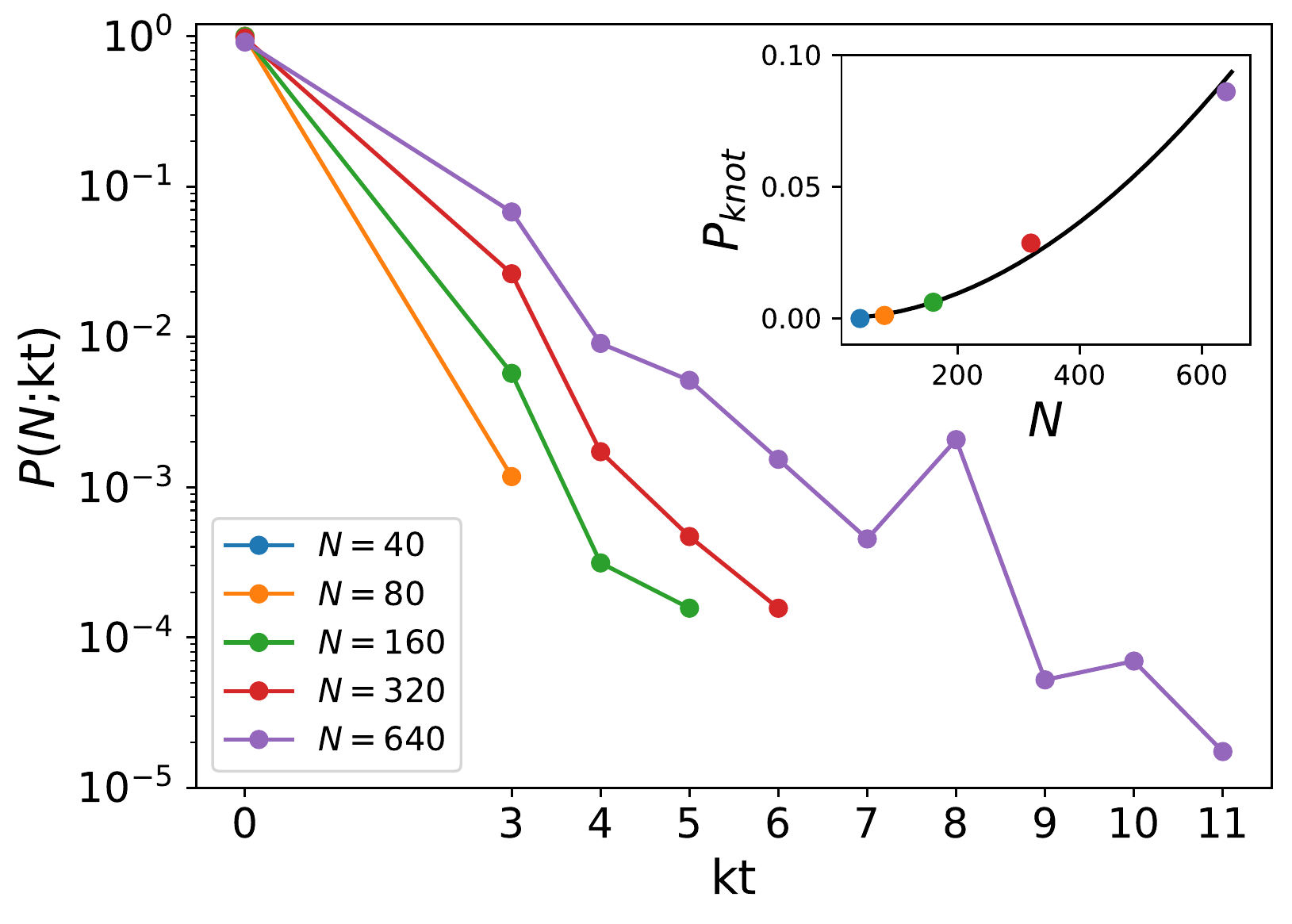}
\caption{
Probability distribution function, $P(N; {\rm kt})$, for knot type ($\rm kt$) detected in $N$-monomer rings (see legend) undergoing continuous SC's.
Rings with $N=40$ do not knot during the simulation.
Inset: Knotting probability $P_{\rm knot}$, Eq.~\eqref{eq:DefineKnottingProb}, as a function of $N$ (symbols) and best fit to power-law behavior (line, Eq.~\eqref{eq:KnottingProbTheo} with Eqs.~\eqref{eq:KnottingProbTheo-N0} and~\eqref{eq:KnottingProbTheo-alpha}).
}
\label{fig:Knots}
\end{figure}

Fig.~\ref{fig:Knots} shows the probability, $P(N; {\rm kt})$, that $N$-monomer rings have given knot type ${\rm kt} = 0_1$ (the unknot), ${\rm kt} = 3_1$ (the trefoil, {\it i.e.} the simplest non trivial knot) and so on for knots of increasing complexity.
Knots of complex shapes are in general rare (the trefoil dominates), yet their frequency increases~\cite{OrlandiniWhittington-RMP2007,Micheletti-PhysRep2011} steadily with $N$
and for $N=640$ we are even able to detect a few, and quite complex, knotted structures with $11$ crossings.

Overall, the cumulative knotting probability
\begin{equation}\label{eq:DefineKnottingProb}
P_{\rm knot}(N) \equiv \sum_{\rm kt=3_1, 4_1, ...} P(N; {\rm kt}) = 1-P(N; {\rm kt}=0_1)
\end{equation}
is well described (symbols {\it vs.} line in the inset of Fig.~\ref{fig:Knots}) by the power law behavior:
\begin{equation}\label{eq:KnottingProbTheo}
P_{\rm knot}(N) = \left( \frac{N}{N_{\rm knot}} \right)^{\alpha_{\rm knot}} \, ,
\end{equation}
with~\cite{NoteOnErrorBars}
\begin{eqnarray}
N_{\rm knot} & = & 2203 \pm 433 \, , \label{eq:KnottingProbTheo-N0} \\
\alpha_{\rm knot} & = & 1.9 \pm 0.2 \, . \label{eq:KnottingProbTheo-alpha}
\end{eqnarray}
The value for $\alpha_{\rm knot}$ (Eq.~\eqref{eq:KnottingProbTheo-alpha}) is compatible with the fact that knots form ``cooperatively'', due to random SC's between pairs of polymer strands.
Furthermore, by extrapolation to large $N$, Eq.~\eqref{eq:KnottingProbTheo} implies that rings with $N\gtrsim N_{\rm knot}$ are always knotted on average.

Interestingly, Eq.~\eqref{eq:KnottingProbTheo} appears in contrast with the study~\cite{lang2012effect} showing that for catenated $N$-monomer rings in solution one finds $P_{\rm knot}(N) \simeq 1 - \exp(-N/N_0) \sim N/N_0$, where $N_0$ is some characteristic polymer length. 
Noticeably, this matches the known conjecture~\cite{OrlandiniWhittington-RMP2007} that the unknotting probability for {\it ideal}, closed lattice polygons decays exponentially with the chain contour length.

We speculate briefly on the discrepancy between this and our result Eq.~\eqref{eq:KnottingProbTheo} with exponent $\alpha_{\rm knot} \simeq 2$ (Eq.~\eqref{eq:KnottingProbTheo-alpha}). 
In order to enforce the strand crossing mechanism, in Ref.~\cite{lang2012effect} rings are simulated via the bond fluctuation model~\cite{CarmesinKremer1988,Paul1991} with the addition of a set of diagonal moves which switch temporarily off all the entanglements (see comments in Sec.~\ref{sec:Comparison2PreviousWorks}).
In this sense, the SC mechanism implemented in~\cite{lang2012effect} is somehow reproducing the features of an ideal polymer and for this reason the reported knotting probability decays exponentially as conjectured~\citep{OrlandiniWhittington-RMP2007} in ideal rings. 
Instead our rings are not ideal (Fig.~\ref{fig:gyration_radius}(b)), two nearby strands are never allowed to overlap and the physical entanglements resulting from the uncrossability~\cite{DoiEdwardsBook,RubinsteinColbyBook} between nearby polymer strands are resolved through a more rigorous mechanism (Fig.~\ref{fig:swapping_moves_polymer}).

\subsection{Structure and dynamics of ring polymers with SC's}\label{sec:Network-Str+Dyn}

\subsubsection{Physical links}\label{sec:Physical-Links}
The physical links between any given pair of rings ${\mathcal R}_1$ and ${\mathcal R}_2$ have been quantified in terms of the corresponding {\it Gauss linking number}~\cite{OrlandiniWhittington-RMP2007,Micheletti-PhysRep2011}:
\begin{equation}\label{eq:GaussLN}
{\mathcal G} \equiv \frac1{4\pi} \oint_{{\mathcal R}_1} \oint_{{\mathcal R}_2} \frac{(\vec r_2 - \vec r_1) \cdot (d{\vec r}_2 \wedge d{\vec r}_1)}{|\vec r_2 - \vec r_1|^3} \, ,
\end{equation}
where $\vec r_1$ (respectively, $\vec r_2$) is the spatial coordinate for a point on the (oriented) contour line formed by ring ${\mathcal R}_1$ (resp., ring ${\mathcal R}_2$) and $d\vec r_1$ (resp., $d\vec r_2$) is the corresponding infinitesimal increment.
As in the case of the Alexander polynomials (used in Sec.~\ref{sec:Knots}), $\mathcal G$ is also a {\it topological invariant}: physically, it represents the number of times (with ``$+$'' or ``$-$'' sign, depending on the reciprocal orientations of the curves) that each curve winds around the other.
For our rings modeled as discretized closed paths on the $3d$ FCC lattice,
Eq.~\eqref{eq:GaussLN} has been evaluated numerically by employing the efficient algorithm by Klenin and Langowski~\cite{KleninLangowski2000}.

\begin{table*}
\begin{tabular}{cccc}
\hline
\hline
\, \, $N$ \, \, & ${\mathcal G}_0$ & $\langle {\rm LD} \rangle$ & $p_{\rm link}$ \\ 
\hline
\\
40 & \, \, $0.164\pm0.004$ \, \, & $0.376\pm0.001$ & $(7.32 \pm 0.02) \cdot 10^{-5}$ \\ 
80 & $0.22\pm0.01$ & $1.143 \pm 0.003$ & $(4.46 \pm 0.01) \cdot 10^{-4}$ \\ 
160 & $0.29\pm0.01$ & \, \, $2.754\pm 0.004$ \, \, & $(2.153 \pm 0.003) \cdot 10^{-3}$ \\ 
320 & $0.36\pm0.01$ & $5.77\pm 0.01$ & $(9.00 \pm 0.02) \cdot 10^{-3}$ \\ 
640 & $0.46\pm0.01$ & $10.92 \pm 0.08$ & $(3.29 \pm 0.02) \cdot 10^{-2}$ \\ 
\hline
\hline
\end{tabular}
\caption{
Network properties in melts of $N$-monomer rings with strand crossings.
(i)
${\mathcal G}_0$: decay length for the probability distribution function of the Gauss linking number, $P(N; {\mathcal G}) \sim e^{-|{\mathcal G}| / {\mathcal G}_0(N)}$ (see Fig.~\ref{fig:Links} and Fig.~\ref{fig:GPDF-noAbsolute} in SM).
(ii) 
$\langle {\rm LD} \rangle$: mean linking degree.
(iii) 
$p_{\rm link}$: mean fraction of pairs of linking rings. 
}
\label{tab:NetworkStructure}
\end{table*}
\begin{figure}
\includegraphics[width=0.45\textwidth]{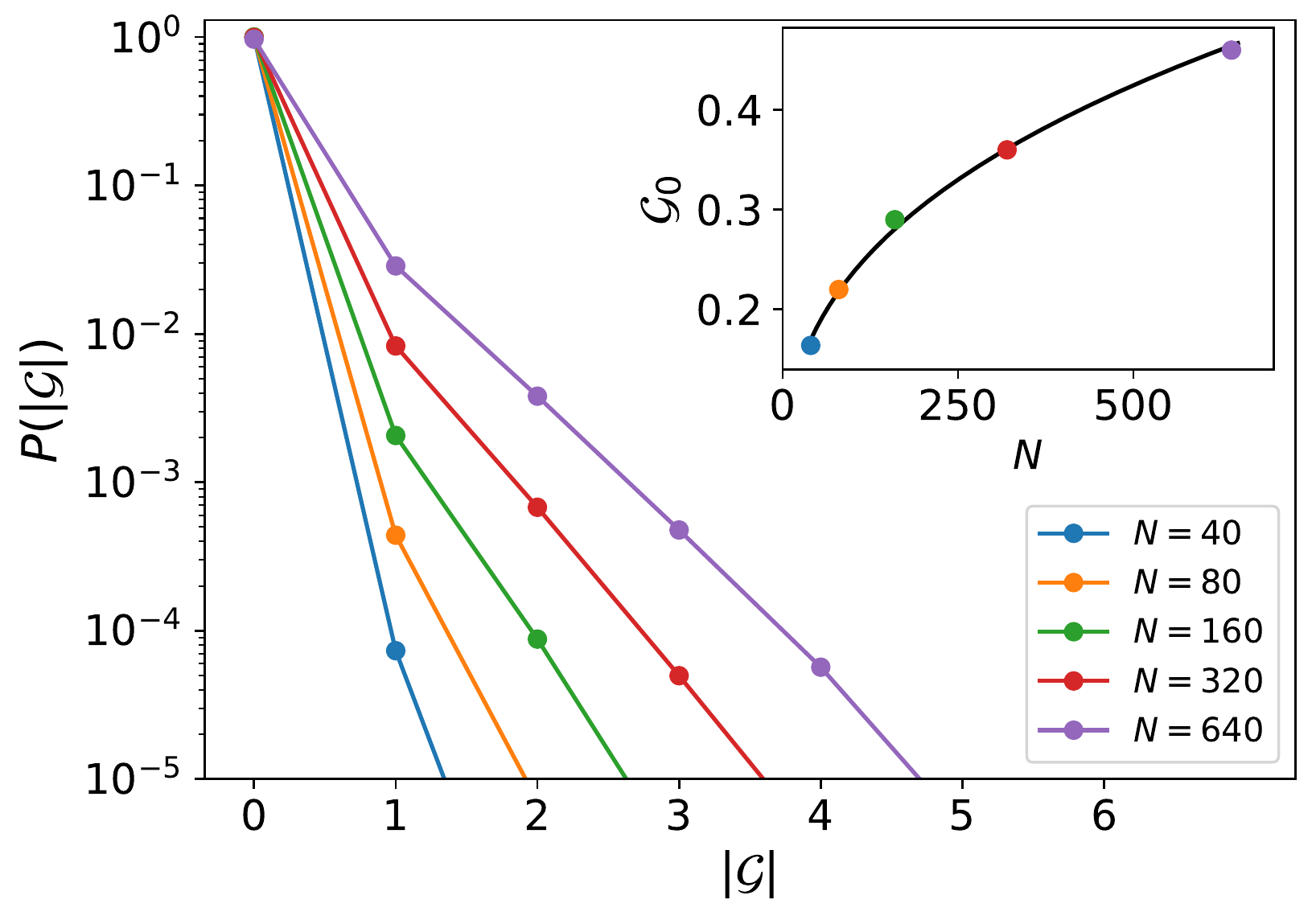}
\caption{
Probability distribution function, $P(|\mathcal{G}|)$, of the absolute value, $|\mathcal G|$, of the Gauss linking number between pairs of $N$-monomer rings.
Inset: Mean absolute Gauss linking number $\langle |{\mathcal G}| \rangle$ {\it vs.} $N$ (symbols) and best fit to power-law behavior (line, Eq.~\eqref{eq:GaussDecayLength} with Eqs.~\eqref{eq:GaussDecayLength-N0} and~\eqref{eq:GaussDecayLength-alpha}).
Color code is as in Fig.~\ref{fig:Knots}.
}
\label{fig:Links}
\end{figure}

To validate the method we verify first that the distribution functions for $\mathcal G$, $P(N; \mathcal G)$, are symmetric around ${\mathcal G}=0$ (Fig.~\ref{fig:GPDF-noAbsolute} in SM). 
Then, for $|\mathcal G| \geq 1$~\cite{PGinG0-Note} $P(N; \mathcal G)$ follows the exponential decay $\sim e^{-|{\mathcal G}| / {\mathcal G}_0(N)}$ (see Fig.~\ref{fig:Links}).
The ``decay length'' $\mathcal G_0(N)$ as a function of $N$ 
(for the specific values, see Table~\ref{tab:NetworkStructure}) is well described (symbols {\it vs.} line in the inset of Fig.~\ref{fig:Links}) by the power law behavior:
\begin{equation}\label{eq:GaussDecayLength}
{\mathcal G}_0(N) = \left( \frac{N}{N_{\rm link}} \right)^{\alpha_{\rm link}} \, ,
\end{equation}
with~\cite{NoteOnErrorBars}
\begin{eqnarray}
N_{\rm link} & = & 5277 \pm 239 \, , \label{eq:GaussDecayLength-N0} \\
\alpha_{\rm link} & = & 0.363 \pm 0.005 \, . \label{eq:GaussDecayLength-alpha}
\end{eqnarray}
The reported value for $\alpha_{\rm link}$, close to the scaling exponent $\nu$ of the gyration radius of the ring (Fig.~\ref{fig:gyration_radius}(a)), is consistent with the intuitive picture that two rings link to each other if the spatial distance between the corresponding centers of mass is of the order or smaller than $R_g(N) \sim N^\nu$ (Eq.~\eqref{eq:Rg2}). 

\begin{figure}
$$
\begin{array}{c}
\includegraphics[width=0.45\textwidth] {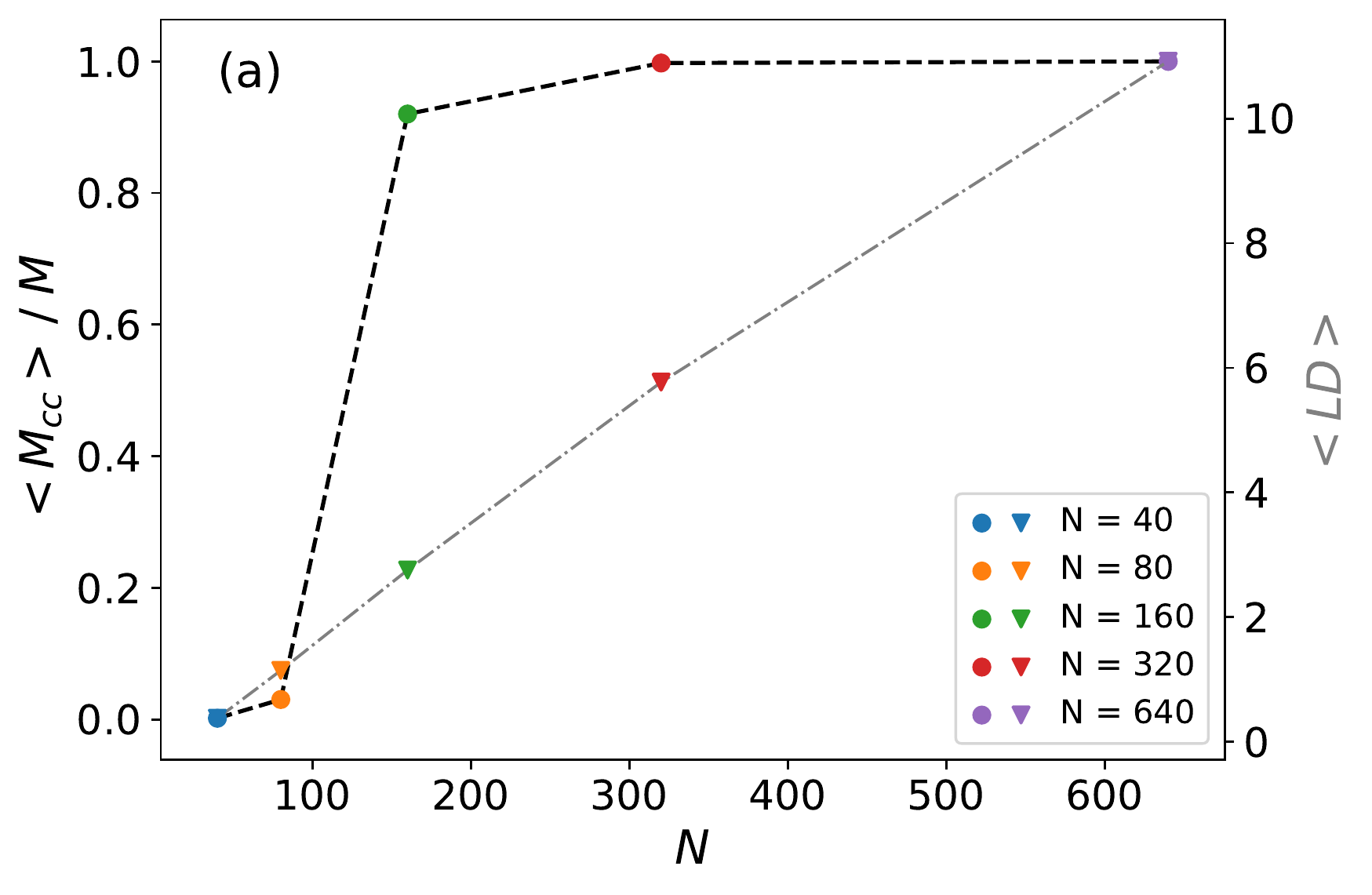} \\
\includegraphics[width=0.45\textwidth] {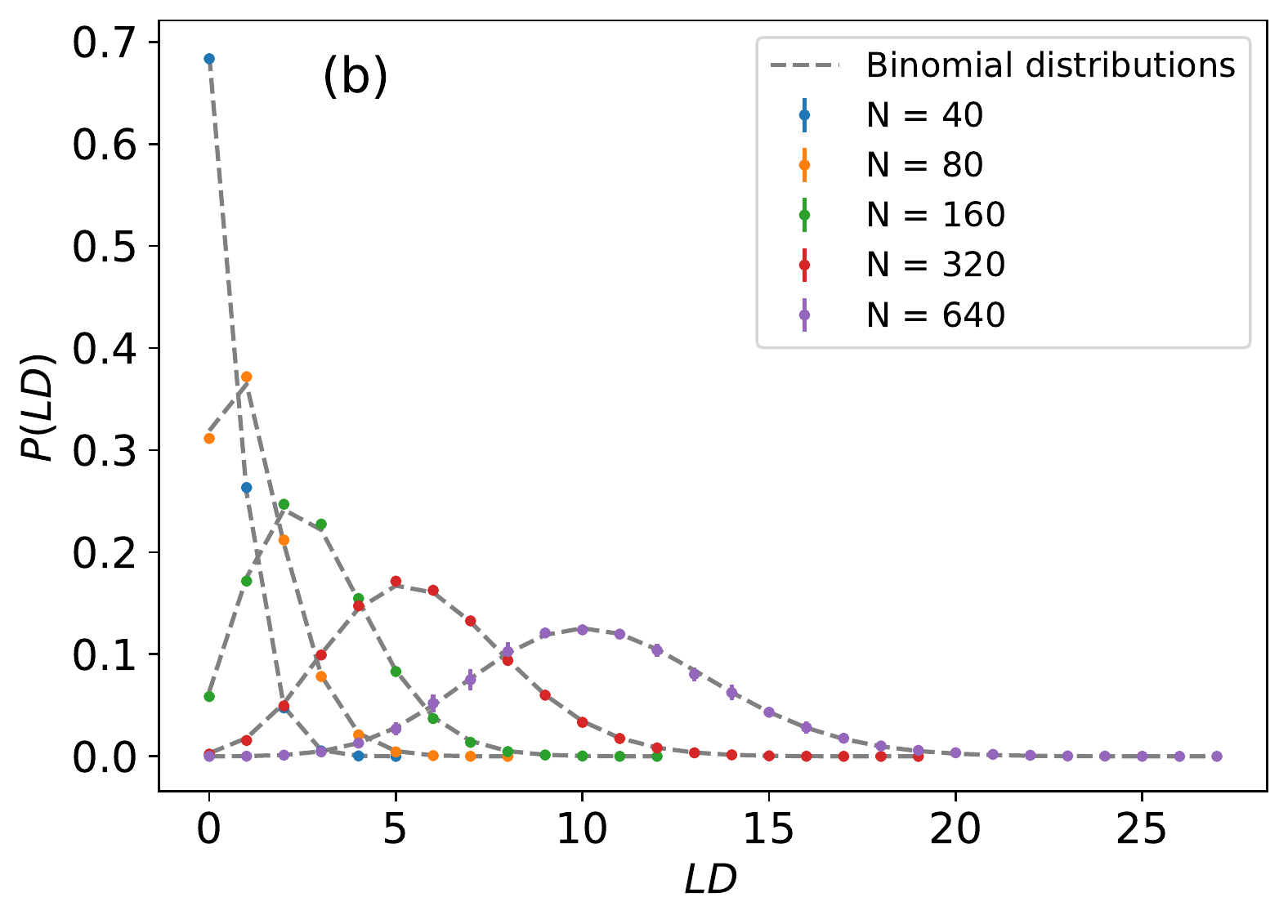} 
\end{array}
$$
\caption{
(a)
Mean ring fraction in the largest {\it connected component}, $\langle M_{\rm cc}(N)\rangle / M$ (black line, left y-axis),
and
mean {\it linking degree}, $\langle {\rm LD}(N) \rangle$ (grey line, right y-axis)
as a function of $N$. 
(b)
The frequency of observing a ring linking to, respectively, ${\rm LD} = 0, 1, ..., M-1$ other rings (symbols) in comparison to the binomial function (Eq.~\eqref{eq:BinomialFunct}) for random graphs.
Color code is as in Fig.~\ref{fig:Knots}.
}
\label{fig:Network-Str+Dyn}
\end{figure}

By using the results on the Gauss linking number, 
we consider
(i) the mean {\it linking degree}, $\langle {\rm LD}(N) \rangle$, defined as the mean number of chains linking to a single ring
and
(ii) the mean ring fraction, $\langle M_{\rm cc}(N)\rangle / M$, belonging to the largest {\it connected component} of chains in the melt.
The results are shown in Fig.~\ref{fig:Network-Str+Dyn}(a).
$\langle {\rm LD}(N) \rangle$ increases linearly~\cite{LangMacromolecules2012} with $N$ and the largest ($\approx 10$) attained value is consistent with the characteristic number of $10-20$ chains~\cite{RosaEveraersPRL2014,PanyukovRubinsteinMacromolecules2016} protruding the volume occupied by a single ring in melt.
We see that for $\langle {\rm LD}(N) \rangle \approx 2$, {\it i.e.} when one ring is connected on average to two other rings, a single giant network is obtained (see also Fig.~\ref{fig:Representation_network} in SM for instantaneous snapshots of the networks for different $N$).

It is interesting to notice that, in agreement with previous studies~\cite{FischerSommerJCP2015,michieletto2015kinetoplast}, the network of connections has the structure of a {\it random graph}, hence the frequency of observing a ring linking to, respectively, ${\rm LD} = 0, 1, ..., M-1$ other rings is accurately described (symbols {\it vs.} lines in Fig.~\ref{fig:Network-Str+Dyn}(b)) by the binomial function:
\begin{equation}\label{eq:BinomialFunct}
P(M, p_{\rm link}; {\rm LD}) = \binom{M-1}{\rm LD} \, p^{\rm LD}_{\rm link} \, (1-p_{\rm link})^{M-1-{\rm LD}} \, .
\end{equation}
Eq.~\eqref{eq:BinomialFunct} is equivalent to the probability that a single node in a random graph made of $M$ nodes is connected to ${\rm LD}$ other nodes,
with $p_{\rm link}(N) = \langle {\rm LD}(N) \rangle / (M-1)$ (see values in Table~\ref{tab:NetworkStructure}) representing the {\it linking probability} or the fraction of distinct node-to-node links out of the $M(M-1)/2$ total possible combinations. 

\begin{figure*}
$$
\begin{array}{cc}
\includegraphics[width=0.46\textwidth]{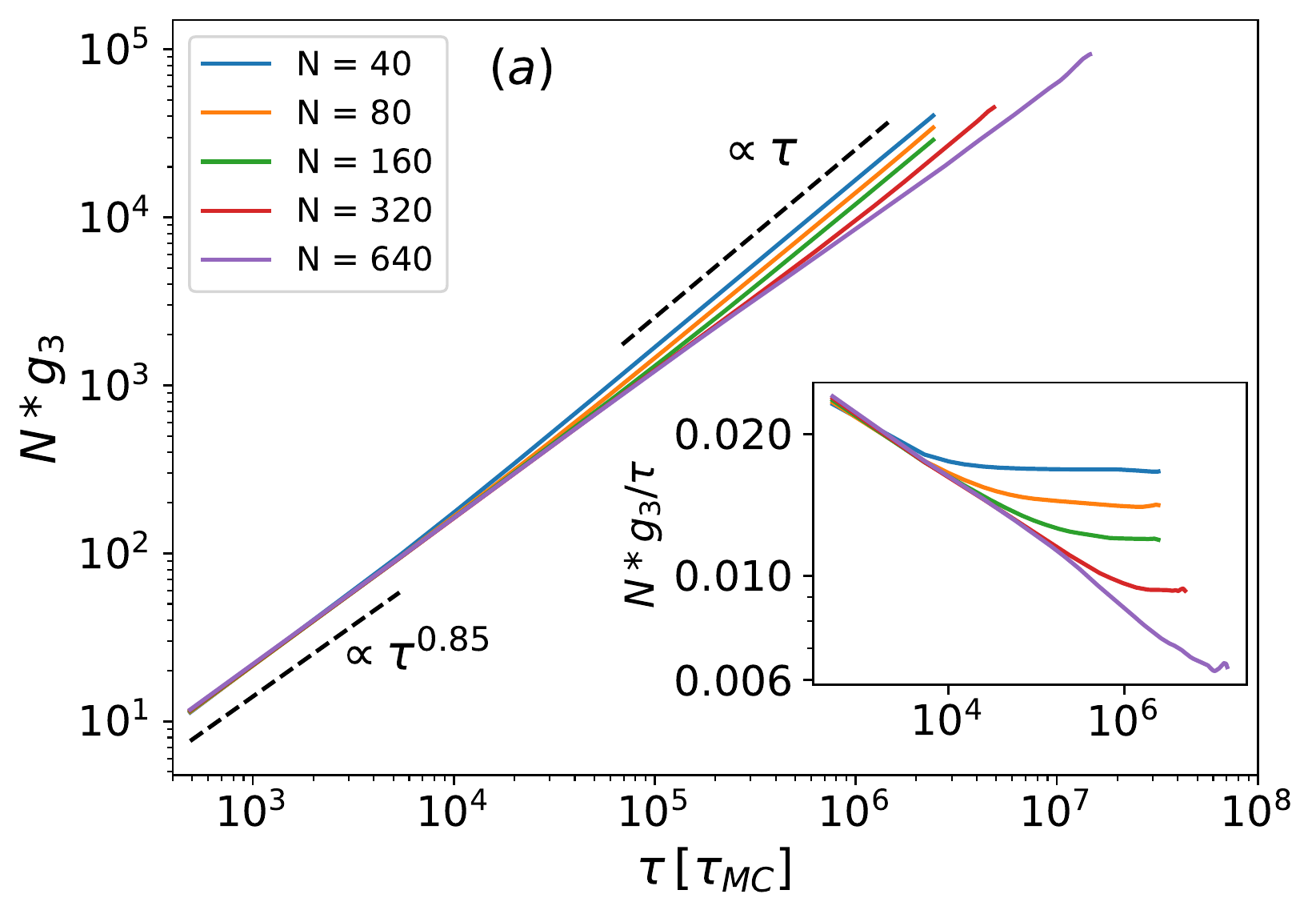} & \includegraphics[width=0.46\textwidth]{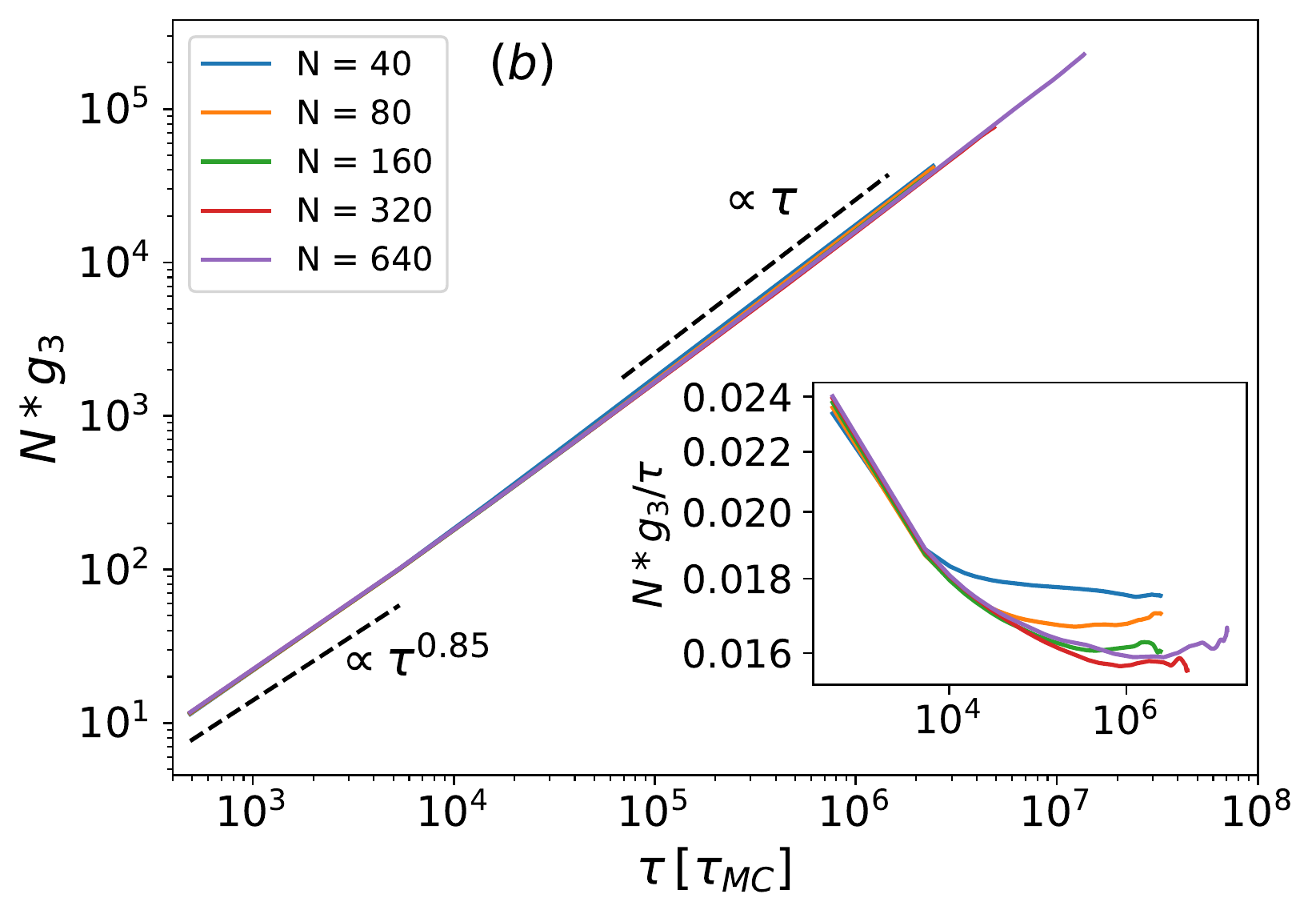} \\
\multicolumn{2}{c}{\includegraphics[width=0.46\textwidth]{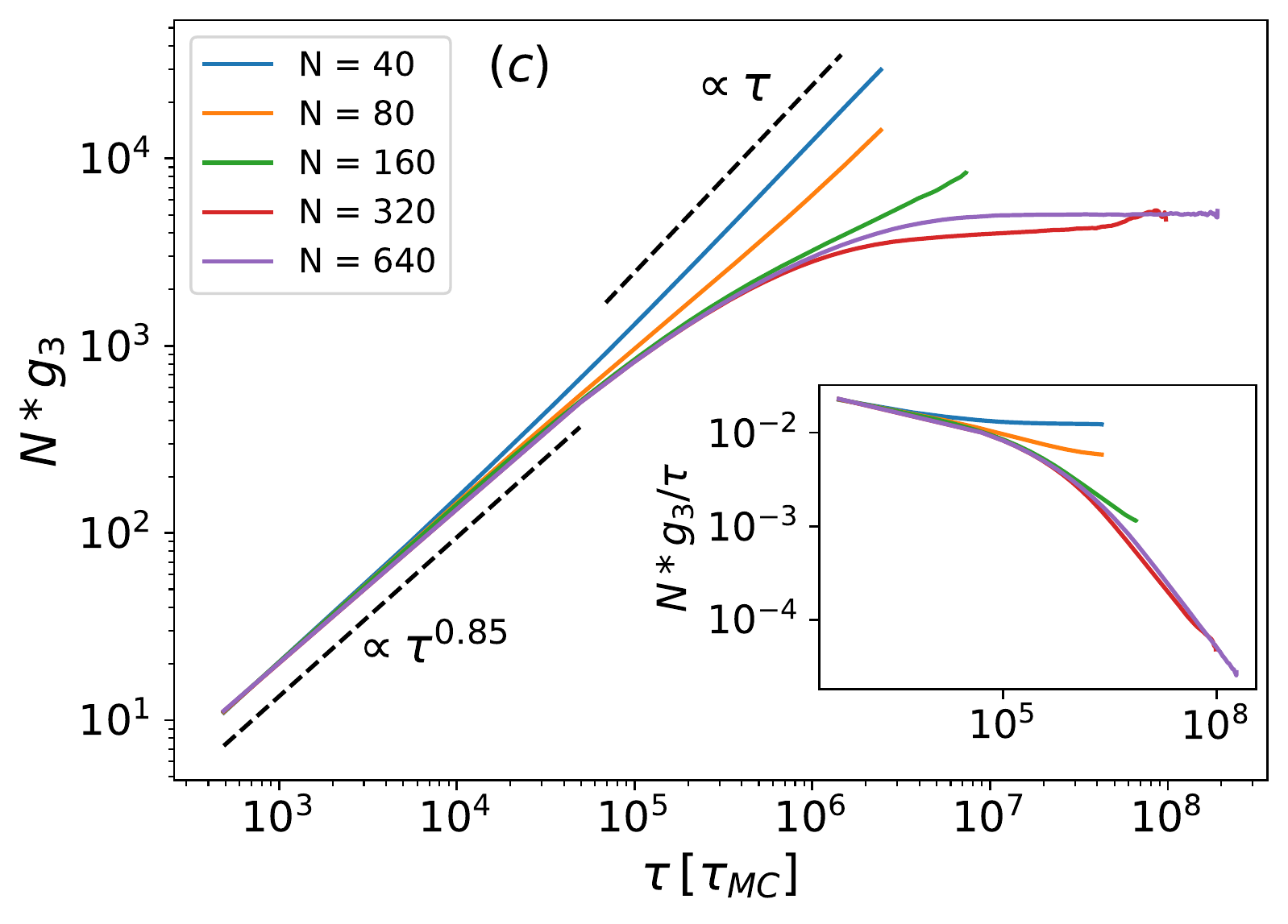}} 
\end{array}
$$
\caption{
Time mean-square displacement, $Ng_3(\tau)$, of the center of mass of $N$-monomer rings in melt. 
(a)
Melts of untangled rings. 
(b)
Melts of rings in the presence of SC's. 
(c)
Melts of permanently catenated rings. 
Curves of different colors are for different $N$ (see legends). 
The insets are for $Ng_3(\tau) / \tau$ as a function of $\tau$.
Color code is as in Fig.~\ref{fig:Knots}. 
}
\label{fig:SingleChainDynamics}
\end{figure*}
%

\subsubsection{Single chain and network dynamics}\label{sec:SingleChainNetwork-Dynamics}
We analyze first polymer dynamics in the different ensembles.
To this purpose, we consider the mean-square displacement of the spatial position, $\vec r_{\rm cm}(t)$, of the centre of mass of the chain~\cite{KremerGrest-JCP1990}:
\begin{equation}\label{eq:g3}
g_3(\tau) \equiv \langle \left( {\vec r}_{\rm cm}(t+\tau) - {\vec r}_{\rm cm}(t) \right)^2 \rangle \, ,
\end{equation}
as a function of time $\tau$. 

Unconstrained motion implies that $g_3(\tau) \propto \tau / N$ in the long-time regime.
Fig.~\ref{fig:SingleChainDynamics}(a) shows that this is not the case for untangled rings, in agreement with the original simulations by Schram and Barkema~\cite{Schram-LatticeModel2018}.
Conversely, introducing SC's into the system (Fig.~\ref{fig:SingleChainDynamics}(b)) 
removes the constraints and accelerates the dynamics to the extent that $g_3(\tau)$ is now proportional to $1/N$ (Fig.~\ref{fig:SingleChainDynamics}(b)).
Then, by the right amount of SC's, it is possible to ``resolve'' the entanglements induced by the presence of uncrossable strands and in this way to fluidize the polymer system.

In agreement with that, by again turning off the SC activity and then ``quenching'' the topology, polymer dynamics slows down dramatically (Fig.~\ref{fig:SingleChainDynamics}(c)) up to the complete arrest (evident in the saturation of $g_3(\tau)$ at large times).
Slow-down for $N=40$ and $N=80$ is due to the the fact that rings have linked into multi-chain structures (Fig.~\ref{fig:Representation_network}(a, b) in SM) which tend to move slower.
Starting from $N=160$ (Fig.~\ref{fig:Representation_network}(c, d, e) in SM) rings are locked together into a single, giant structure and, therefore, unable to perform large scale diffusion, hence the reported saturation.
This effect confirms experimental reports~\cite{SpakowitzPRL2018} of a rubber-like plateau in the storage modulus of topoII-inactivated solutions of catenated DNA rings.
Accordingly, the relative motion displayed by rings with $N=160$ (green line, Fig.~\ref{fig:SingleChainDynamics}(c)) is the consequence of the fact that a non negligible amount of unconcatenated rings is still undergoing random diffusion (see Fig.~\ref{fig:Representation_network}(c) and Fig.~\ref{fig:g3-single-quenched}(c) in SM). 
Notice that these dynamic effects appear on time scales larger than the imposed (Sec.~\ref{sec:PolymerModel}) time scale $\lambda_{\rm SC}^{-1} = 10^{4} \tau_{\rm MC}$ for SC's:
on time scales $\tau \lesssim \lambda_{\rm SC}^{-1}$, the three ensembles show the same subdiffusive behavior $\sim \tau^{0.85}$ characteristic~\cite{Schram-LatticeModel2018} of untangled rings.

\begin{figure}
\includegraphics[width=0.46\textwidth]{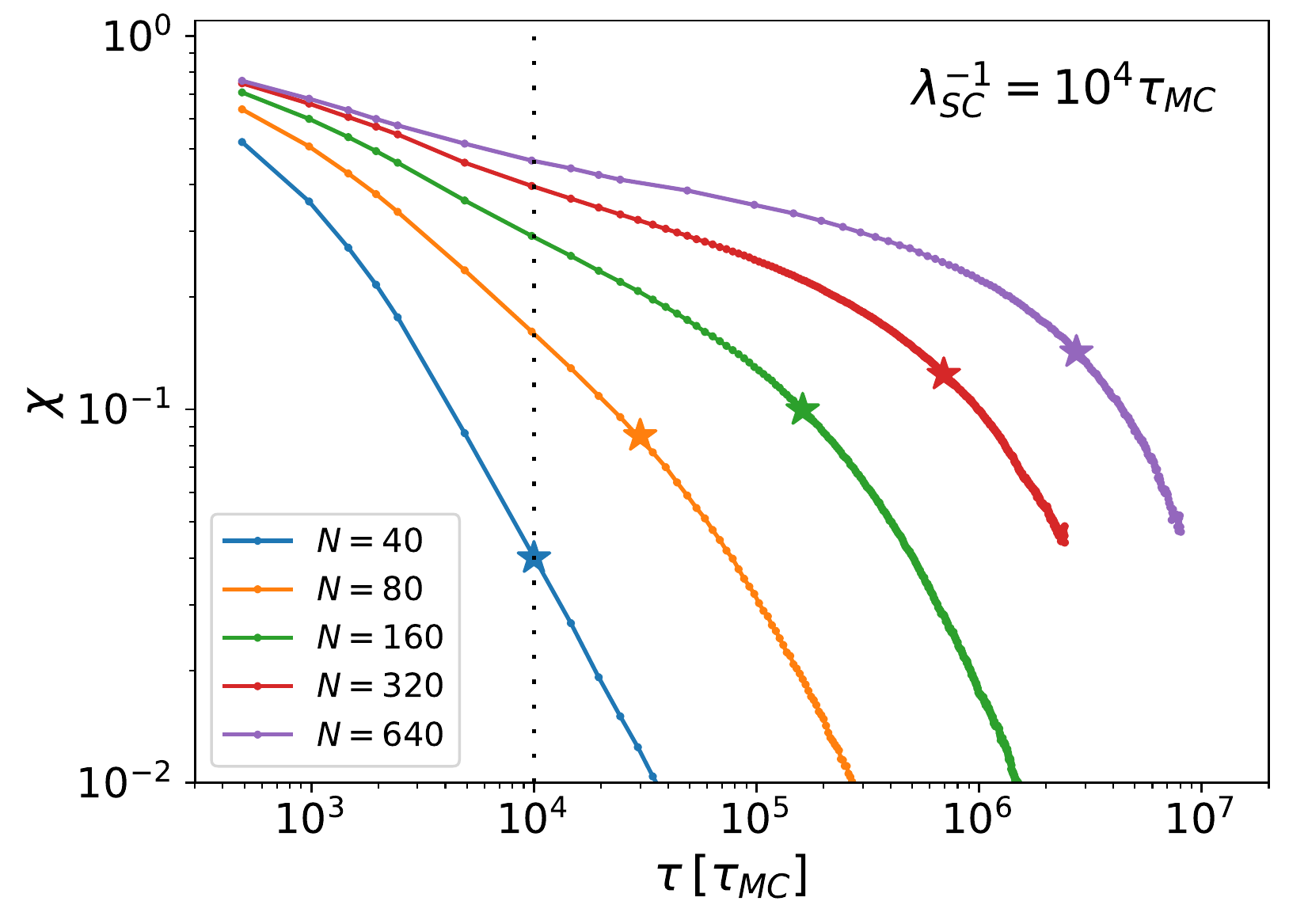}
\caption{
Time auto-correlation function $\chi(\tau)$ for two rings remaining linked on time scale $=\tau$ in the presence of SC's (Eq.~\eqref{eq:LinkLink-TimeCF}).
The vertical dotted line corresponds to the time scale $\lambda^{-1}_{\rm SC} =10^4\tau_{\rm MC}$ for SC.
The ``$\star$'s'' mark the positions of the polymer self-diffusion times $\tau_d(N)$ (see Table~\ref{tab:ComputationalCost}).
Color code is as in Fig.~\ref{fig:Knots}. 
}
\label{fig:LinkingTimeACF}
\end{figure}

To complement the analysis on ring dynamics (Fig.~\ref{fig:SingleChainDynamics}) in the presence of active SC's,
we characterize now the interplay between ring motion and the fluidization process induced by the SC mechanism from the point of view of the formed polymer network.
To this purpose, we introduce the characteristic function $C^{\rm link}_{ij}(t) = 1/0$ between the pair of rings $i$ and $j$ being linked/unlinked at time $t$, 
and calculate the corresponding time auto-correlation function:
\begin{equation}\label{eq:LinkLink-TimeCF}
\chi(\tau) \equiv \frac{\langle C^{\rm link}_{ij}(t+\tau) \, C^{\rm link}_{ij}(t)\rangle}{\langle C^{\rm link}_{ij}(t)^2\rangle} \, ,
\end{equation}
where the average is taken over all possible pairs $i$ and $j$.
The results for rings made of $N$ monomers are shown in Fig.~\ref{fig:LinkingTimeACF}. 
Qualitatively, we identify three regimes:
(i)
Below the SC time scale $\lambda^{-1}_{\rm SC} = 10^4\tau_{\rm MC}$, $\chi(\tau)$ displays power law decay.
(ii)
This is followed by a second regime which, by increasing $N$, becomes slower than the first one and attains a quasi-plateau.
Intuitively, this is due to the fact that on such time scales both linking and unlinking events may happen, while at times $\tau < \lambda^{-1}_{\rm SC}$ we expect on average only a single unlinking event.
(iii)
Finally, on time scales larger than the ring self-diffusion time $\tau_d(N)$ (corresponding to the time scale for the polymer to spread over a distance the size of its own mean gyration radius, $g_3(\tau_d(N)) \equiv \langle R_g^2(N)\rangle$, see Table~\ref{tab:ComputationalCost}), the two rings occupy, on average, distinct regions in space and $\chi(\tau)$ decays as an exponential.
The ``persistent'' regime (ii) valid for long chains is particularly noteworthy, because it suggests that with SC's at work rings coalesce into a ``dynamic'' gel-like structure.

\section{Discussion}\label{sec:Discussion}
Our melts of rings with active SC's form transient networks (Fig.~\ref{fig:LinkingTimeACF}) which, in spite of the non negligible amount of introduced linking (Fig.~\ref{fig:Links}), move faster than in the untangled case (Fig.~\ref{fig:SingleChainDynamics}, panel (a) vs. panel (b)).
Physically this happens because SC's operate at a reasonably fast rate ($\lambda_{\rm SC}^{-1}=10^4 \, \tau_{\rm MC}$, see Sec.~\ref{sec:PolymerModel}), guaranteeing rapid linking/unlinking events which maintain rings only ``loosely'' entangled with each other.

By the same argument one may imagine that, by opportunely slowing down the SC rate, it ought to be possible to produce systems of (temporarily) interlocked rings whose dynamics is actually {\it slower} than in untangled melts.
Intuitively this situation can be realized by choosing $\lambda_{\rm SC}^{-1}$ to be of the same order or larger than the self-diffusion time $\tau_d$ of rings in untangled melts since, supposedly, during this time scale a single polymer has interacted with the chains to which it is effectively able to link.

\begin{figure}
\includegraphics[width=0.45\textwidth] {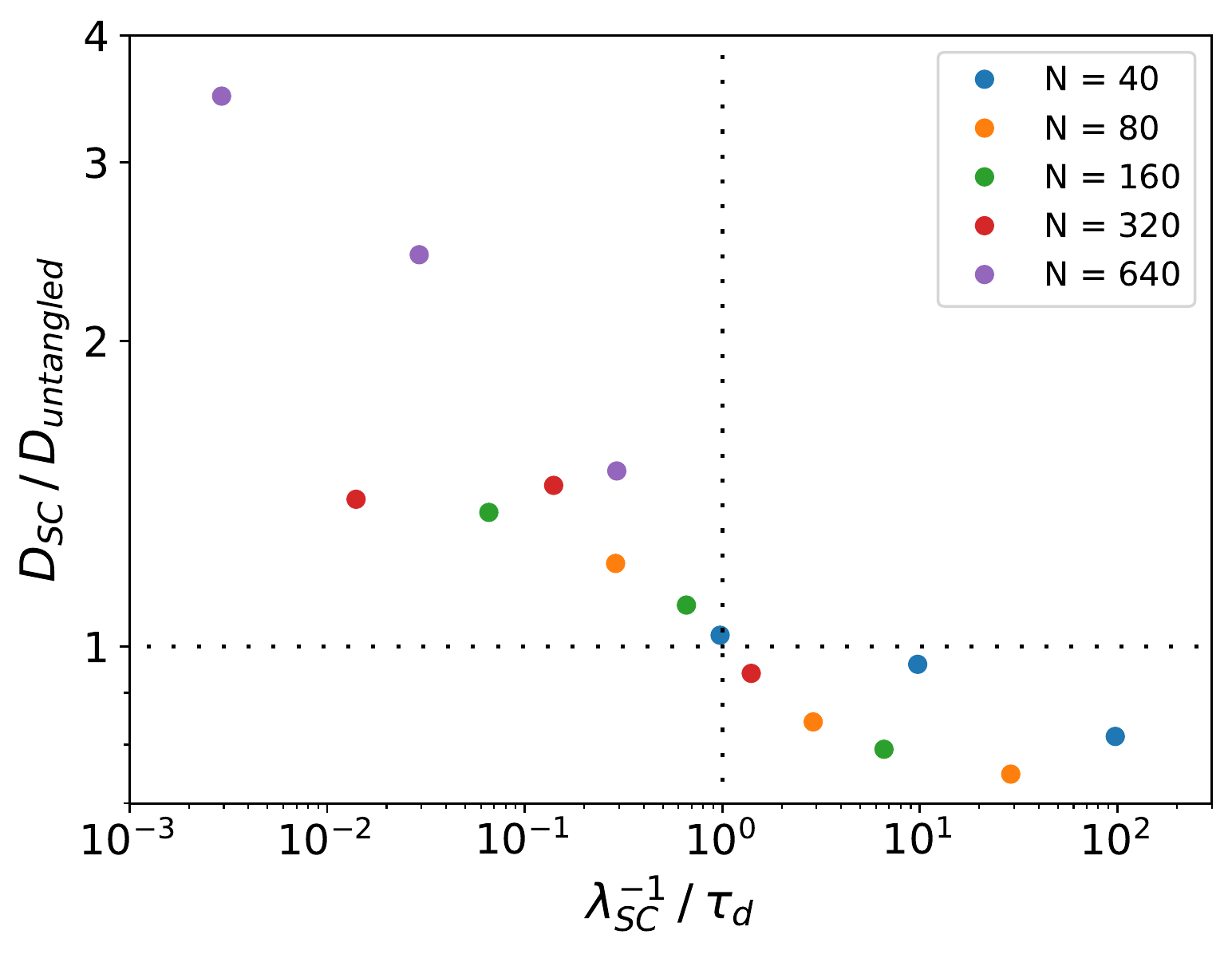}
\caption{
Asymptotic diffusion coefficient of rings with active SC's normalized to the corresponding values in untangled melts as a function of the inverse of the SC rate, $\lambda^{-1}_{\rm SC} \, / \, \tau_d(N)$, normalized to the ring self-diffusion time in untangled melts (see Table~\ref{tab:ComputationalCost}).
Color code is as in Fig.~\ref{fig:Knots}.
}
\label{fig:Diff-for-DifferentLambdas}
\end{figure}

To validate this idea (which may be also tested experimentally, for instance by resorting to DNA rings~\cite{SpakowitzPRL2018}), we performed new simulations for the same melts of rings but with the two different rates
$\lambda_{\rm SC}^{-1} = 10^{5} \, \tau_{\rm MC}$
and
$\lambda_{\rm SC}^{-1} = 10^{6} \, \tau_{\rm MC}$,
{\it i.e.} ten and one hundred times slower than the previous one.
We have then estimated the asymptotic diffusion coefficients of the rings by best fits of the terminal behaviors of the corresponding mean-square displacements, $g_3(\tau) / \tau$ (see Fig.~\ref{fig:comparison_rate_main} in SM), normalized to time $\tau$.
The results (normalized to the corresponding values for untangled melts) {\it vs.} the inverse of the SC rate, $\lambda^{-1}_{\rm SC} \, / \, \tau_d(N)$, normalized to the polymer self-diffusion times in untangled melts are shown in Fig.~\ref{fig:Diff-for-DifferentLambdas}.
The plots confirm our expectations: slow SC rates result in melts with slower relaxation dynamics compared to the untangled case.
Notice that while the precise value of the SC rate affects the dynamics of the melt, static quantities like the gyration radius of the ring or the Gauss linking number (see, respectively, Figs.~\ref{fig:Comparison_rates_GR} and~\ref{fig:Comparison_rates_P_G} in SM) do not change for the different set-up's.

It is also worth noticing that, even in those cases where SC accelerates dynamics with respect to the untangled case, the asymptotic behavior is preceded (see Fig.~\ref{fig:comparison_rate_main} in SM) by a time regime where the action of SC's make the rings temporarily slower.
It is not difficult to understand the reason.
Take, for instance, the blue and orange curves for the ``$N=640$''-rings in Fig.~\ref{fig:comparison_rate_main} in SM on the time scales $\tau / \tau_{\rm MC} \lesssim 10^5$.
On the same time scales, the time auto-correlation function $\chi(\tau)$ for the link between two rings (see the violet line in the bottom panel of Fig.~\ref{fig:ChiTau-DifferentLambdas} in SM) displays a very slow decay, meaning that the slow down compared to the untangled case is arguably due to the slow dynamics of the linked rings.

\section{Conclusions}\label{sec:Concls}
Motivated by recent experiments~\cite{SpakowitzPRL2018} employing the enzyme topoII to induce the fluidization of entangled polymer solutions of DNA rings,
we have introduced a dynamic Monte Carlo computational scheme for polymer chains on the FCC lattice which takes explicitly into account the action of the enzyme by controlling the rate at which two nearby polymer strands are able to cross through each other.
By applying then the model to ring polymers made of $N$ monomers and in melt conditions,
we discuss how the strand crossing mechanism influences both the static and the dynamic properties of the chains.

At stationary conditions ring polymers swell with respect to the untangled ({\it i.e.}, unknotted and unconcatenated) case and stay non-ideal (Fig.~\ref{fig:gyration_radius}).
On the other hand, they tend to become increasingly knotted (Fig.~\ref{fig:Knots}) and to form a macroscopic network of linked chains (Fig.~\ref{fig:Links} and Fig.~\ref{fig:Network-Str+Dyn}).
Interestingly, the probability that rings remain unknotted (Fig.~\ref{fig:Knots}, inset) appears to decay with $N$ {\it faster} than the exponential function predicted for ideal rings with unconstrained topology,
and this finding was explained based on the consideration that knots form through the random crossings between {\it pairs} of polymer strands.

On the dynamics side, we show (Fig.~\ref{fig:SingleChainDynamics}) that the ability to produce strand crossings make polymers faster and that large rings tend to ``glue'' together into a permanent gel as soon as crossings are not let anymore.
Yet an acceleration of the dynamics is not true {\it in general}, but only when the rate for strand crossings is fast enough.
In the opposite case the dynamics of the melt may be even slower than the untangled case (Fig.~\ref{fig:Diff-for-DifferentLambdas}), a prediction which might be put at test by using, again, DNA rings in the presence of topoII.

We conclude on a technical remark.
Notice that the model presented here is for flexible chains (Sec.~\ref{sec:Properties-PolymerSols}) while polymers in general, and DNA in particular~\cite{MarkoSiggia1995}, are more like semi-flexible {\it i.e.} locally stiff~\cite{RubinsteinColbyBook}.
The inclusion of a bending penalty term in our model is not presenting particular technical difficulties and its consequences on the topological properties of ring polymers will be examined in future studies.

{\it Acknowledgements} -- The authors would like to acknowledge the networking support by the ``European Topology Interdisciplinary Action'' (EUTOPIA) CA17139.

\bibliography{biblio}

\begin{thebibliography}{47}
\bibitem [{Not()}]{NoteOnErrorBars}%
  \BibitemOpen
  \href@noop {} {}\bibinfo {note} {Mean values and corresponding error bars for
  $N_{\rm knot}$ and $\alpha_{\rm knot}$ (Eq.~\eqref{eq:KnottingProbTheo}) were
  calculated by averaging distinct estimates of these parameters. These
  estimates were obtained by separate best fits of the data to
  Eq.~\eqref{eq:KnottingProbTheo} in the linear-linear and log-log
  representations. The same procedure was used for $N_{\rm link}$ and
  $\alpha_{\rm link}$ (Eq.~\eqref{eq:GaussDecayLength}).}\BibitemShut {Stop}%
\bibitem [{PGi()}]{PGinG0-Note}%
  \BibitemOpen
  \href@noop {} {}\bibinfo {note} {Notice that since $\mathcal G$ has been
  computed between all possible ring pairs, the value for $P(N; {\mathcal
  G}=0)$ does depend on the total number of chains in the sample (in an
  infinite system, there is an infinite number of unlinked chains). Conversely,
  two chains are linked only if they are spatially close to each other and so
  each value $P(N; {\mathcal G}\geq 1)$ is expected to correspond to ``bulk''
  properties.}\BibitemShut {Stop}%
\end{thebibliography}

\clearpage

\widetext
\clearpage
\begin{center}
\textbf{\Large Computer simulations of melts of ring polymers with non-conserved topology: A dynamic Monte Carlo lattice model \\ \vspace*{1.5mm} -- Supplemental Figures --} \\
\vspace*{5mm}
Mattia Alberto Ubertini, Angelo Rosa
\vspace*{10mm}
\end{center}
\balancecolsandclearpage

\setcounter{equation}{0}
\setcounter{figure}{0}
\setcounter{table}{0}
\setcounter{page}{1}
\setcounter{section}{0}
\makeatletter
\renewcommand{\theequation}{S\arabic{equation}}
\renewcommand{\thefigure}{S\arabic{figure}}
\renewcommand{\thetable}{S\arabic{table}}
\renewcommand{\thesection}{S\arabic{section}}

\makeatletter
\@fpsep\textheight
\makeatother


%
\begin{figure*}[p!]
$$
\begin{array}{cc}
\includegraphics[width=0.44\textwidth]{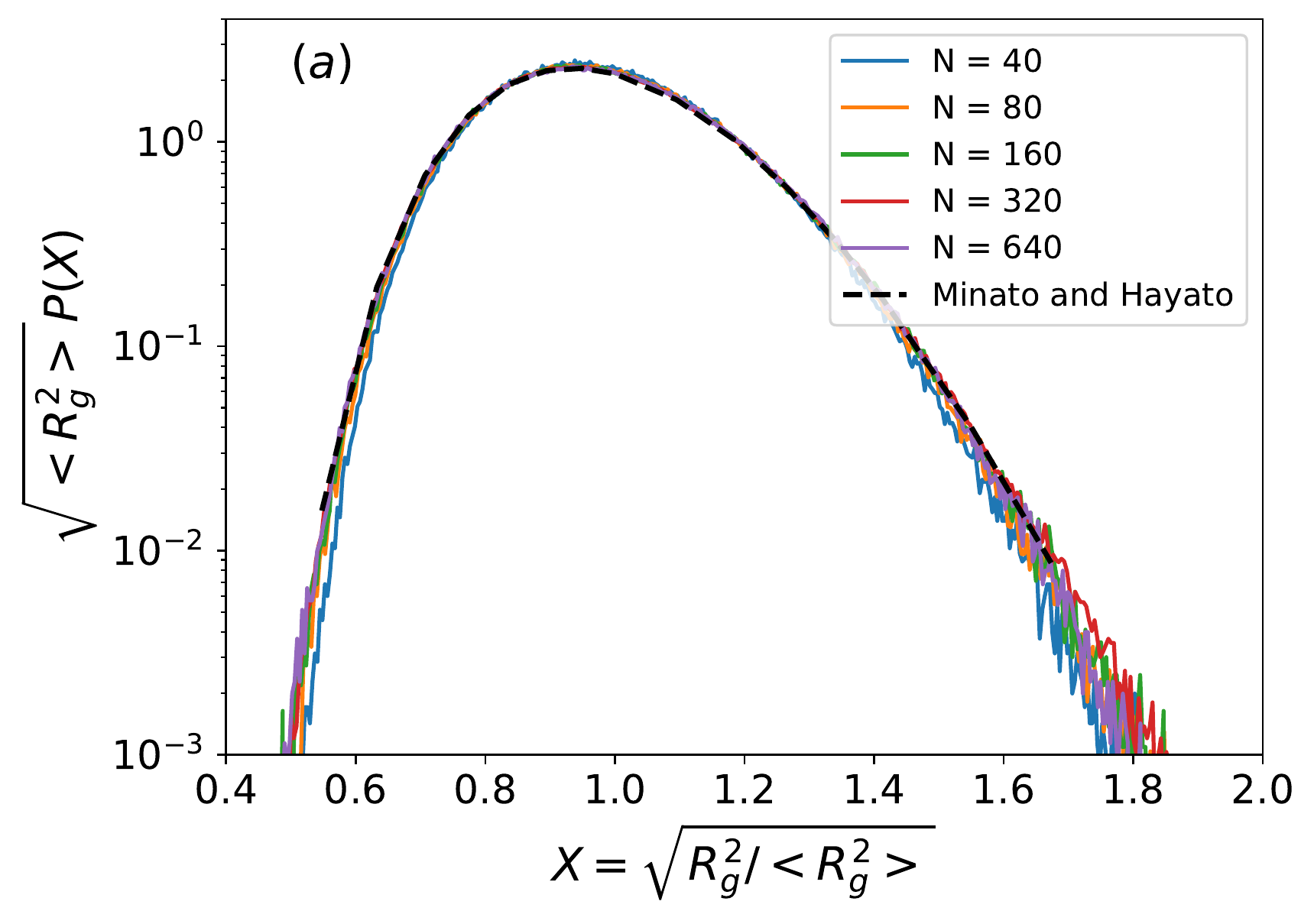} & \includegraphics[width=0.45\textwidth]{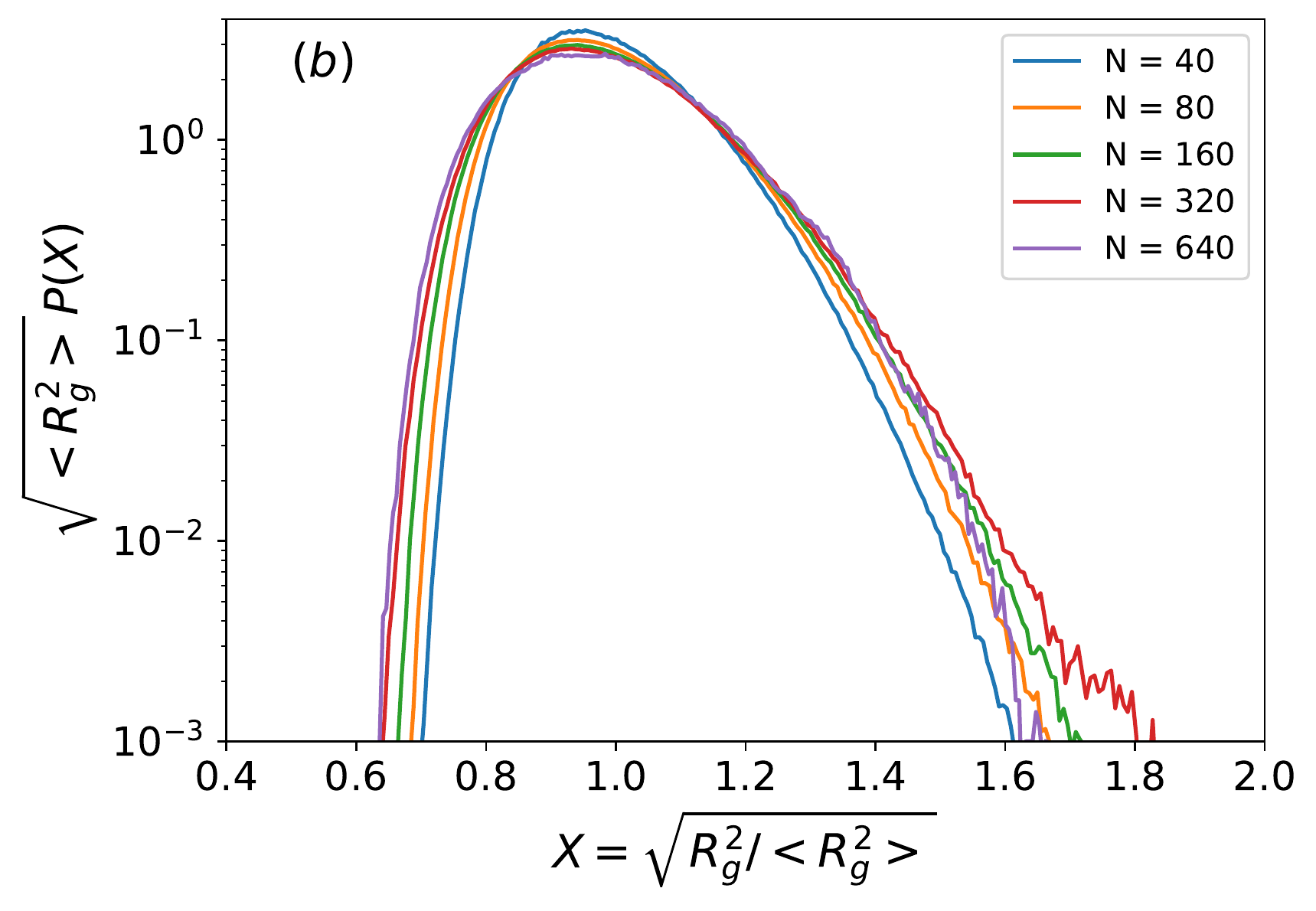} \\
\includegraphics[width=0.45\textwidth]{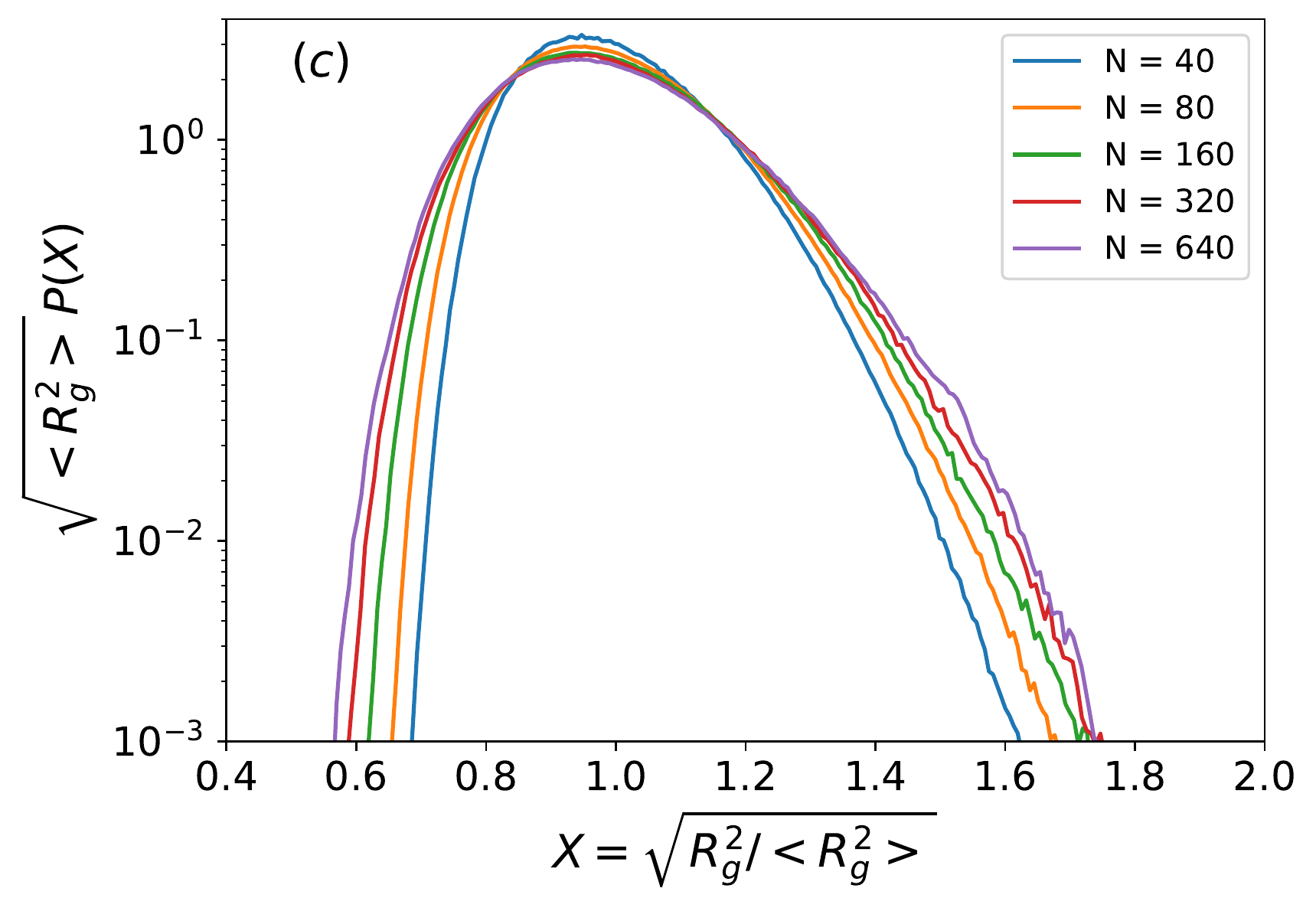} & \includegraphics[width=0.45\textwidth]{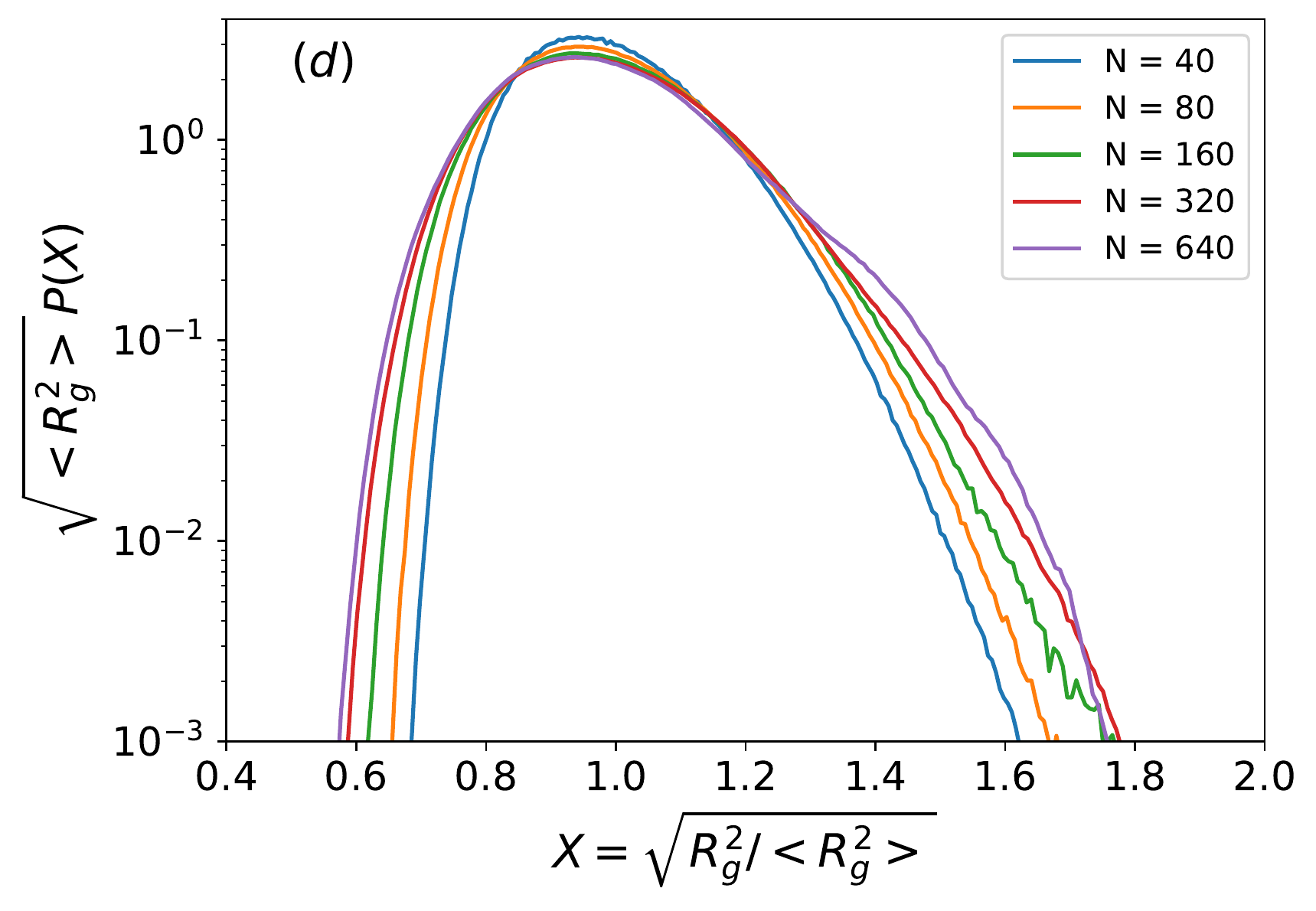}
\end{array}
$$
\caption{
Probability distribution functions, $P(R_g / \sqrt{\langle R_g^2 \rangle})$, of the gyration radius of $N$-monomer rings (see legends for details).
Results for:
(a)
Ideal rings. The dashed line corresponds to the exact expression by Minato and Hatano~\cite{minato1977distribution}.
(b)
Melts of untangled rings.
(c)
Melts of rings with strand crossings.
(d)
Melts of permanently catenated rings.
}
\label{fig:Rg-PDFs}
\end{figure*}
\begin{figure*}[p!]
\includegraphics[width=0.50\textwidth]{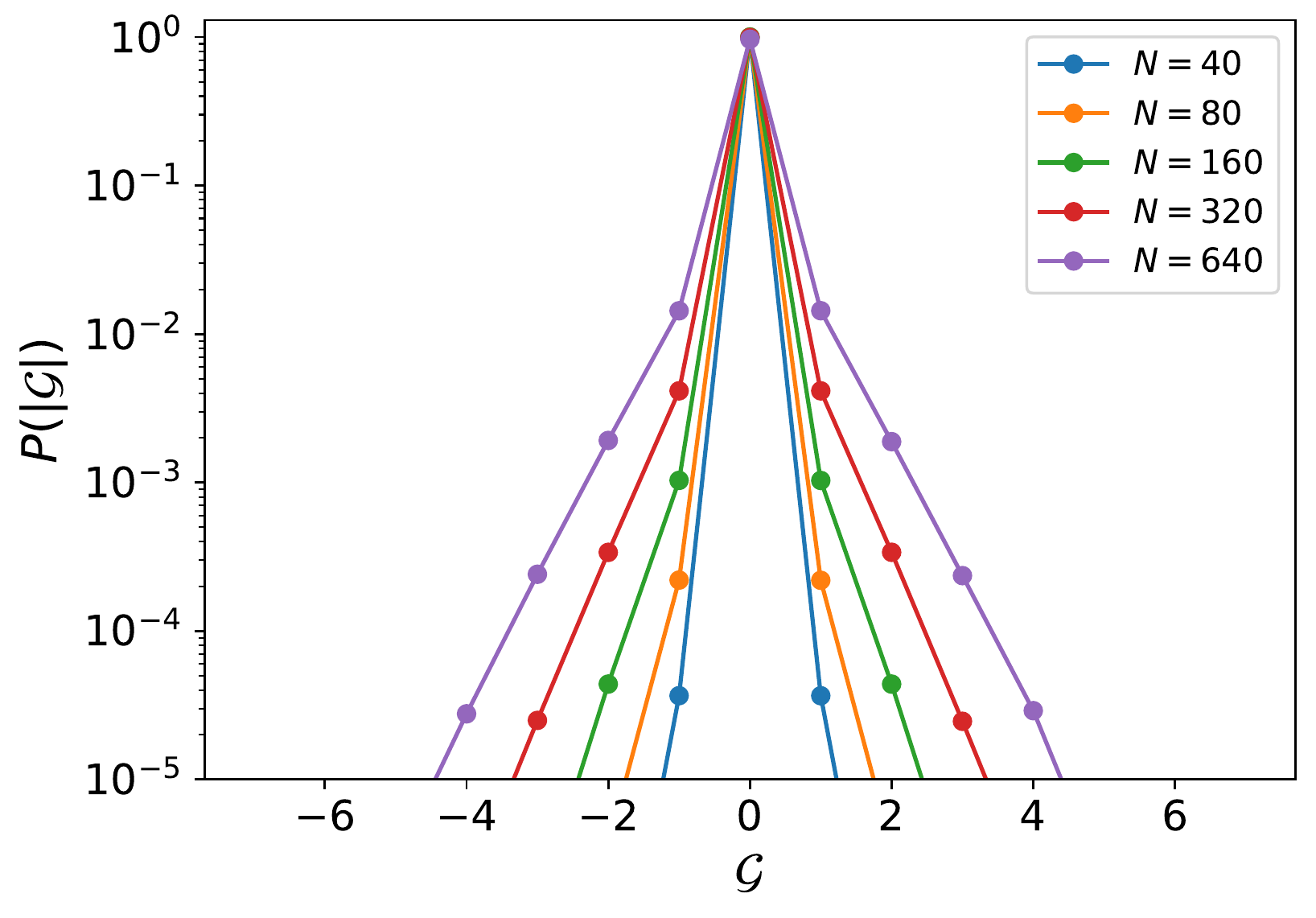}
\caption{
Probability distribution function, $P(\mathcal{G})$, of the Gauss linking number $\mathcal G$ between pairs of $N$-monomer rings.
The function is perfectly symmetric around $\mathcal G=0$, which validates the approach used to detect $\mathcal G$ and based on the algorithm~\cite{KleninLangowski2000}.
}
\label{fig:GPDF-noAbsolute}
\end{figure*}
%


%
\begin{figure*}[p!]
\label{fig:Network}
$$
\begin{array}{cc}
\includegraphics[width=0.45\textwidth]{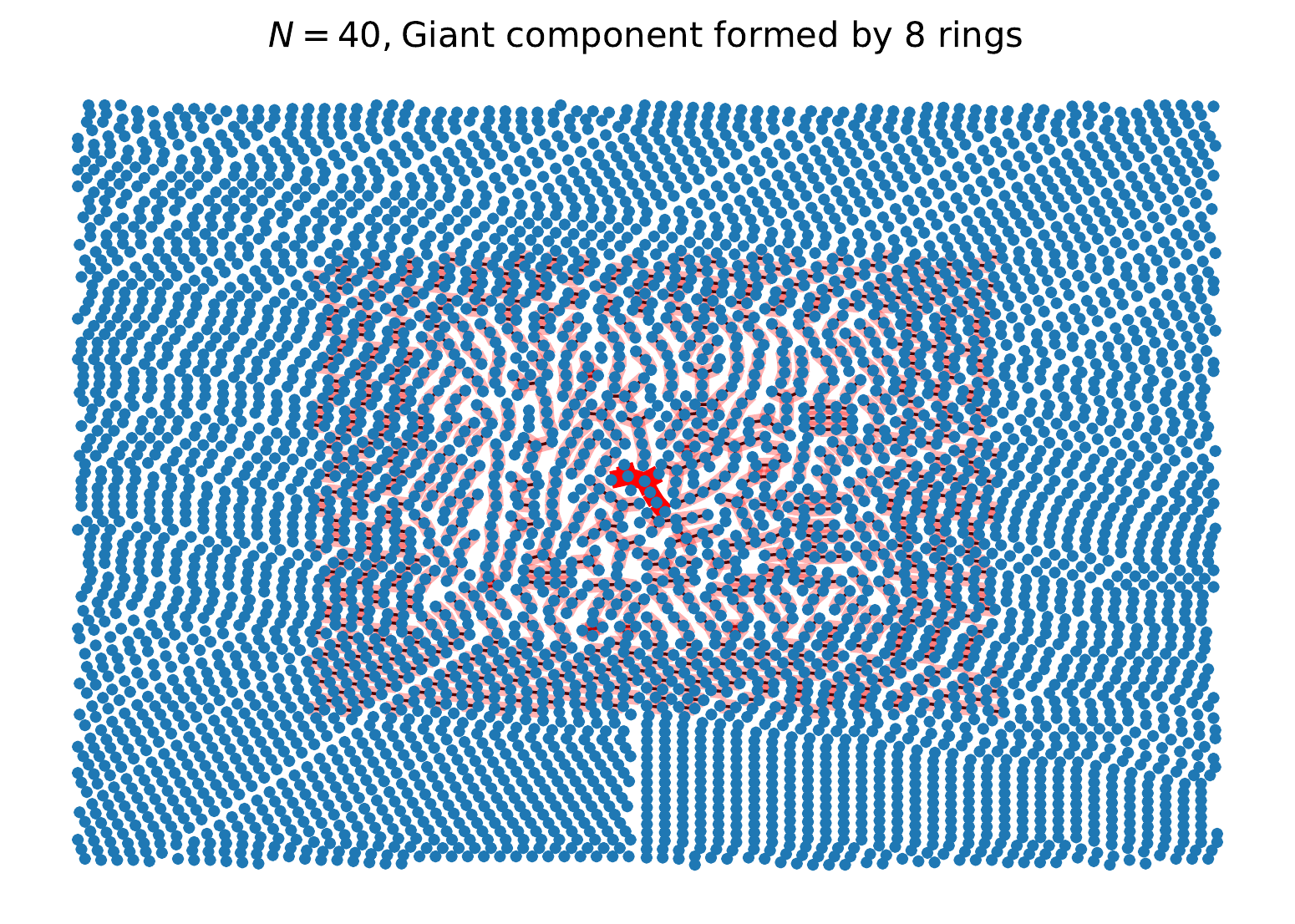} & \includegraphics[width=0.45\textwidth]{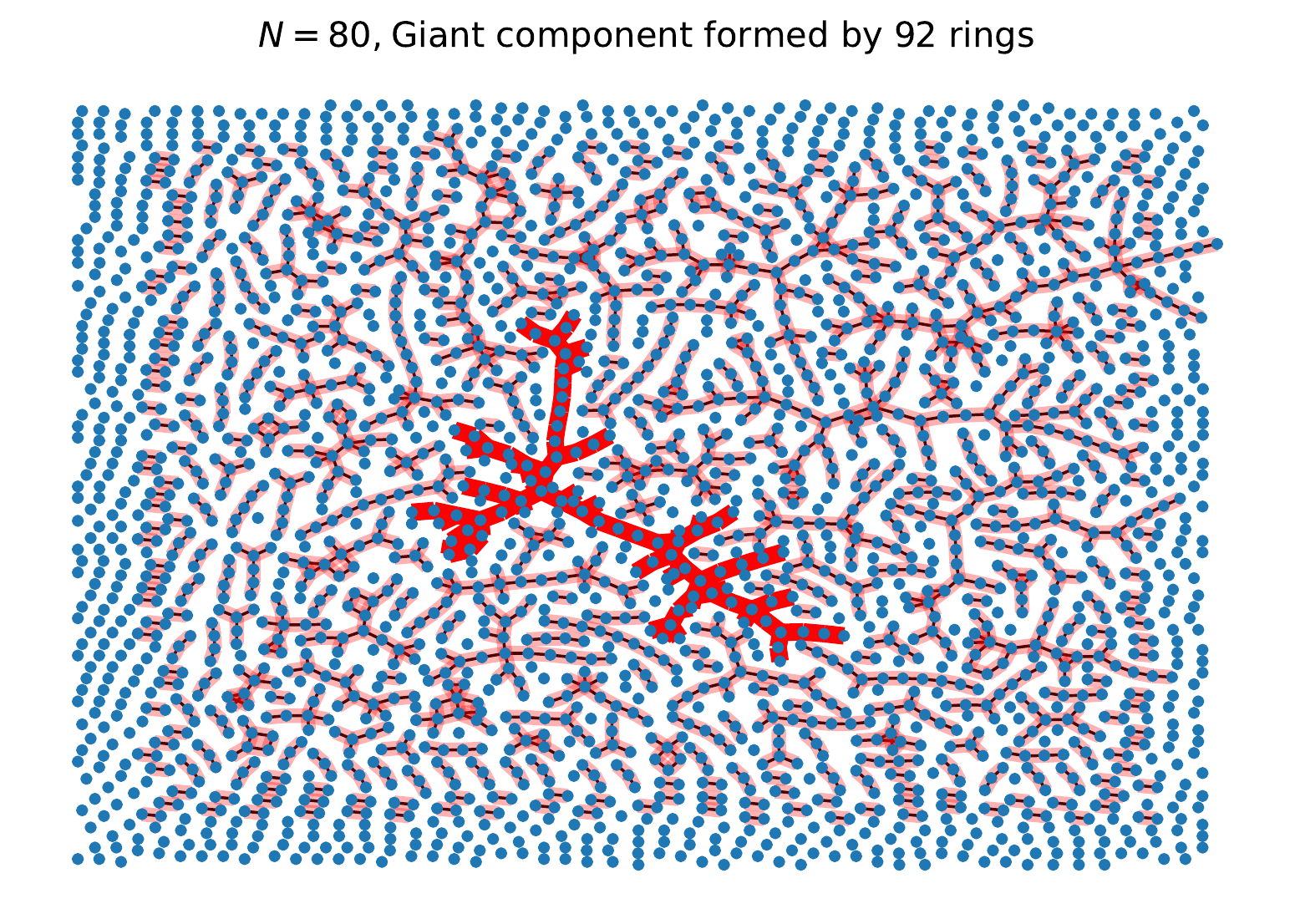} \\
\includegraphics[width=0.45\textwidth]{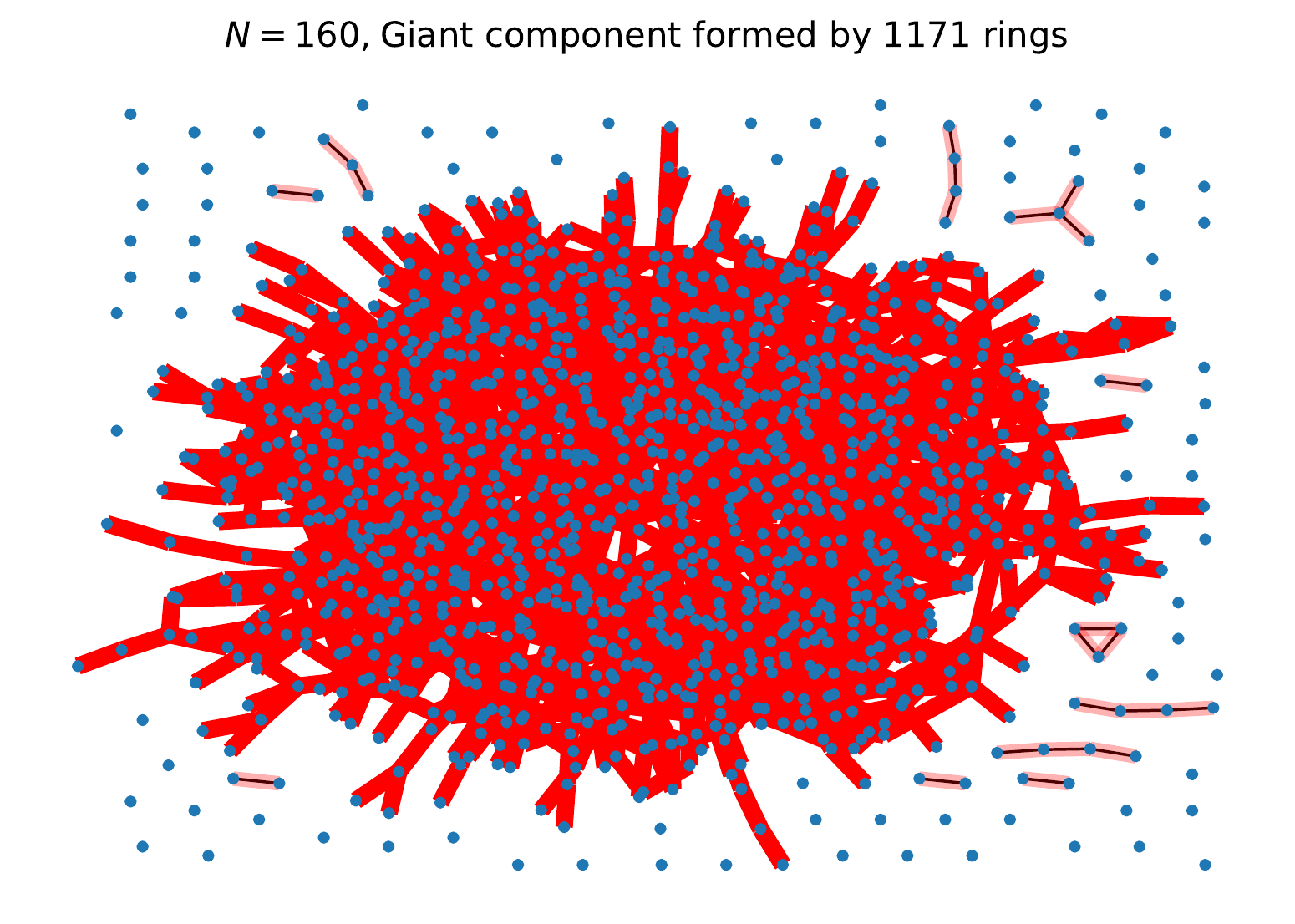} & \includegraphics[width=0.45\textwidth]{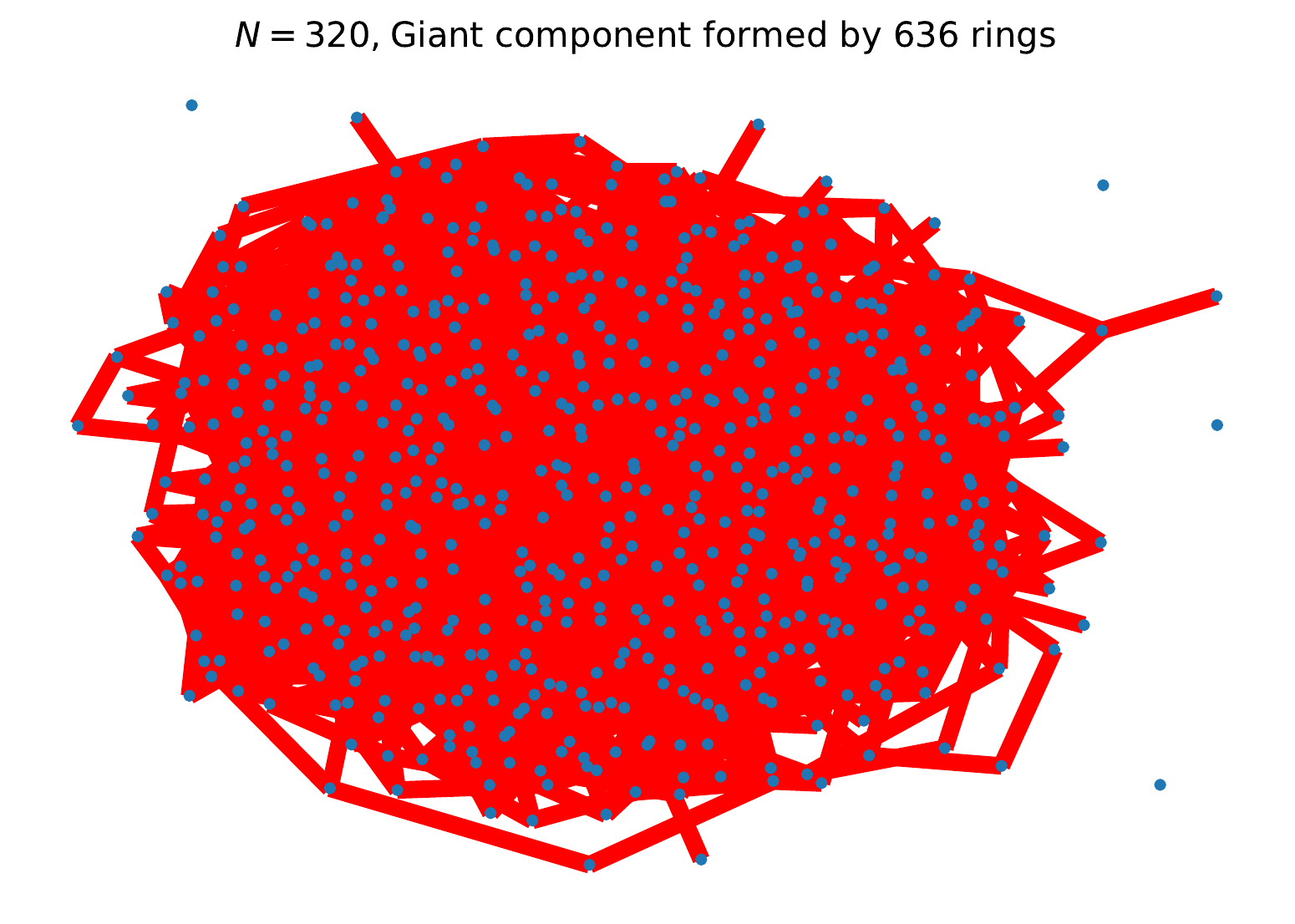} \\
\multicolumn{2}{c}{\includegraphics[width=0.45\textwidth]{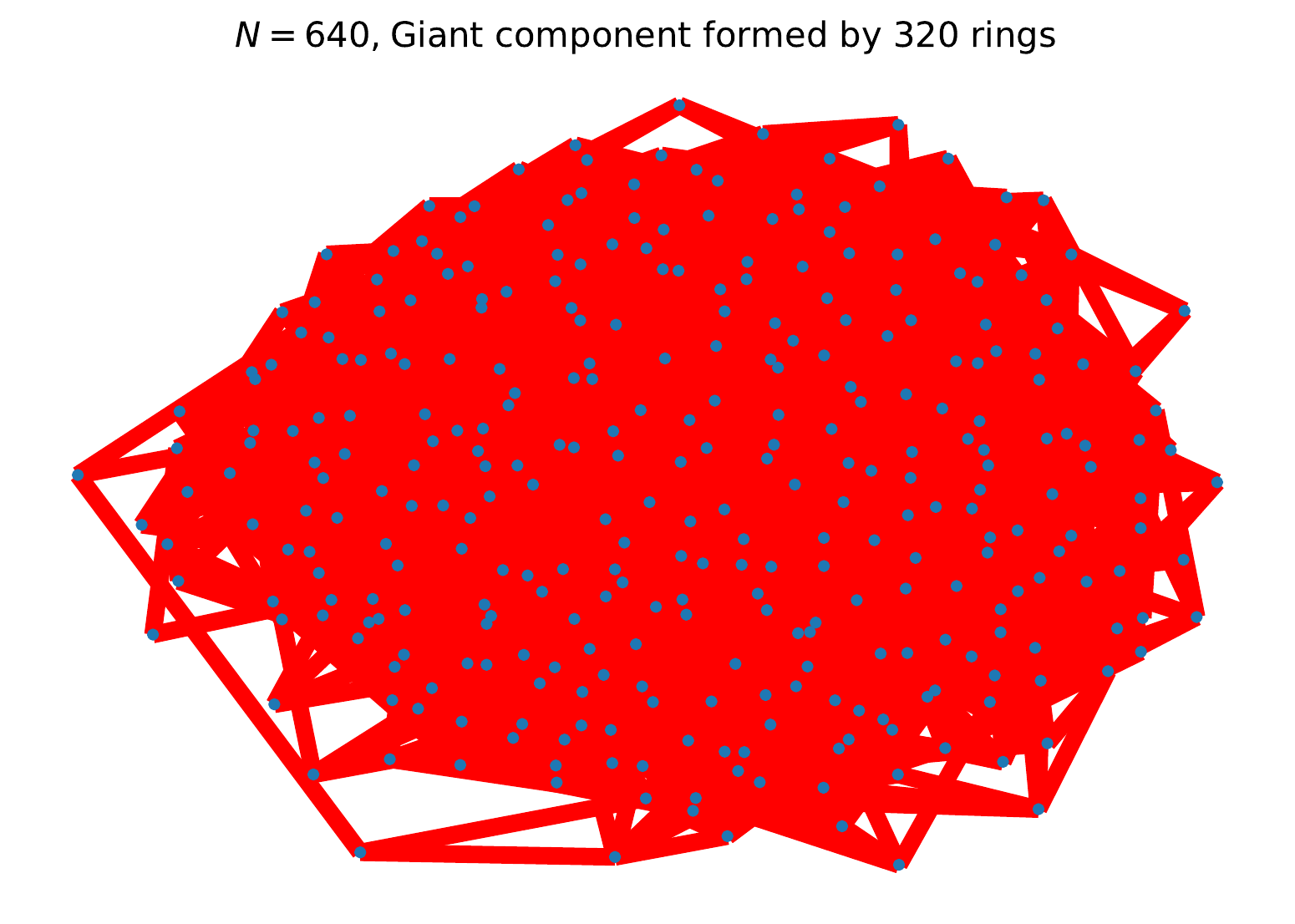}}
\end{array}
$$
\caption{
Schematic illustrations of network structures of the system of rings with dynamical topology~\cite{ClusterNote}.
In each plot, any blue dot represents a ring of the system, rings which are linked are connected through the drawn light red curves.
The thicker red curve represents the biggest cluster in the network.
Any picture represents a single snapshot taken from the trajectories of the system of rings with dynamical topology.
}
\label{fig:Representation_network}
\end{figure*}
\begin{figure*}[p!]
$$
\begin{array}{cc}
\includegraphics[width=0.45\textwidth]{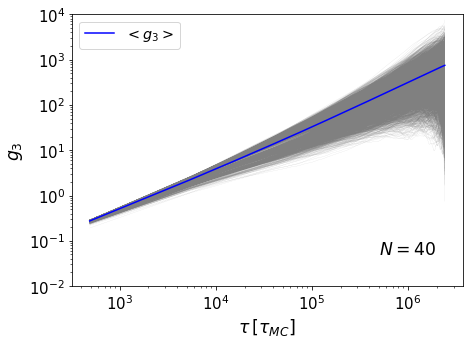} & \includegraphics[width=0.45\textwidth]{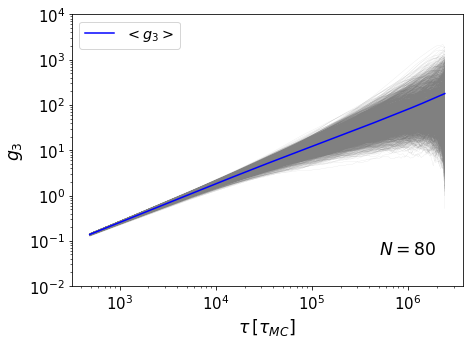} \\
\includegraphics[width=0.45\textwidth]{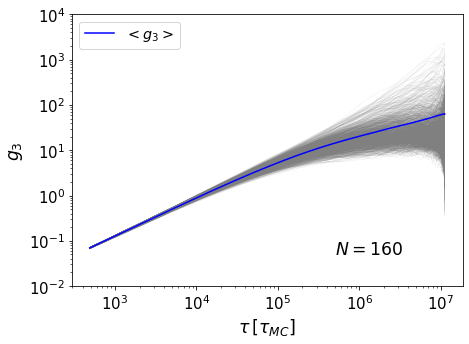} & \includegraphics[width=0.45\textwidth]{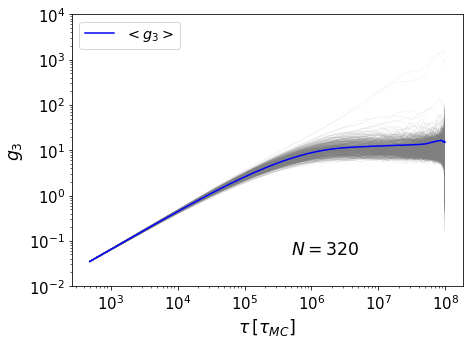} \\
\multicolumn{2}{c}{\includegraphics[width=0.45\textwidth]{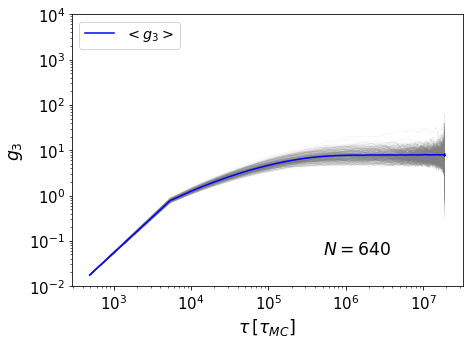}}
\end{array}
$$
\caption{
Details for the time mean-square displacement of the chain centre of masses, $g_3(\tau)$, in melts of permanently catenated rings.
Gray lines are for displacements of single rings, the blue line is the average over the entire ensemble of rings.
}
\label{fig:g3-single-quenched}
\end{figure*}
\begin{figure*}[p!]
$$
\begin{array}{cc}
\includegraphics[width=0.45\textwidth]{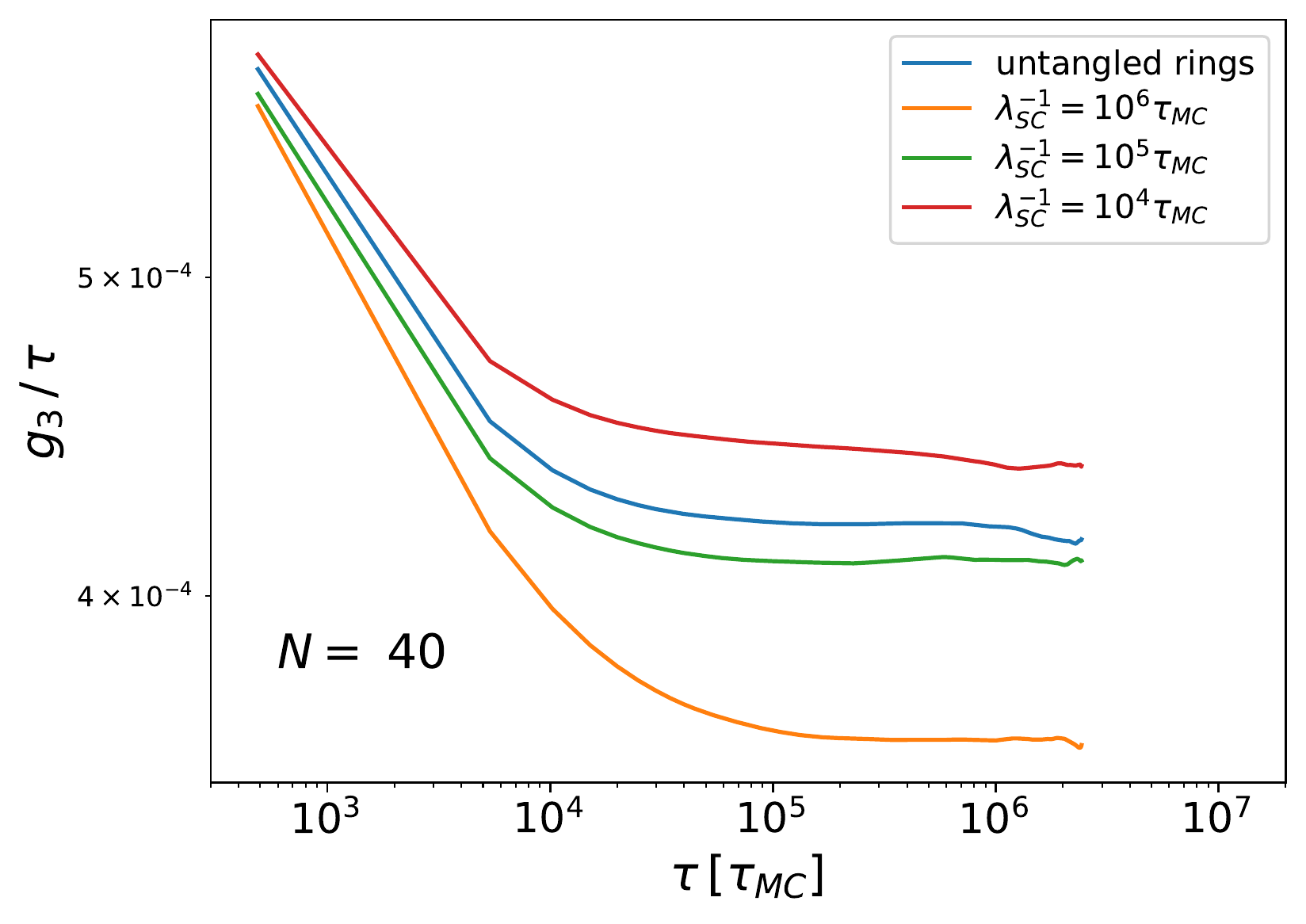} & \includegraphics[width=0.45\textwidth]{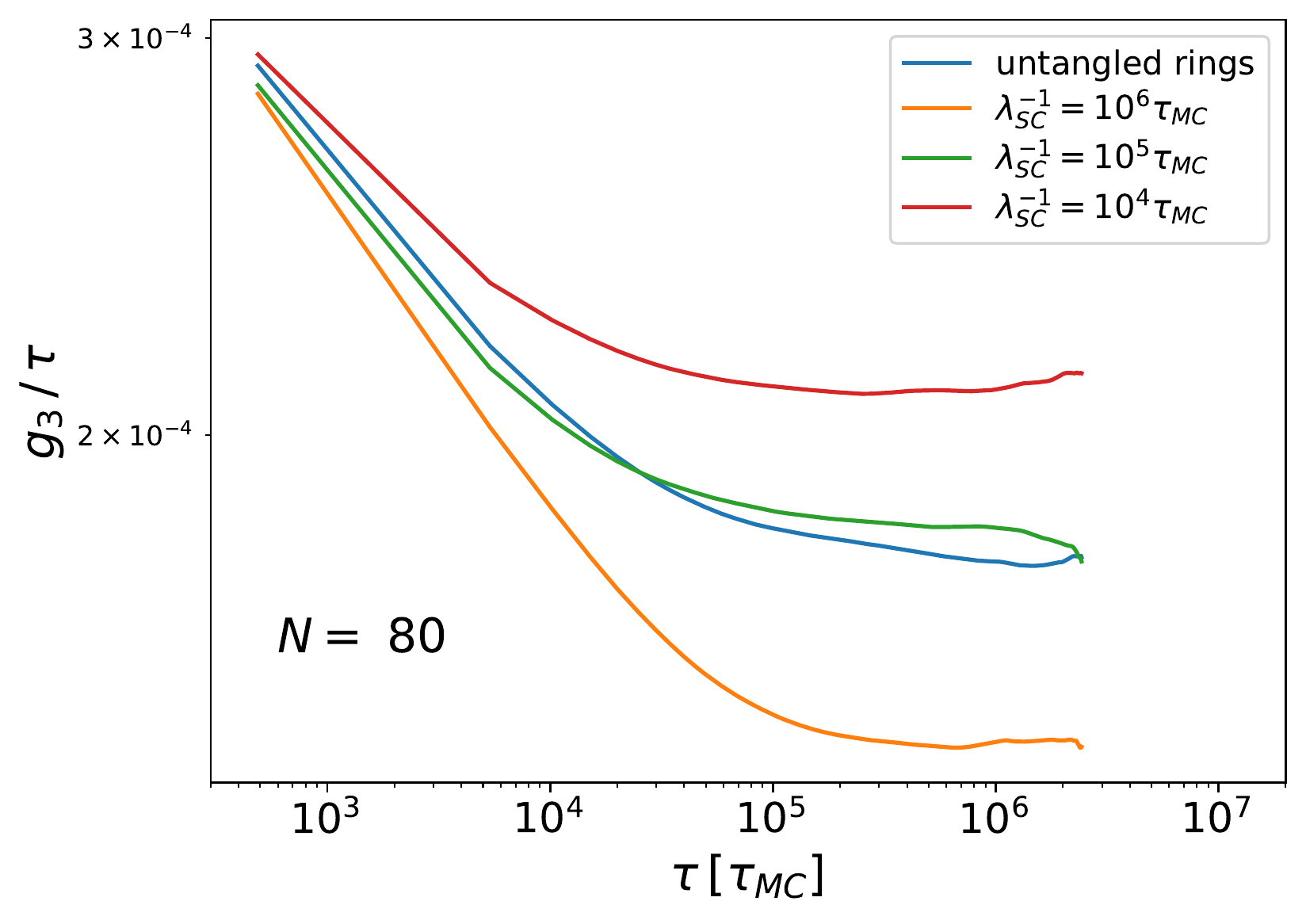} \\
\includegraphics[width=0.45\textwidth]{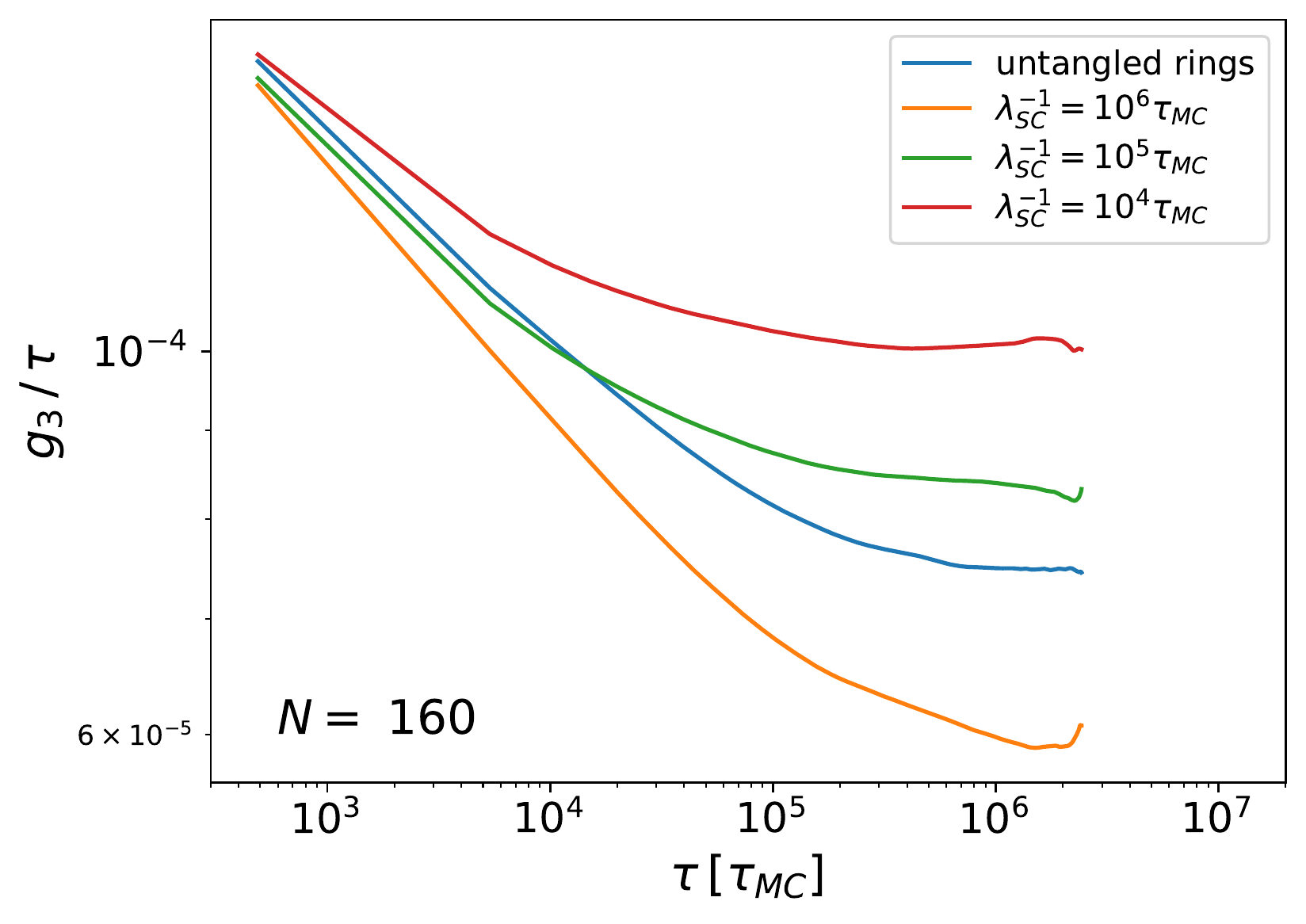} & \includegraphics[width=0.45\textwidth]{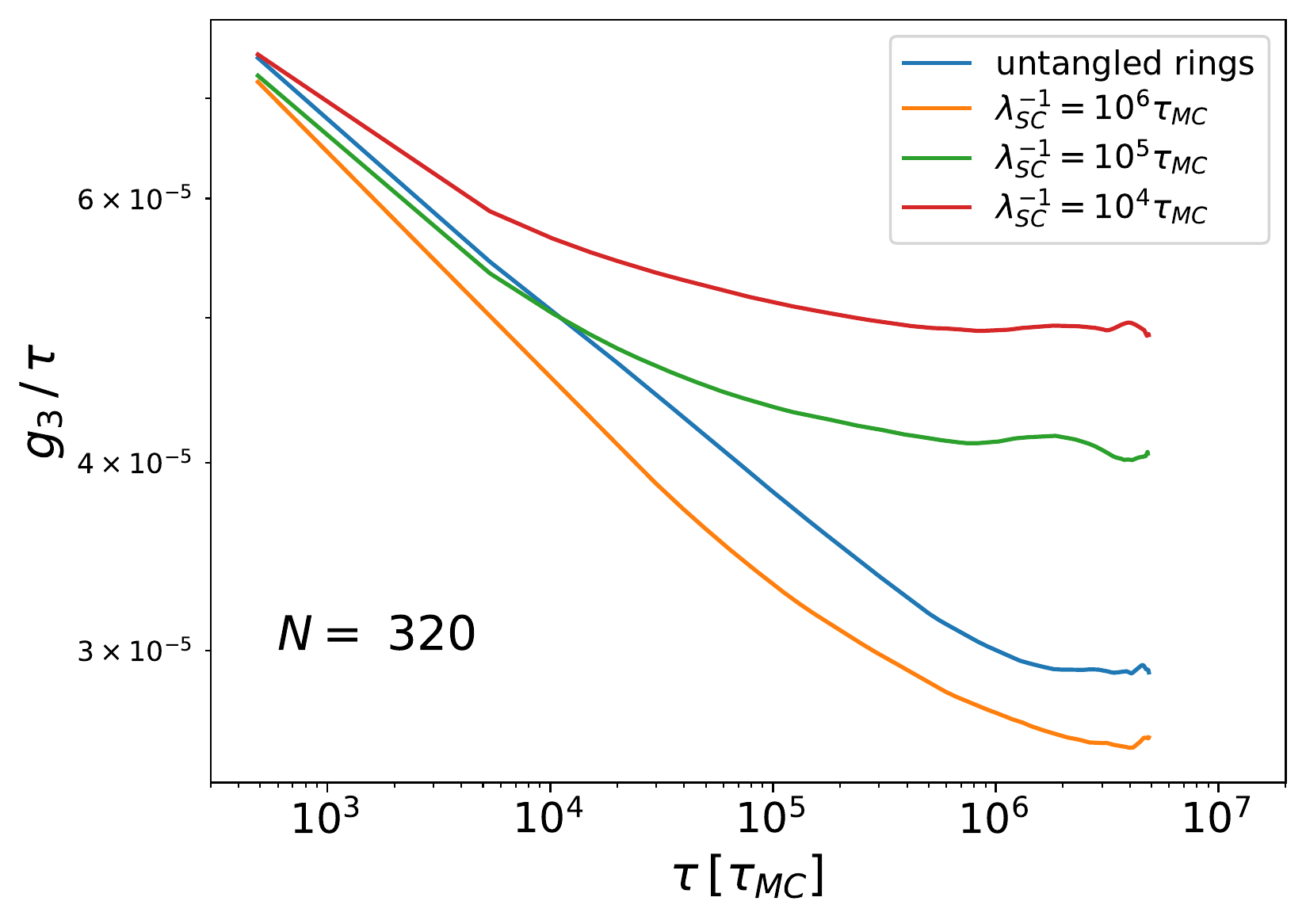} \\
\multicolumn{2}{c}{\includegraphics[width=0.45\textwidth]{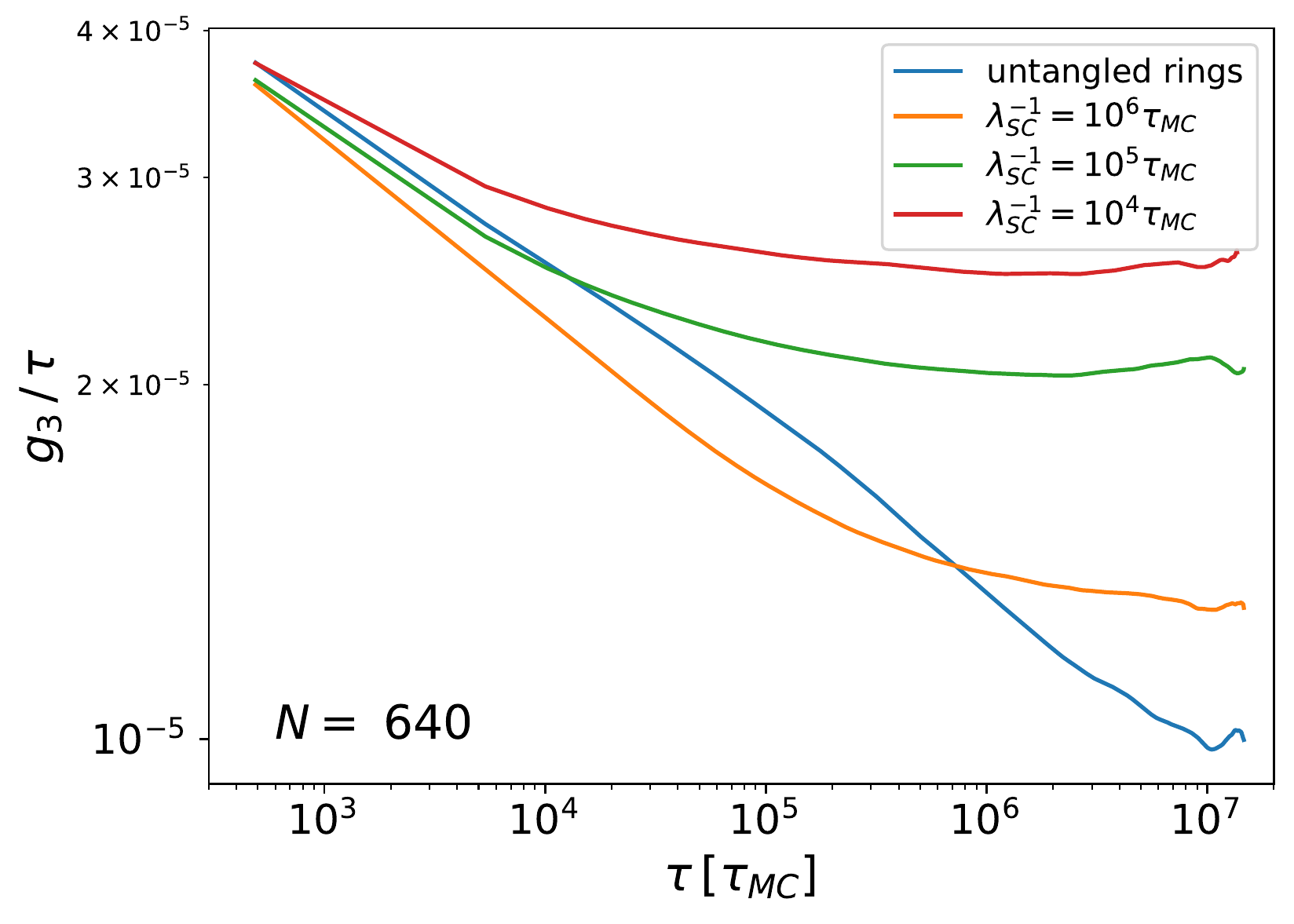}}
\end{array}
$$
\caption{
Time mean-square displacement of the centre of mass of $N$-monomer rings, $g_3(\tau) / \tau$, normalized to the time $\tau$.
Results for different SC rates $\lambda_{\rm SC}$ (see legends) are compared to melts of untangled rings. 
}
\label{fig:comparison_rate_main}
\end{figure*}
\begin{figure*}[p!]
\includegraphics[width=0.45\textwidth]{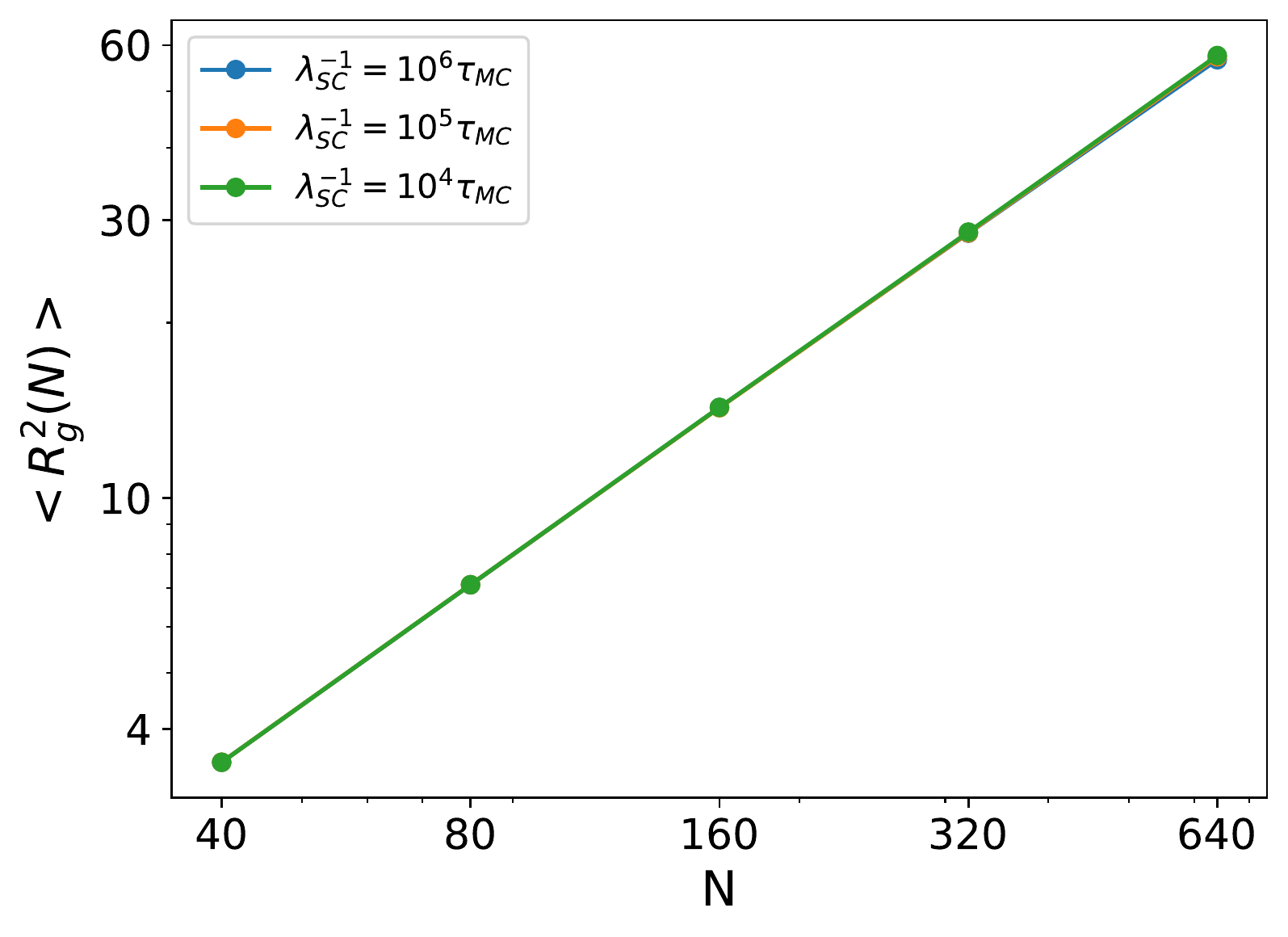} 
\caption{
Mean-square gyration radius, $\langle R_g^2(N) \rangle$, of ring polymers as a function of the number of bonds, $N$, and for different SC rates $\lambda_{\rm SC}$ (see legend).
}
\label{fig:Comparison_rates_GR}
\end{figure*}
\begin{figure*}[p!]
$$
\begin{array}{cc}
\includegraphics[width=0.45\textwidth]{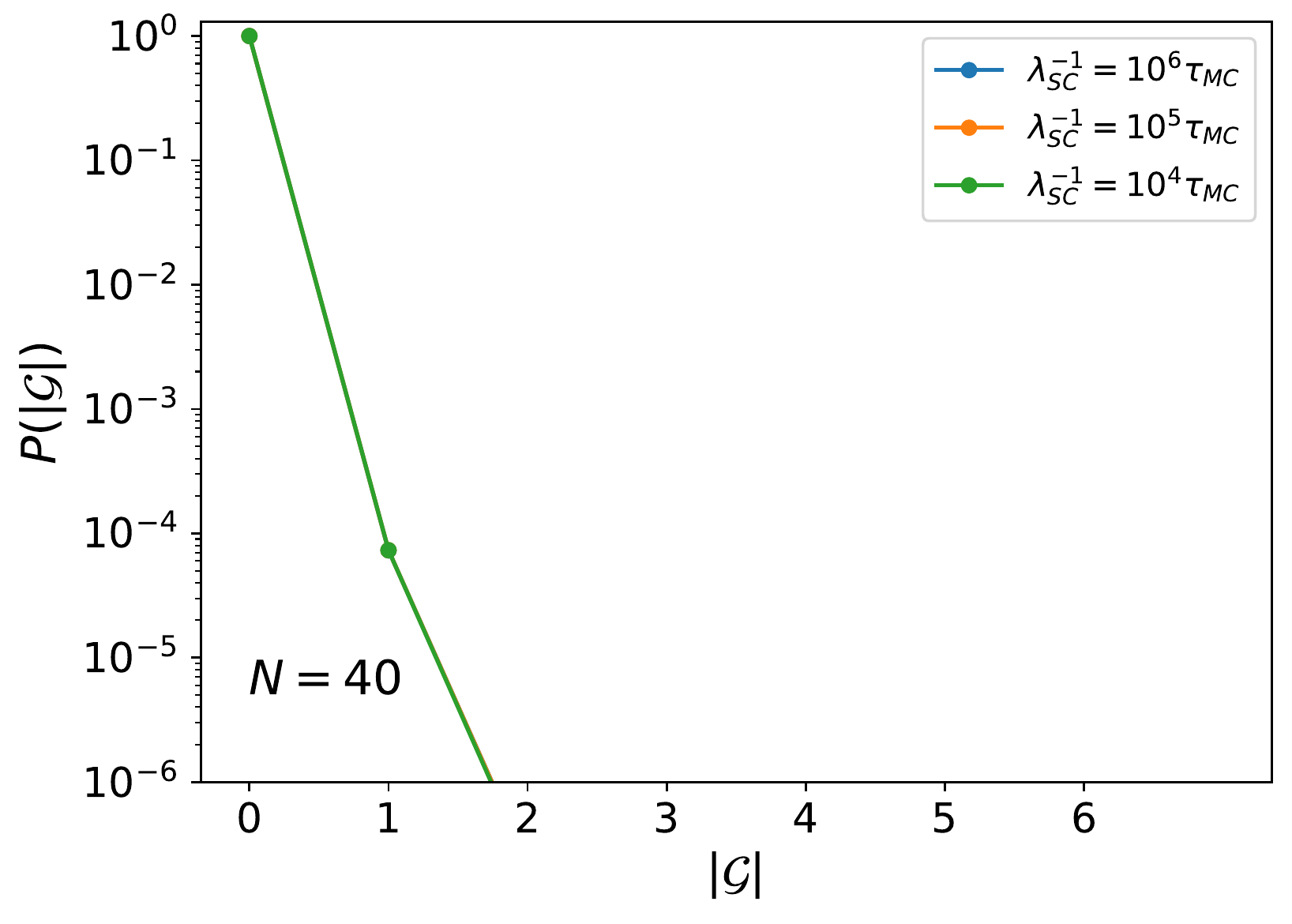} & \includegraphics[width=0.45\textwidth]{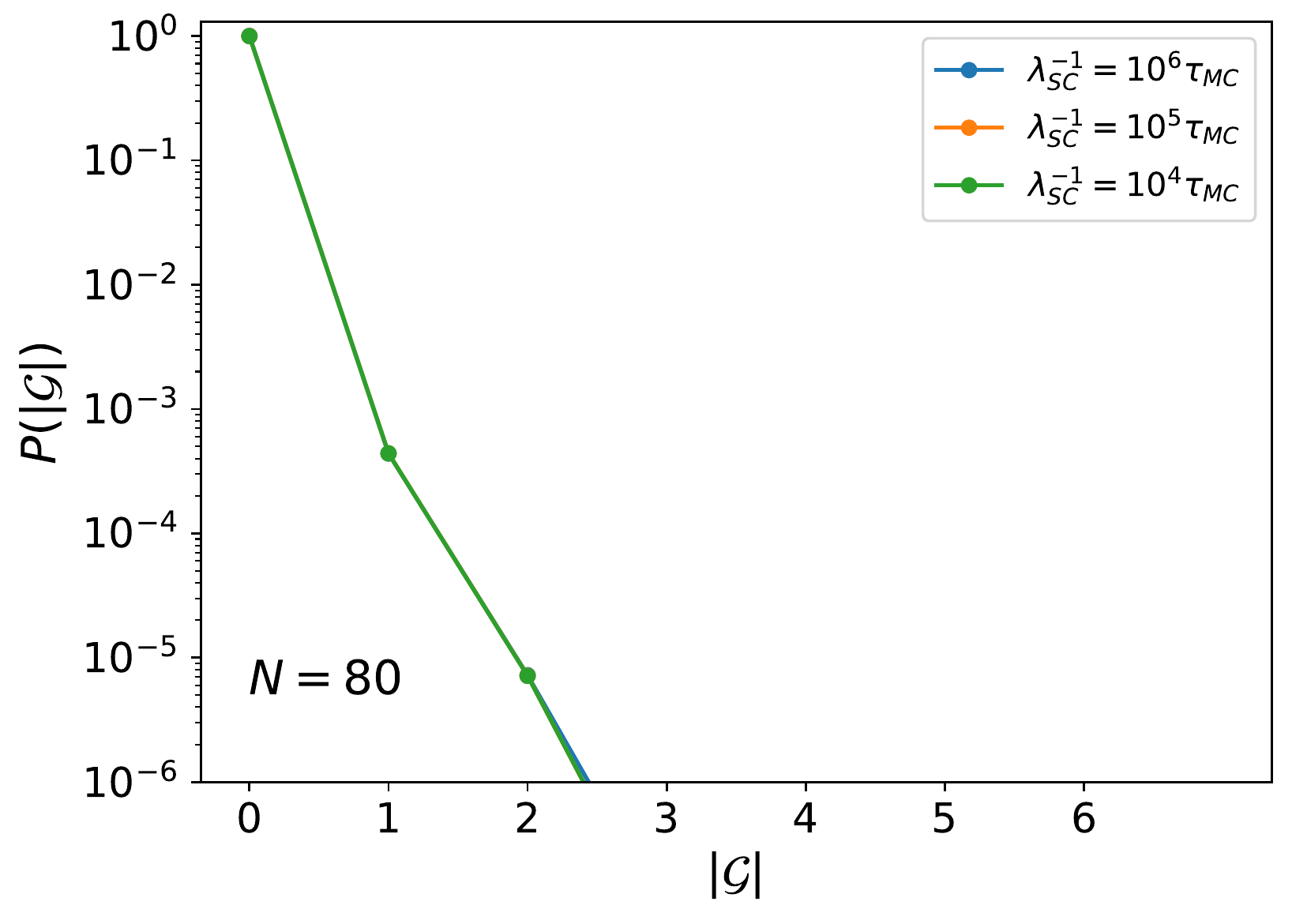} \\
\includegraphics[width=0.45\textwidth]{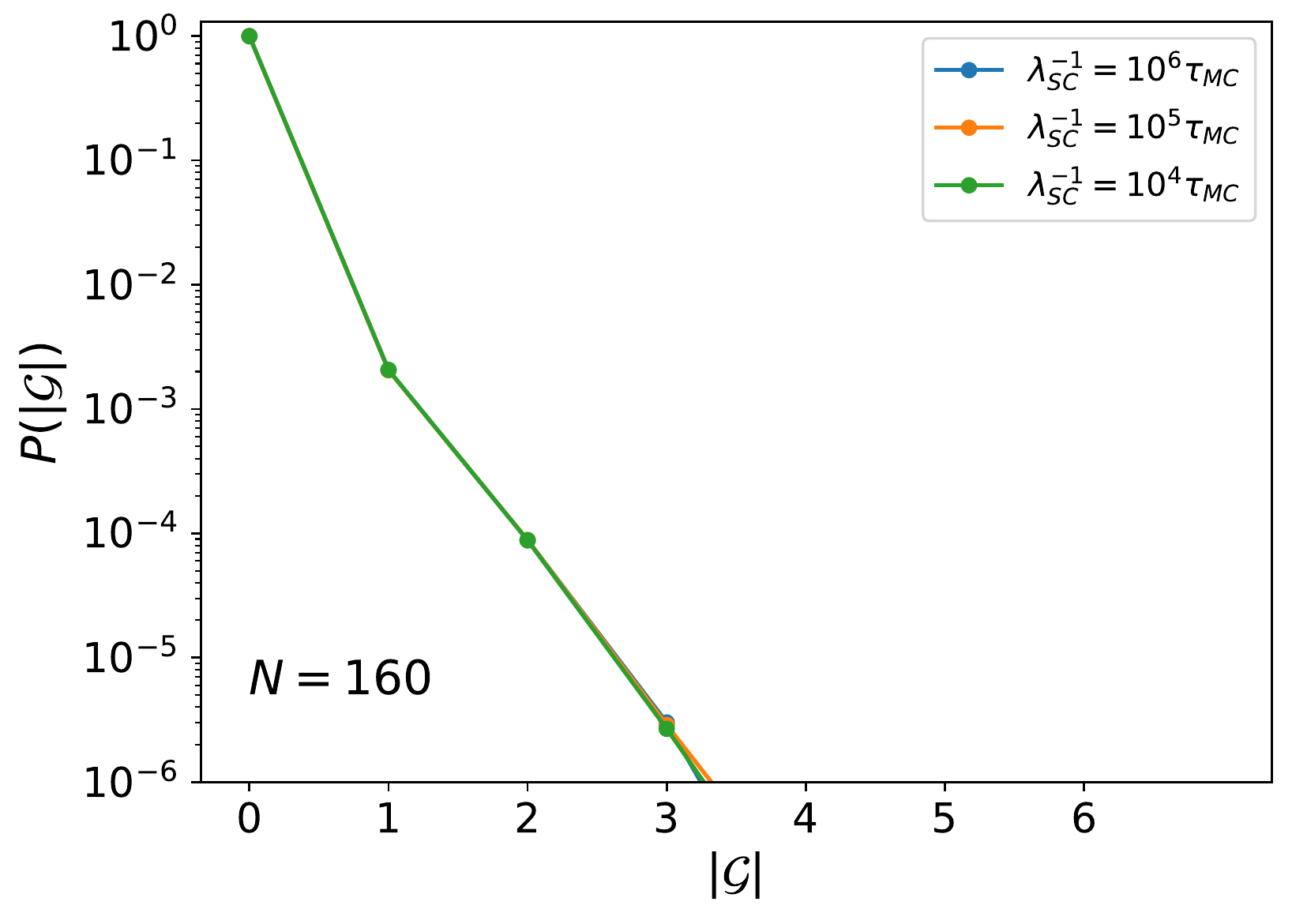} & \includegraphics[width=0.45\textwidth]{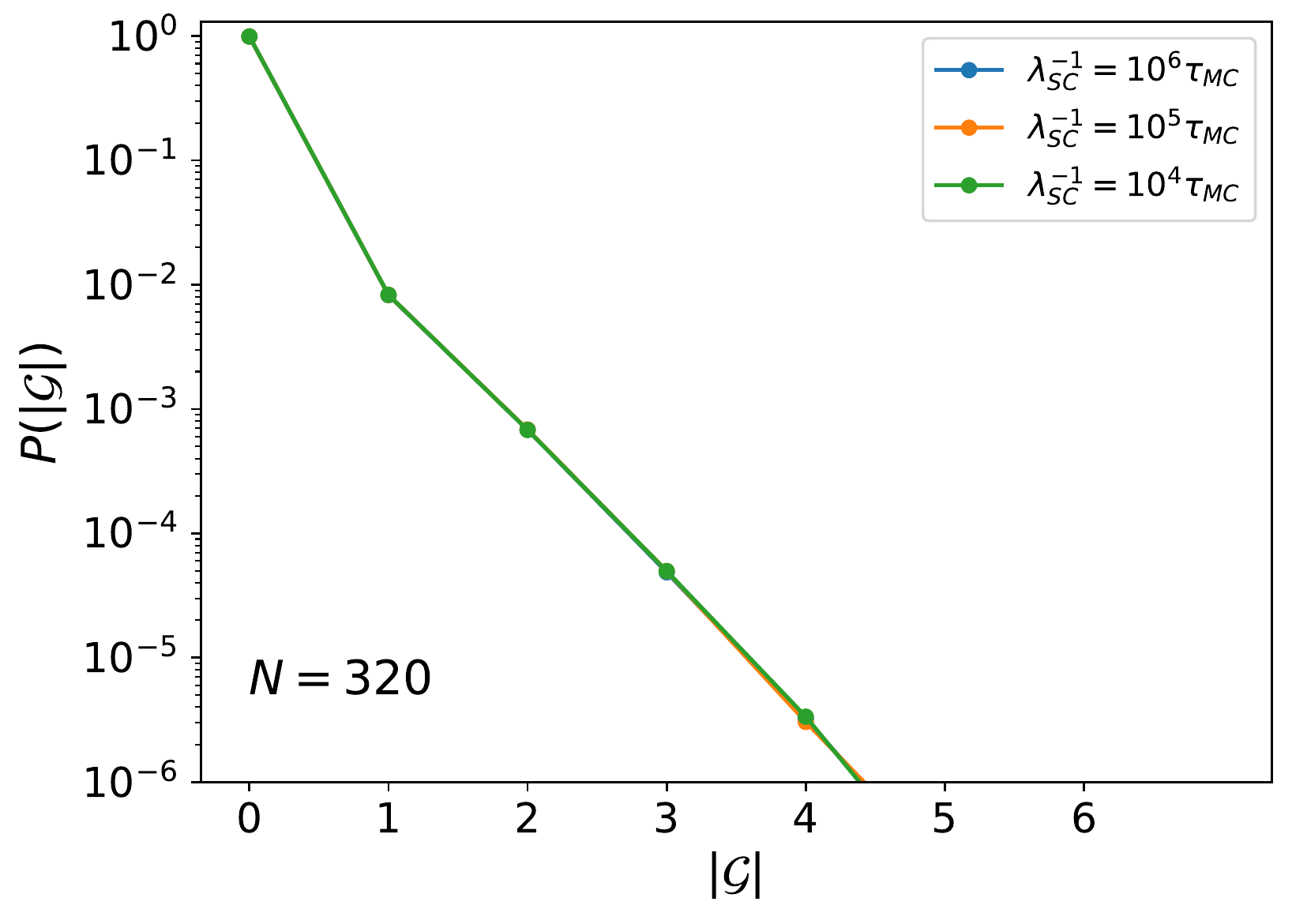} \\
\multicolumn{2}{c}{\includegraphics[width=0.45\textwidth]{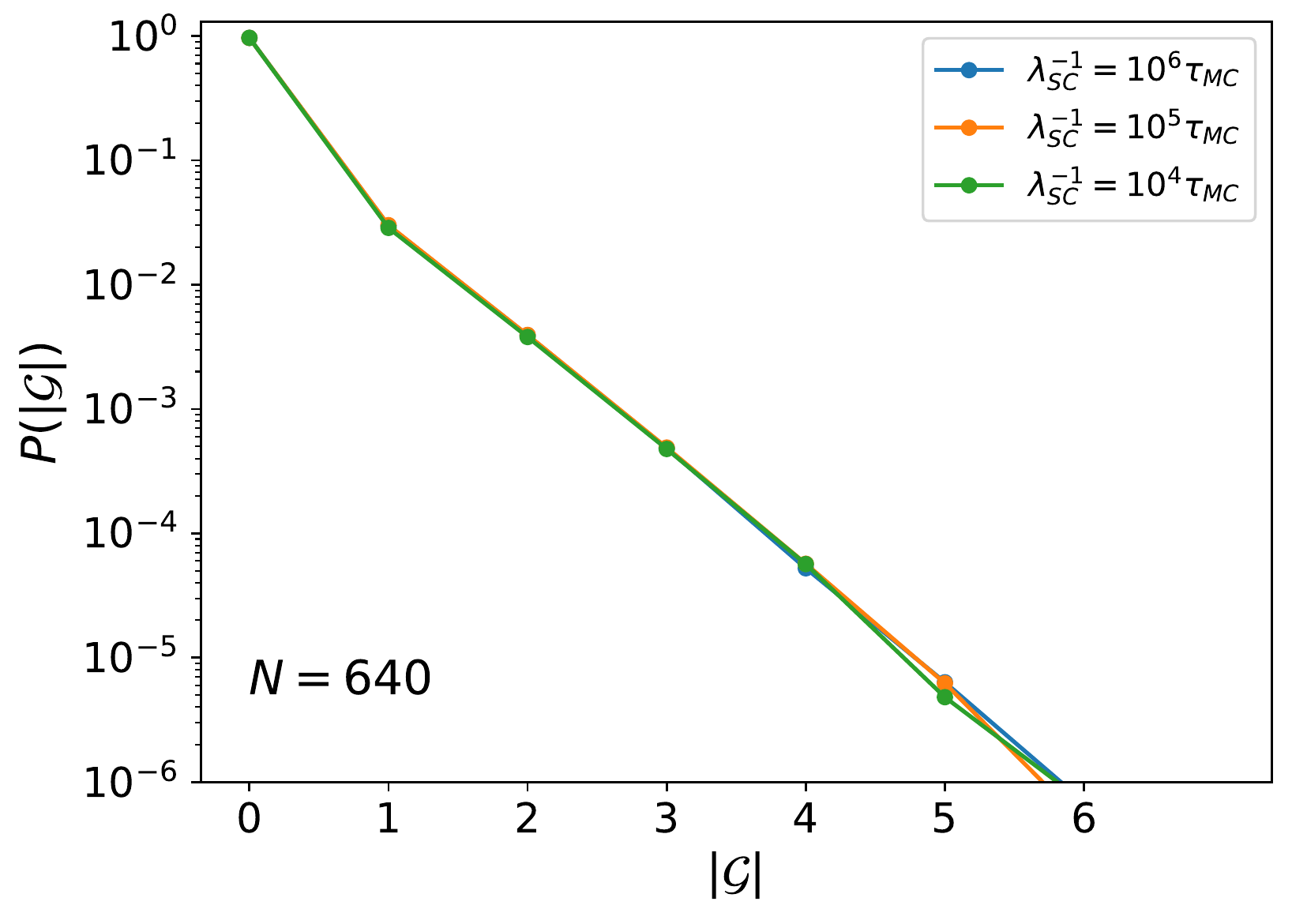}} \\
\end{array}
$$
\caption{
Probability distribution functions of the absolute Gauss linking number, $|\mathcal{G}|$, for different ring sizes, $N$, and for different SC rates $\lambda_{\rm SC}$ (see legends).
}
\label{fig:Comparison_rates_P_G}
\end{figure*}
\begin{figure*}[p!]
$$
\begin{array}{c}
\includegraphics[width=0.45\textwidth]{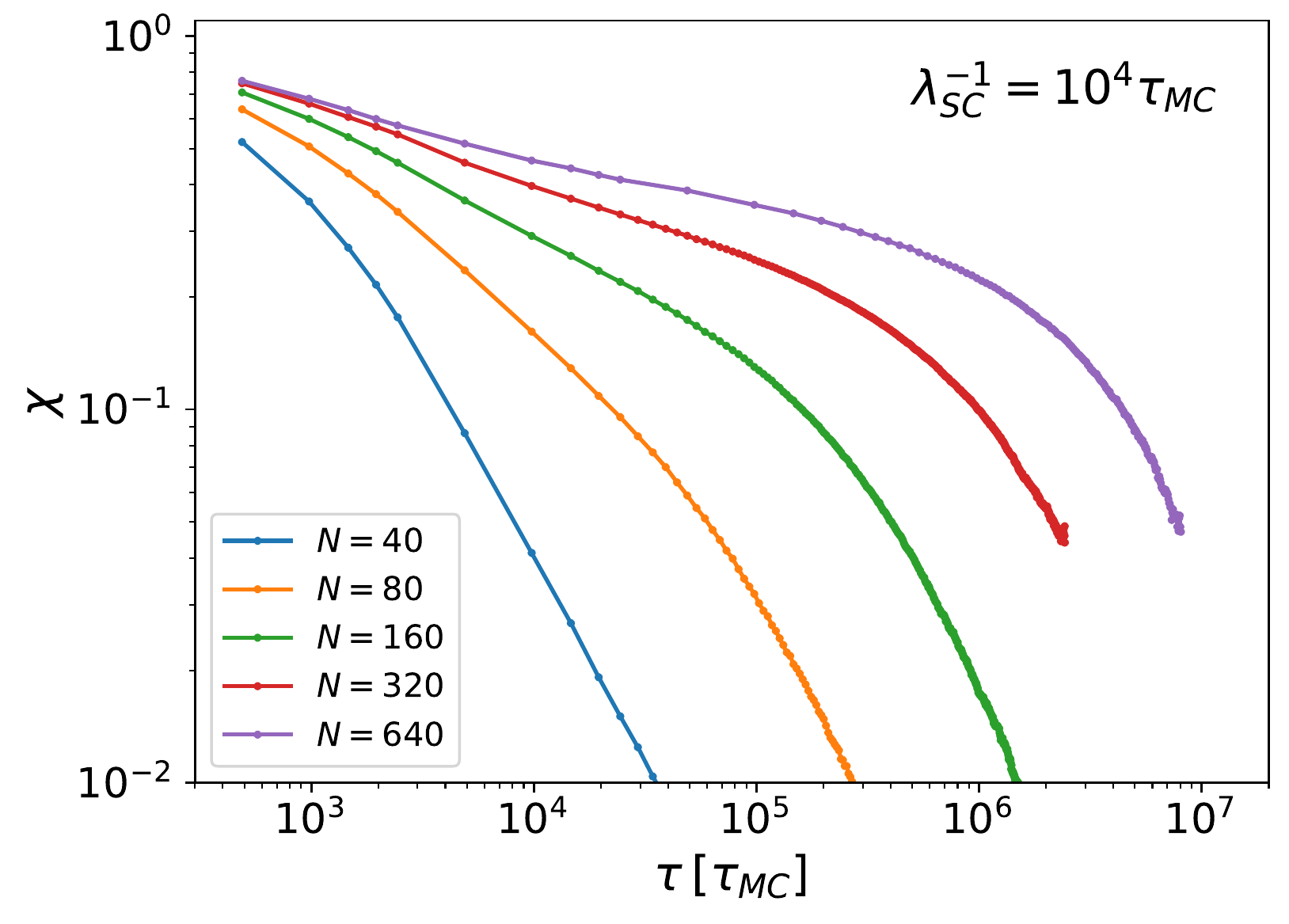} \\
\includegraphics[width=0.45\textwidth]{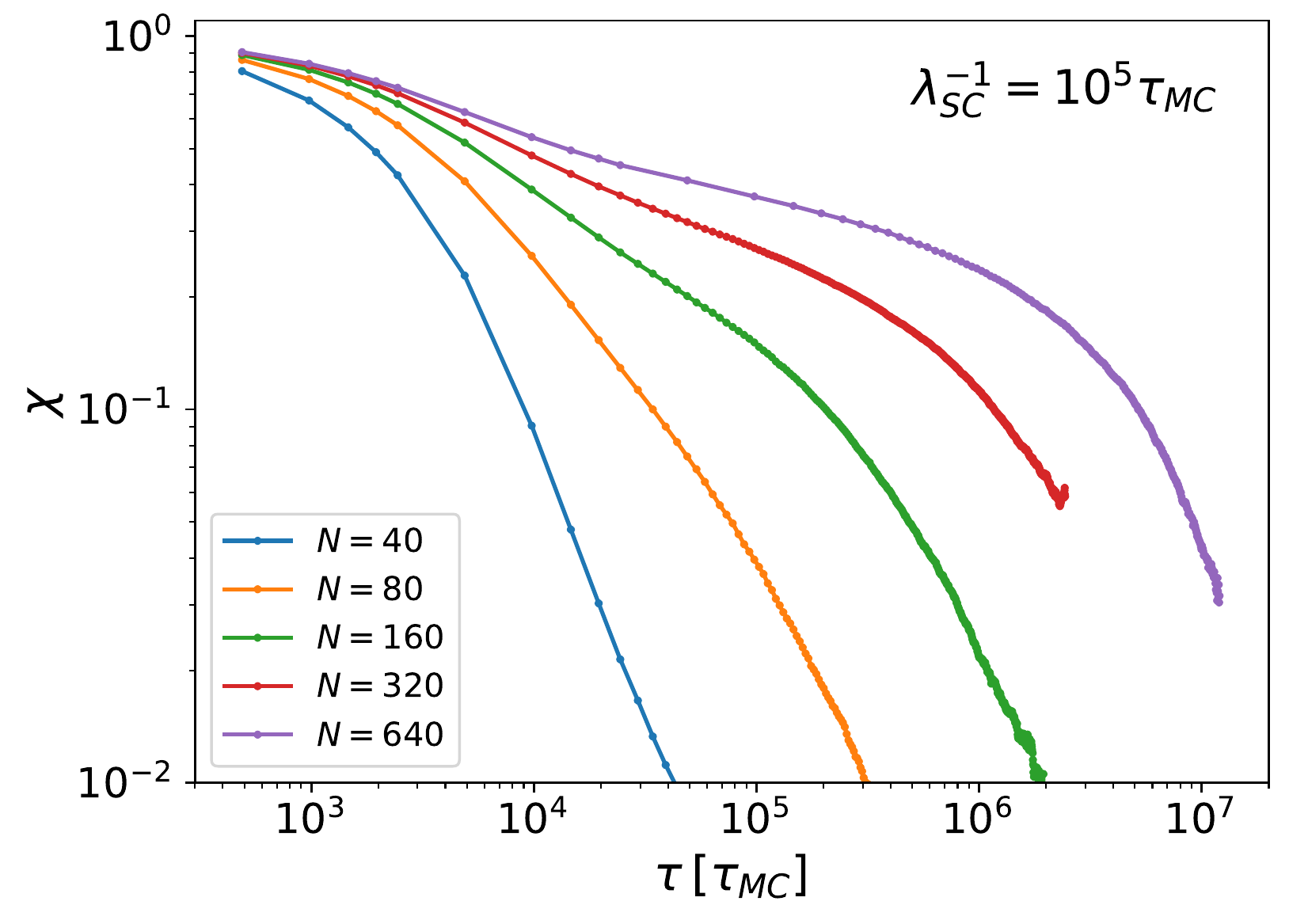} \\
\includegraphics[width=0.45\textwidth]{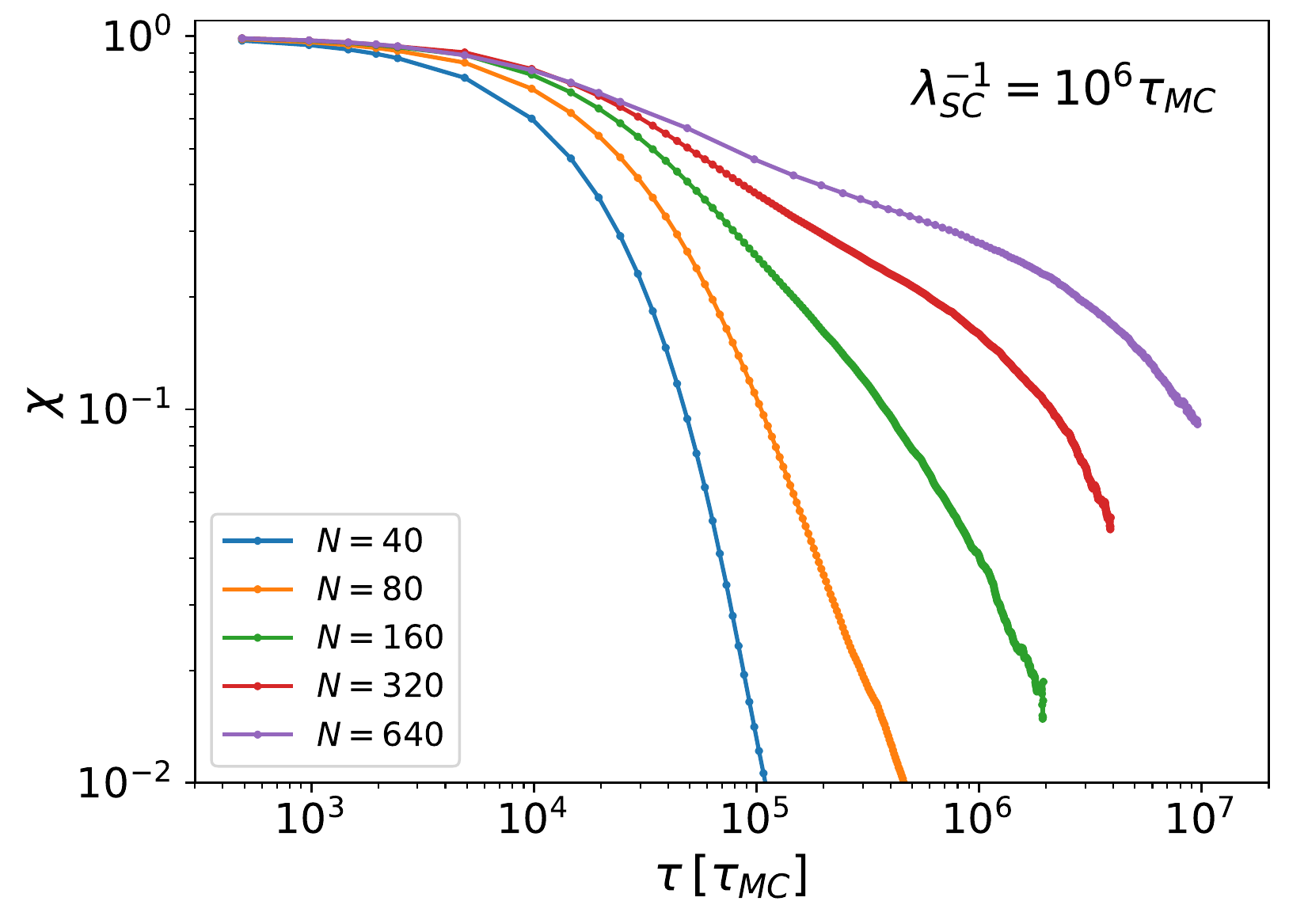} 
\end{array}
$$
\caption{
Time correlation function $\chi(\tau)$ for two rings remaining linked on time span $=\tau$ in the presence of SC's (Eq.~\eqref{eq:LinkLink-TimeCF} in the main text).
Results for:
(top) $\lambda_{\rm SC}^{-1} = 10^4 \, \tau_{\rm MC}$,
(middle) $\lambda_{\rm SC}^{-1} = 10^5 \, \tau_{\rm MC}$,
(bottom) $\lambda_{\rm SC}^{-1} = 10^6 \, \tau_{\rm MC}$.
}
\label{fig:ChiTau-DifferentLambdas}
\end{figure*}
%

\end{document}